\documentclass[a4paper,11pt]{article}
\pdfoutput=1 
\usepackage{physics}

\usepackage{jheppub} 
\usepackage[T1]{fontenc} 
\usepackage{slashed}
\newcommand{\be}{\begin{equation}}
\newcommand{\ee}{\end{equation}}

\usepackage{ucs}
\usepackage{hyperref}
\usepackage[utf8x]{inputenc}
\usepackage{amsmath,bm}
\usepackage{amsfonts}
\usepackage{comment}
\usepackage{amssymb}
\usepackage{array}
\usepackage{epsfig}
\usepackage{slashed}
\usepackage{lipsum}
\usepackage{hyperref}
\hypersetup{
	colorlinks=true,
	linkcolor=blue,
	filecolor=red,
	urlcolor=blue,
	citecolor=red
}

\usepackage{float}
\usepackage[T1]{fontenc} 

\usepackage{tikz}
\usetikzlibrary{shapes,positioning,matrix,fit,arrows}
\tikzset{every picture/.style={/utils/exec={\sffamily}}}
\usetikzlibrary{shapes.geometric, arrows}
\tikzstyle{startstop} = [rectangle, rounded corners, minimum width=3cm, minimum height=1cm,text centered, draw=black, fill=white!30]
\tikzstyle{io} = [trapezium, trapezium left angle=70, trapezium right angle=110, minimum width=3cm, minimum height=1cm, text centered, draw=black, fill=blue!30]
\tikzstyle{process} = [rectangle, minimum width=3cm, minimum height=1cm, text centered, draw=black, fill=white!30]
\tikzstyle{decision} = [diamond, minimum width=3cm, minimum height=1cm, text centered, draw=black, fill=green!30]
\tikzstyle{arrow} = [thick,->,>=stealth]
\tikzstyle{block} = [rectangle, draw, fill=blue!20, text width=15em, text centered, rounded corners, minimum height=4em]

\title{\large Mapping Large N Slightly Broken Higher Spin (SBHS) theory correlators to Free theory correlators}

\author{Prabhav Jain,}
\author{Sachin Jain,}
\author{Bibhut Sahoo,}
\author{Dhruva K.S,}
\author{Aashna Zade}

\affiliation{Indian Institute of Science Education and Research,\\ Dr Homi Bhabha Road, Pashan, Pune, India}

\emailAdd{prabhav.jain@students.iiserpune.ac.in}
\emailAdd{sachin@iiserpune.ac.in}
\emailAdd{bibhut.chandansahoo@students.iiserpune.ac.in}
\emailAdd{k.s.dhruva@students.iiserpune.ac.in}
\emailAdd{aashna.anilzade@students.iiserpune.ac.in}

\abstract{We develop a systematic method to constrain any n-point correlation function of spinning operators in Large N Slightly Broken Higher Spin (SBHS) theories. As an illustration of the methodology, we work out the three point functions which reproduce the previously known results. We then work out the four point functions of spinning operators. We show that the correlation functions of spinning operators in the interacting SBHS theory take  a remarkably simple form and that they can be written just in terms of the free fermionic and critical bosonic theory correlators. They also interpolate nicely between the results in these two theories. When expressed in spinor-helicity variables we obtain an anyonic phase which nicely interpolates between the  free fermionic and critical bosonic results which makes 3D bosonization manifest. Further, we also obtain a form for five and higher point functions as well by performing a similar analysis.}

\begin{document} 
\maketitle
\flushbottom

\section{Introduction}
\label{sec:intro}
Bootstrapping is an important idea which enables us to compute physical quantities of interest based on symmetries and general considerations without requiring much of the microscopic details of the theory. Recently, the conformal bootstrap programme \cite{Rychkov:2016iqz,Poland:2022qrs,Hartman:2022zik} and S-matrix bootstrap \cite{Kruczenski:2022lot} have played a significant role in our understanding of conformal field theory as well as quantum field theories in general. The conformal bootstrap programme has been developed mainly in position space  and in Mellin space \cite{Gopakumar:2016wkt}. The S-matrix bootstrap, on the other hand is naturally developed in momentum space. These two programmes have been developed independently without much overlap. One possible way to bridge this gap is by studying conformal field theory in momentum space. Momentum space CFT analysis is important as it  plays an important role in the context of cosmological correlation functions \cite{Maldacena:2011nz}, in condensed matter physics applications, its connection to one higher dimensional amplitudes in curved space and the flat space S-matrix \cite{Maldacena:2011nz,Penedones:2010ue,Raju:2012zr}. Even though it is important to understand CFT correlation functions in momentum space, there exist very limited results. \cite{Maldacena:2011nz,Coriano:2013jba,Bzowski:2013sza,Bzowski:2015pba,Bzowski:2017poo,Bzowski:2018fql,Kundu:2014gxa,Pajer:2020wxk,Gillioz:2019lgs,Isono:2019ihz,Baumann:2019oyu,Jain:2020rmw,Jain:2020puw,Jain:2021wyn,Jain:2021vrv,Caron-Huot:2021kjy,Jain:2021gwa,Gandhi:2021gwn}.

One of the reasons why conformal field theory in momentum space has got very limited attention is its technical difficulty even at the level of the three point functions. One can also ask, what new things might one learn by thinking in Fourier space? As has been established in recent developments, at least in three dimensions, momentum space analysis has already led to interesting new insights \footnote{Interestingly, momentum space CFT correlators have their own life which cannot be understood as a Fourier transform of position space CFT correlators. For example, in momentum space and in spinor helicity variables, one can obtain a larger class of CFT correlation functions which are not consistent with the position space OPE analysis. However, they play a significant role in connection with cosmological correlation functions in the $\alpha$ vaccum \cite{Jain:2022uja} and its connection to scattering amplitudes in one higher dimension \cite{Jain:2022ujj}.} into the structure of CFT correlators \cite{Farrow:2018yni,Caron-Huot:2021kjy,Jain:2021vrv,Jain:2021qcl,Jain:2021gwa,Gandhi:2021gwn,Jain:2021whr}. In an early work \cite{Giombi:2011rz,Maldacena:2012sf} by performing a position space analysis, it was shown that the three point functions of conserved currents in three dimensions generally have three structures which are  the free bosonic, the free fermionic and a parity odd part which can't be obtained from the free theories. However momentum space analysis revealed that the parity odd part can be obtained by a simple transformation \cite{Jain:2021gwa} \footnote{The epsilon transform \cite{Caron-Huot:2021kjy,Jain:2021gwa} maps the parity even part of the correlation function to the parity odd part and viceversa.} of the parity even part. Further it was shown that all three different structures can be constructed \footnote{This statement is true inside the triangle $s_i+s_j\ge s_k$ for $i,j,k$ taking any of the values $1,2,3$. Outside the triangle, we do require at least two structures, the free bosonic and free fermionic structures.} just from the free bosonic or just the free fermionic theory results \cite{Jain:2021whr}. Even though the computation of the three point functions has seen some progress, very limited results exist for four point functions. There has been a lack of systematic analysis of four point CFT correlation functions \footnote{See\cite{Bzowski:2019kwd} for some recent development for the four point function of scalar operators in momentum space.} and all the more less for spinning ones. Any development of the four point functions would be very useful. In this paper we consider a particular class of CFTs, the slightly broken higher spin theories \cite{Maldacena:2012sf,Giombi:2011kc,Aharony:2011jz}. We constrain the form of spinning four point functions and also show that we can extend the analysis to the five point level and beyond using momentum space or spinor helicity considerations. We show that the momentum space considerations are particularly useful in the context of theories with slightly broken higher spin symmetries. 

Examples of slightly broken higher spin theories are given by Chern-Simons matter theories at large N. Chern-Simons gauge field coupled to  matter in the fundamental representation has been the subject of intense research in the recent past \cite{Sezgin:2002rt, Klebanov:2002ja, Giombi:2009wh,
  Benini:2011mf, Giombi:2011kc, Aharony:2011jz, Maldacena:2011jn,
  Maldacena:2012sf, Chang:2012kt, Jain:2012qi, Aharony:2012nh,
  Yokoyama:2012fa, GurAri:2012is, Aharony:2012ns, Jain:2013py,
  Takimi:2013zca, Jain:2013gza, Yokoyama:2013pxa, 
  Jain:2014nza,  Gurucharan:2014cva,Bardeen:2014qua, Bardeen:2014paa, Dandekar:2014era,
  Frishman:2014cma, Moshe:2014bja, Aharony:2015pla, Inbasekar:2015tsa,
  Bedhotiya:2015uga,  Minwalla:2015sca, Gur-Ari:2015pca,
  Radicevic:2015yla, Geracie:2015drf, Aharony:2015mjs,
 Radicevic:2016wqn, Karch:2016sxi,   Hsin:2016blu, Yokoyama:2016sbx, Gur-Ari:2016xff, Murugan:2016zal,
  Seiberg:2016gmd, Giombi:2016ejx,
  Karch:2016aux, Giombi:2016zwa, Wadia:2016zpd, Aharony:2016jvv,
  Giombi:2017rhm,  Charan:2017jyc, Benini:2017dus, Sezgin:2017jgm, Nosaka:2017ohr,
  Komargodski:2017keh, Giombi:2017txg, Gaiotto:2017tne,
  Jensen:2017dso, Jensen:2017xbs, Inbasekar:2020hla, Gomis:2017ixy, Inbasekar:2017ieo,
  Inbasekar:2017sqp, Cordova:2017vab, Benini:2017aed,
  Aitken:2017nfd, Jensen:2017bjo, Karch:2018mer, Chattopadhyay:2018wkp,
  Turiaci:2018nua, Choudhury:2018iwf,  Aharony:2018npf,
  Yacoby:2018yvy, Aitken:2018cvh, Aharony:2018pjn, Dey:2018ykx,
  Chattopadhyay:2019lpr, Dey:2019ihe, Halder:2019foo, Aharony:2019mbc,
  Li:2019twz, Jain:2019fja, Inbasekar:2019wdw, Inbasekar:2019azv,
  Jensen:2019mga, Kalloor:2019xjb, Ghosh:2019sqf, 
  Jain:2020rmw, Minwalla:2020ysu, Jain:2020puw, Mishra:2020wos,Minwalla:2022sef,Jain:2022izp}. These are interesting models as several large-N exact computations can be performed, \footnote{There has been an exact computation of the partition function, 2-2 S-matrix, some three point and four point correlation functions of operators etc..., please see \cite{Sezgin:2002rt, Klebanov:2002ja, Giombi:2009wh,
  Benini:2011mf, Giombi:2011kc, Aharony:2011jz, Maldacena:2011jn,
  Maldacena:2012sf, Chang:2012kt, Jain:2012qi, Aharony:2012nh,
  Yokoyama:2012fa, GurAri:2012is, Aharony:2012ns, Jain:2013py,
  Takimi:2013zca, Jain:2013gza, Yokoyama:2013pxa, 
  Jain:2014nza,  Gurucharan:2014cva,Bardeen:2014qua, Bardeen:2014paa, Dandekar:2014era,
  Frishman:2014cma, Moshe:2014bja, Aharony:2015pla, Inbasekar:2015tsa,
  Bedhotiya:2015uga,  Minwalla:2015sca, Gur-Ari:2015pca,
  Radicevic:2015yla, Geracie:2015drf, Aharony:2015mjs,
 Radicevic:2016wqn, Karch:2016sxi,   Hsin:2016blu, Yokoyama:2016sbx, Gur-Ari:2016xff, Murugan:2016zal,
  Seiberg:2016gmd, Giombi:2016ejx,
  Karch:2016aux, Giombi:2016zwa, Wadia:2016zpd, Aharony:2016jvv,
  Giombi:2017rhm,  Charan:2017jyc, Benini:2017dus, Sezgin:2017jgm, Nosaka:2017ohr,
  Komargodski:2017keh, Giombi:2017txg, Gaiotto:2017tne,
  Jensen:2017dso, Jensen:2017xbs, Inbasekar:2020hla, Gomis:2017ixy, Inbasekar:2017ieo,
  Inbasekar:2017sqp, Cordova:2017vab, Benini:2017aed,
  Aitken:2017nfd, Jensen:2017bjo, Karch:2018mer, Chattopadhyay:2018wkp,
  Turiaci:2018nua, Choudhury:2018iwf,  Aharony:2018npf,
  Yacoby:2018yvy, Aitken:2018cvh, Aharony:2018pjn, Dey:2018ykx,
  Chattopadhyay:2019lpr, Dey:2019ihe, Halder:2019foo, Aharony:2019mbc,
  Li:2019twz, Jain:2019fja, Inbasekar:2019wdw, Inbasekar:2019azv,
  Jensen:2019mga, Kalloor:2019xjb, Ghosh:2019sqf, 
  Jain:2020rmw, Minwalla:2020ysu, Jain:2020puw, Mishra:2020wos,Minwalla:2022sef,Jain:2022izp} and references therein.}they show strong weak field theory/field theory duality which are called Bose-Fermi dualities. These theories also provide a concrete example of non super symmetric Gauge/Gravity duality. Several exact computation suggest that these theories may be integrable. In the context of correlation functions of spinning operators, direct perturbative computations have been done for a few three and four point functions in a special kinemetic regime \cite{Aharony:2012nh,GurAri:2012is,Bedhotiya:2015uga,Turiaci:2018nua,Kalloor:2019xjb}.
Recently, using purely conformal field theory arguments it was shown that in spinor helicity variables that the three point functions of spinning operators take on a very interesting form and appear to have an anyonic phase \cite{Gandhi:2021gwn} \footnote{Interestingly, in \cite{Skvortsov:2018uru} it was shown that the parity violating term in Chiral Higher Spin theory correlator appears from a certain EM duality.} which was earlier observed in the context of $2\rightarrow 2$ scattering. Since three point functions are universal and are completely fixed by conformal symmetry, we required hardly any input from the Chern-Simons matter theories\footnote{The only input from the CS matter theory required is the dependence of coupling constants. This leads to an interesting anyonic phase which appeared previously in the calculation of a $2 $ to $2$ scattering amplitude. This anyonic phase appears in spinor helicity variables and reveals anyonic features of the CFT correlation functions. A deeper understanding of the same phenomena would be a interesting future work.}. However at the level of the four point functions, any such results would require a lot more information from the specific theory at hand \footnote{See \cite{Gandhi:2021gwn} for a very naive bootstrap analysis using momentum space analysis which indicates that the four point function as well takes a very simple form in the spinor helicity variables.}. At the level of the three point functions, in \cite{Maldacena:2011jn,Maldacena:2012sf} slightly broken HS symmetry was used to constrain the three point functions in these theories.
  Subsequently, in \cite{Li:2019twz,Jain:2020rmw,Jain:2020puw,Silva:2021ece}, four point correlation functions of the form $\langle J_s OOO\rangle$ were explored.
  In this paper, we develop a methodology to solve  slightly broken HS equations  which in principle can be used to solve any n-point function  of spinning operators in terms of the free theory correlators. 

The rest of the paper is organised as follows:\\ In section \ref{sum1}, we briefly describe the theory that we shall be interested in. In section \ref{sum2} we briefly review and summarise the answers that we obtain in this paper. Section \ref{Mapmth} describes the steps that we use in the rest of the paper to solve slightly broken HS theories. In section \ref{3ptfunctions}, we demonstrate our methodology with the help of several examples of three point functions and reproduce known results. In section \ref{4ptex} we solve the four point functions. In section \ref{WtFr1} we describe how the Ward-Takahashi identity for slightly broken HS theory can be understood in terms of the free theory Ward-Takahashi identities. This also suggests that slightly broken HS algebra can be understood in terms of the free theory HS algebra. We explore this in one of the appendices. In section \ref{dis} we summarize our results and discuss various future directions. We also have several appendices which are useful for our main text. In appendix \ref{qbtheory}, we briefly discuss another SBHS theory of interest, that is, the quasi bosonic theory. In appendix \ref{HScurrentAlgebraAppendix}, we provide the current algebra and the non conservation equations for the CS matter theories. In appendix \ref{sphident}, we briefly review the spinor helicity formalism for 3d CFTs. In appendix \ref{epstrans}, we discuss the epsilon transformation in spinor helicity variables. In appendix \ref{2pthse}, we show how we can use the SBHS symmetry to compute two point functions. In appendix \ref{naivebtst}, we perform a naive conformal block decomposition which yields expressions for four point functions in the SBHS theory. In appendix \ref{4PointDetails}, we provide the details of solving the HSE for the four point functions. In appendix \ref{5ptappendix}, we attempt to extend our results to the five point case. In appendix \ref{sec:npointappendix}, we present a conjecture for the general form of n point functions in the SBHS theory. Finally, in appendix \ref{SBHSinTermsofFree}, we attempt to write the SBHS algebra in terms of the exactly conserved HS algebra.

\section{CS Matter theories: A short summary}\label{sum1}
Here we will briefly review Chern-Simons matter theories. There are two classes of these theories, namely the Quasi fermion (QF) and the Quasi bosonic theories. In this work, we mainly deal with the QF theory. The details of these theories can be found for example in \cite{Aharony:2018pjn}.

\subsection{Quasi Fermionic theory}
Quasi fermionic theory refers to two different theories, namely CS gauge field coupled to a fermion or CS gauge field coupled to a critical boson. We review this below.
	\subsection*{Fermionic theory coupled to CS field}
	The fermionic theory coupled to $SU(N_f)$ Chern-Simons gauge field has the following action
	\begin{align}
		S=\int d^3x\left[\bar\psi\gamma_\mu D^\mu\psi+i\epsilon^{\mu\nu\rho}\frac{\kappa_f}{4\pi}\text{Tr}(A_\mu\partial_\nu A_\rho-\frac{2i}{3}A_\mu A_\nu A_\rho)\right]    
	\end{align}
	We are interested in the limit as $N_f\rightarrow \infty$ and $\kappa_f\rightarrow \infty$ such that $\lambda_f=\frac{N_f}{\kappa_f}$ is held fixed.
	The spectrum of operators consists of exactly conserved spin 1 and spin 2 currents and in general spin-s currents $J_s$ with scaling dimensions $\Delta_s=s+1+{\mathcal O}(\frac{1}{N_f})$ for $s\ge 3$. The scalar operator has conformal dimension $\Delta=2+\mathcal O\left(\frac {1}{N_f}\right)$ and is parity odd. 
	
	\subsection*{Critical bosonic theory coupled to Chern-Simons theory in $d=3$}
	
	Let us consider the critical bosonic theory coupled to Chern-Simons $SU(N_b)$ gauge field. The critical theory is obtained by adding an interaction of the kind $\sigma_b\bar\phi\phi$ where $\sigma_b$ is an auxiliary field to the free bosonic Lagrangian. The theory has the following action
\begin{align}
\label{CBCSaction}
    S=\int d^3x\left[D_\mu\bar\phi D^\mu\phi+i\epsilon^{\mu\nu\rho}\frac{\kappa_b}{4\pi}\text{Tr}(A_\mu\partial_\nu A_\rho-\frac{2i}{3}A_\mu A_\nu A_\rho)+\sigma_b\bar\phi\phi\right]
\end{align}
 Again, we are interested in the limit as $N_b\rightarrow \infty$ and $\kappa_b\rightarrow \infty$ such that $\lambda_b=\frac{N_b}{\kappa_b}$ is held fixed. The spin-1 and spin-2 conserved currents have scaling dimensions 2 and 3 respectively. Apart from these, there is an infinite tower of slightly broken higher spin currents. The conformal dimension of the spin $s$ current $\Delta=s+1+\mathcal O\left(\frac{1}{N}\right)$. The scalar operator has conformal dimension $\Delta =2+\mathcal{O}(\frac{1}{N})$ and is parity even.
\subsection{Slightly broken Higher spin symmetry}	The free theories have exactly conserved higher spin currents $\partial.J_s=0$ for all s. For CS matter theories, we have currents which are not exactly conserved, that is
$\partial.J_s\neq 0$ for $s>2.$ In this paper we are going to use these symmetries to constrain the form of correlation functions of arbitrary spinning operators following \cite{Maldacena:2011jn,Maldacena:2012sf}. More precisely we are going to use
\begin{align}\label{HSEIT}
\sum_{i=1}^{n}\langle J_{s_1}(x_1) \ldots[Q_{s},J_{s_i}(x_i) ] \ldots J_{s_n}(x_n)\rangle=\int_x\langle \partial.J_s(x)J_{s_1}(x_1)\ldots J_{s_i}(x_i)  \ldots J_{s_n}(x_n)\rangle
\end{align}
where $Q_s$ is the charge associated with current $J_s$ and $\partial.J_s(x)\neq 0$. A more detailed form will be discussed in the subsequent sections. In this paper following \cite{Maldacena:2012sf} we shall use \eqref{HSEIT} to solve for the correlation functions.
\subsection*{Some useful definitions}\label{params}
	The coupling constant in the CS gauge field coupled to a fermion in the limit $N_f\rightarrow \infty, \kappa_f\rightarrow \infty,$ is defined as follows,
	\begin{equation}
		\lambda_f=\frac{N_f}{\kappa_f}
	\end{equation}
We now introduce a few other useful variables which will help simplify our expressions 
\cite{Aharony:2012nh,GurAri:2012is}
\begin{align}\label{cplconst}
	\tilde{N}=2N_f\frac{\sin\pi\lambda_f}{\pi\lambda_f},\quad\tilde{\lambda}=\tan\left(\frac{\pi\lambda_f}{2}\right)
\end{align} 
In the main text we will be working only in the quasi-fermionic theory. In some normalisation we can fix $\lambda_f$ to take values $0\le \lambda_f\le 1.$
We discuss the Quasi Bosonic theory in appendix \ref{qbtheory}
. We will also frequently use the abbreviations listed in the table below:
\begin{table}[h!]
  \begin{center}
    \begin{tabular}{|c|l|}
    \hline    
      \textbf{Abbreviation} & \textbf{Full Form} \\
      \hline
    SBHS & Slightly broken higher spin\\
      \hline
    HSE & Higher spin equation\\
      \hline
    FB & Free Boson\\
      \hline
    FF & Free Fermion\\
      \hline
    CB & Critical Boson\\
      \hline
    CF & Critical Fermion\\
      \hline
      QB & Quasi Boson\\
      \hline
      QF & Quasi Fermion\\
      \hline
    \end{tabular}
    \caption{Abbreviations}
  \end{center}
\end{table}
\section{Summary of results}\label{sum2}
In this section we summarise the results we have obtained in this paper. Let us describe some of the notation that is going to be useful.
\subsection*{Notation}
Here we introduce some notation for the correlators which we will use to state our results.
For any correlator we define
\begin{table}[h!]
  \begin{center}
    \label{tab:tablecorr}
    \begin{tabular}{|c|l|}
    \hline    
      \textbf{Notation} & \textbf{Description} \\
      \hline
      $\langle ...\rangle_\text{QF}$ & In quasi-fermionic theory\\
      \hline
      $\langle ...\rangle_\text{FF/FB}$ & In free fermionic/bosonic theory\\
      \hline
      $\langle ...\rangle_\text{CB}$ & In critical bosonic theory\\
      \hline
      $\langle ...\rangle_{\text{odd}}$ & Parity odd correlator\\
      \hline
      $\langle ...\rangle_\text{FF+FB}$ & $\langle ...\rangle_\text{FF}+\langle ...\rangle_\text{FB}$\\
      \hline
      $\langle ...\rangle_\text{FF-FB}$ & $\langle ...\rangle_\text{FF}-\langle ...\rangle_\text{FB}$\\
      \hline
    \end{tabular}
    \caption{Notation for correlators}
  \end{center}
\end{table}
\subsection*{Epsilon transform}
We will also be using an operation known as the \textit{epsilon transform} very frequently. We denote the epsilon transform \ref{epstrans} of $X$ as $\epsilon\cdot X$\footnote{ 
As an example we write the epsilon transforms of $\langle JJO\rangle$ and $\langle TTT\rangle$ as follows,
\begin{align}
\langle\epsilon\cdot J_{\mu} (p_1)J_{\nu}(p_2)O(p_3) \rangle&=\frac{\epsilon_{\mu a b}\,p_{1a}}{p_1}\langle J_{b} (p_1)J_{\nu}(p_2)O(p_3) \rangle\nonumber\\
\langle\epsilon\cdot T_{\mu \nu}(p_1)T_{\alpha\beta}(p_2)T_{\gamma\theta}(p_3)\rangle&=\frac{\epsilon_{\mu a b}p_{1a}}{p_1}\langle T_{b \nu}(p_1)T_{\alpha\beta}(p_2)T_{\gamma\theta}(p_3)\rangle
\end{align}}. The epsilon transform maps a parity even/odd correlation function to a parity odd/even correlator \cite{Caron-Huot:2021kjy,Jain:2021gwa}.
We also make the following useful definition\footnote{When all the spins are non-zero the critical bosonic and free bosonic correlators are identical. When we have some scalar operator then the correlation functions are legendre transforms of each other.} where,
\begin{equation}
    \epsilon\cdot X_{\mu\alpha_1...\alpha_n}=\frac{\epsilon_{\mu ab}p_a}{p}X_{b\alpha_1...\alpha_n}
\end{equation}
In position space, the epsilon transform is defined as \cite{Caron-Huot:2021kjy,Jain:2021gwa}
\begin{align}\label{oeps}
    \langle \epsilon.J^{\mu_1\mu_2\cdots \mu_{s_1}}(y_1) J_{s_2}(y_2) J_{s_3}(y_3)\rangle = \epsilon^{(\mu_1}_{\sigma\alpha}  \int \frac{d^3 x_1}{|y_1-x_1|^2}~~\partial_{x_1}^\sigma  \langle J^{\alpha \mu_2\cdots\mu_{s_1})}(x_1) J_{s_2}(y_2) J_{s_3}(y_3)\rangle
\end{align}
In momentum space the relation is simpler and can be written as  \cite{Jain:2021gwa}
\small
\begin{align}\label{3ptmapaf}
    \langle \epsilon.J^{\mu_1\mu_2\cdots \mu_{s_1}}(p_1) J_{s_2}(p_2) J_{s_3}(p_3)\rangle =\frac{1}{p_1}\epsilon^{\mu_1 p_1 \alpha }\langle J_{\alpha}^{ \mu_2 \cdots \mu_{s_1}}(p_1)J^{\nu_1 \cdots \nu_{s_2}}(p_2)J^{\rho_1 \cdots \rho_{s_3}}(p_3) \rangle+& (\mu_1 \leftrightarrow \mu_2)+\cdots+(\mu_1 \leftrightarrow \mu_{s_1})
  \end{align}
  \normalsize
  Converting to spinor helicity variables \cite{Jain:2021vrv,Jain:2021gwa}, $\langle\epsilon\cdot J_{s_1}J_{s_2}J_{s_3}\rangle$ becomes $\pm i \langle J_{s_1}J_{s_2}J_{s_3}\rangle$ depending on the helicity. 

Having discussed various notations, let us now summarise the results in this paper. To do that we first summarise the known two and three point functions and then we come to the four point functions.		
\subsection{2 point function}
 We write the general two point function as,
 \begin{equation}
     \langle J_{s}J_{s}\rangle_\text{QF}=\tilde{N}\langle J_{s}J_{s}\rangle_{\text{FF}}+\tilde{N}\tilde{\lambda}\langle J_{s}J_{s}\rangle_{\text{odd}}
 \end{equation}
 It can be shown that \cite{Caron-Huot:2021kjy,Jain:2021gwa} the parity odd part can be written in terms of the parity even part and it turns out that
 \begin{equation}
     \langle J_{s}J_{s}\rangle_\text{QF}=\tilde{N}\langle J_{s}J_{s}\rangle_{\text{FF}}+\tilde{N}\tilde{\lambda}\langle \epsilon\cdot J_{s}J_{s}\rangle_{\text{FF}}
 \end{equation}
 which upon converting to spinor helicity variables gives, 
  \cite{Gandhi:2021gwn}
 \begin{align}
     \langle J_s^{-}J_s^{-}\rangle_{QF}=\frac{N e^{i\pi\lambda_f}}{\pi\lambda_f}\langle J_{s}^{-}J_{s}^{-}\rangle_{FF}
 \end{align}
 where we have used \eqref{cplconst}.
	\subsection{3-point functions}
		It was shown in \cite{Maldacena:2012sf} that for the case of three point functions in the quasi fermionic theory we have\footnote{Our conventions are such that our scalar operator is related to the one in \cite{Maldacena:2012sf} as $O=\frac{O_{\text{MZ}}}{1+\tilde{\lambda}^2}$}
    \begin{align}\label{3ptansmz}
	    \langle J_sOO\rangle_\text{QF}&=\tilde{N}(1+\tilde{\lambda}^2)\langle J_sOO\rangle_\text{FF}\nonumber\\
	    \langle J_{s_1}J_{s_2}O\rangle_\text{QF}&=\tilde{N}\langle J_{s_1}J_{s_2}O\rangle_\text{FF}+\tilde{N}\tilde{\lambda}\langle J_{s_1}J_{s_2}O\rangle_\text{CB}\nonumber\\
	    \langle J_{s_1}J_{s_2}J_{s_3}\rangle_\text{QF}&=\frac{\tilde{N}}{1+\tilde{\lambda}^2}\bigg[ \langle J_{s_1}J_{s_2}J_{s_3}\rangle_\text{FF}+\tilde{\lambda}\langle J_{s_1}J_{s_2}J_{s_3}\rangle_\text{odd}+\tilde{\lambda}^2\langle J_{s_1}J_{s_2}J_{s_3}\rangle_\text{FB} \bigg]
	\end{align}
	In the above expression the momentum labels and the indices have been suppressed for clarity and they can be restored appropriately.

	However, it was realized \cite{Jain:2021gwa} that we can write the odd piece in terms of the FF and FB correlators and the final answer turns out to be
		\begin{align}\label{msexp}
		\langle J_{s_1}OO \rangle_\text{QF}&=\tilde{N}(1+\tilde{\lambda}^2)\langle J_{s_1}OO \rangle_\text{FF}\nonumber\\
			\langle J_{s_1}J_{s_2}O \rangle_\text{QF}&=\tilde{N}\langle J_{s_1}J_{s_2}O \rangle_\text{FF}+\tilde{N}\tilde{\lambda}\langle \epsilon\cdot J_{s_1}J_{s_2}O\rangle_\text{FF}\nonumber\\
			\hspace{-1cm}\langle J_{s_1}J_{s_2}J_{s_3}\rangle_\text{QF}&=\frac{\tilde{N}}{1+\tilde{\lambda}^2}\bigg[ \langle J_{s_1}J_{s_2}J_{s_3}\rangle_\text{FF}+\tilde{\lambda}\langle \epsilon\cdot J_{s_1}J_{s_2}J_{s_3}\rangle_\text{FF-FB}+\tilde{\lambda}^2\langle J_{s_1}J_{s_2}J_{s_3}\rangle_\text{FB}\bigg]
		\end{align}
		We now convert  the last equation of \eqref{msexp} to spinor helicity variables.  For brevity we will work with all helicities as minus. In all minus helicity we have 
		$\langle \epsilon\cdot J_{s_1}J_{s_2}J_{s_3}\rangle_\text{FF-FB}=i \langle  J_{s_1}J_{s_2}J_{s_3}\rangle_\text{FF-FB}$.
		
	Then combining the FF and FB terms, and substituting $\tilde{\lambda}=\tan{\frac{\pi\lambda_f}{2}}$ we get,
		\begin{align}\label{3ptsphcs}
		    \langle J_{s_1}J_{s_2}J_{s_3}\rangle_\text{QF}=&\frac{\tilde{N}}{2}\bigg[ \langle J_{s_1}J_{s_2}J_{s_3}\rangle_{\text{FF+FB}}+e^{-i\pi\lambda_f}\langle J_{s_1}J_{s_2}J_{s_3}\rangle_{\text{FF-FB}} \bigg]\nonumber\\
		    =&\tilde{N}e^{-\frac{i\pi\lambda_f}{2}}\bigg[ \cos{\frac{\pi\lambda_f}{2}}\langle J_{s_1}J_{s_2}J_{s_3}\rangle_\text{FF}+i\sin{\frac{\pi\lambda_f}{2}}\langle J_{s_1}J_{s_2}J_{s_3}\rangle_\text{FB} \bigg]
		\end{align}
	For, $\lambda_f=0$, we get the FF correlator and for $\lambda_f=1$, we get the FB correlator.	
	\subsubsection*{3-point functions in terms of homogeneous and non-homogeneous decomposition}\label{hnh}
		It is very interesting to see that the QF correlators can be written in terms of just the free theory correlators. However for the three-point case 
		it was shown in \cite{Jain:2021whr}
		that we can make a further stronger claim by representing the correlators in terms of the homogeneous and non-homogeneous parts \cite{Jain:2021whr}.
		
		It was shown in \cite{Jain:2021whr} that when the triangle inequality
		\begin{equation}\label{tineq}
			s_i+s_j\geq s_k
		\end{equation}
		 is satisfied, we can define the \textit{homogeneous/non-homogeneous} parts of a correlator and break up the known free theory correlators as
		\begin{align}\label{nohomo}
			\langle J_{s_1}J_{s_2}J_{s_3}\rangle_\text{FB}=\langle J_{s_1}J_{s_2}J_{s_3}\rangle_\text{nh}+\langle J_{s_1}J_{s_2}J_{s_3}\rangle_\text{h}\nonumber\\
			\langle J_{s_1}J_{s_2}J_{s_3}\rangle_\text{FF}=\langle J_{s_1}J_{s_2}J_{s_3}\rangle_\text{nh}-\langle J_{s_1}J_{s_2}J_{s_3}\rangle_\text{h}
		\end{align}
	We invert these relations to get
		\begin{align}
		\langle J_{s_1}J_{s_2}J_{s_3}\rangle_\text{nh}&=\frac{1}{2}\langle J_{s_1}J_{s_2}J_{s_3}\rangle_\text{FB+FF}\nonumber\\
	\langle J_{s_1}J_{s_2}J_{s_3}\rangle_\text{h}&=\frac{1}{2}\langle J_{s_1}J_{s_2}J_{s_3}\rangle_\text{FB-FF}
		\end{align}
		Thus inside the triangle inequality we can express our result for a general spinning correlator in spinor helicity variables \eqref{3ptsphcs} as follows 
		\begin{align}\label{3ptbf}
			\langle J_{s_1}J_{s_2}J_{s_3}\rangle_\text{QF}&=\tilde{N}\langle J_{s_1}J_{s_2}J_{s_3}\rangle_\text{nh}+\tilde{N}e^{-i\pi\lambda_f}\langle J_{s_1}J_{s_2}J_{s_3}\rangle_\text{h}
		\end{align}
	which is an even stronger statement than \eqref{msexp} since the homogeneous and non-homogeneous parts can be computed in just the free bosonic theory or just in the free fermionic theory \cite{Gandhi:2021gwn}. When we are outside the triangle inequality, such that \eqref{tineq} does not hold, the only contribution is from the non-homogeneous parts \cite{Jain:2021whr}, i.e. both the parity even structures and the parity odd structure are non-homogeneous. Also
		 \begin{align}
		     \langle J_{s_1}J_{s_2}J_{s_3}\rangle_{\text{nh,FB}}\ne
		     \langle J_{s_1}J_{s_2}J_{s_3}\rangle_{\text{nh,FF}}
		 \end{align}
		   In that case the distinction in \eqref{nohomo} no longer holds and we can only represent in terms of the free theories as in \eqref{3ptsphcs} in spinor helicity variables.
		
		\subsection{4-point functions}
		Now, we turn our attention to the case of 4-point correlators. 
	For general 4-point correlators, we obtain the following form in momentum space \footnote{As will be discussed in next few sections, the result presented in this section may not be the unique solution to the SBHS equations. However, as will be clear, the structure of equations are very tight as solving for say $\langle JJJJ \rangle$ does require information about $\langle JJTO \rangle$, $\langle JJOO \rangle.$ To solve for $\langle JJOO \rangle$ we need to know the form of  $\langle TOOO \rangle.$ To solve for $\langle TOOO \rangle$ we need to know $\langle OOOO \rangle.$ Thus, we see that the solutions are highly interconnected and even if there are more solutions to the SBHS equation, they will be extremely constrained.}
	\begin{align}\label{4ptans121}
	\langle OOOO\rangle_\text{QF}&=\tilde{N}(1+\tilde{\lambda}^2)^2\langle OOOO\rangle_\text{FF}\cr
		\langle J_sOOO\rangle_\text{QF}&=\tilde{N}(1+\tilde{\lambda}^2)(\langle J_sOOO\rangle_\text{FF}+\tilde{\lambda}\langle J_sOOO\rangle_\text{CB})\cr
		\langle  J_{s_1}J_{s_2}OO\rangle_{QF}&=\tilde{N}\langle  J_{s_1}J_{s_2}OO\rangle_\text{FF}+\tilde{N}\tilde{\lambda}\langle \epsilon\cdot J_{s_1}J_{s_2}OO\rangle_\text{FF-CB}+\tilde{N}\tilde{\lambda}^2\langle  J_{s_1}J_{s_2}OO\rangle_\text{CB} \cr
		\langle J_{s_1}J_{s_2}J_{s_3}O\rangle_\text{QF}&=\tilde{N}\langle J_{s_1}J_{s_2}J_{s_3}O\rangle_\text{FF}+\tilde{N}\tilde{\lambda}\langle J_{s_1}J_{s_2}J_{s_3}O\rangle_\text{CB}\cr
		\langle J_{s_1}J_{s_2}J_{s_3}J_{s_4}\rangle_{\text{QF}}&=\frac{\tilde{N}}{(1+\tilde\lambda^2)}\bigg[\langle J_{s_1}J_{s_2}J_{s_3}J_{s_4}\rangle_\text{FF}+\tilde\lambda \langle\epsilon \cdot J_{s_1}J_{s_2}J_{s_3}J_{s_4}\rangle_\text{FF-CB}+\tilde\lambda^2\langle J_{s_1}J_{s_2}J_{s_3}J_{s_4}\rangle_\text{CB}\bigg]\cr
		\end{align}
		To get a more intuitive form of these correlators, we convert the above expressions to spinor helicity variables. The general expression for an arbitrary correlator in spinor helicity looks like \footnote{In one particular helicity configuration. },	
		\begin{align}
		\langle J_sOOO\rangle_\text{QF}&=\frac{\tilde{N}}{\cos^3{\frac{\pi\lambda_f}{2}}}(\cos{\frac{\pi\lambda_f}{2}}\langle J_sOOO\rangle_{\text{FF}}+\sin{\frac{\pi\lambda_f}{2}}\langle J_sOOO\rangle_{\text{CB}})\cr
		\langle  J_{s_1}J_{s_2}OO\rangle_{QF}&=\frac{\tilde{N}}{\cos^2{\frac{\pi\lambda_f}{2}}}e^{-\frac{i\pi \lambda_f}{2}}\big[\cos{\frac{\pi\lambda_f}{2}}\langle  J_{s_1}J_{s_2}OO\rangle_\text{FF}+i\sin{\frac{\pi\lambda_f}{2}}\langle  J_{s_1}J_{s_2}OO\rangle_\text{CB}\big]\cr
		\langle J_{s_1}J_{s_2}J_{s_3}O\rangle_\text{QF}&=\frac{\tilde{N}}{\cos{\frac{\pi\lambda_f}{2}}}(\cos{\frac{\pi\lambda_f}{2}}\langle J_{s_1}J_{s_2}J_{s_3}O\rangle_{FF}+\sin{\frac{\pi\lambda_f}{2}}\langle J_{s_1}J_{s_2}J_{s_3}O\rangle_{CB})\cr
		\langle J_{s_1}J_{s_2}J_{s_3}J_{s_4}\rangle_{\text{QF}}&=\tilde{N}e^{-\frac{i\pi \lambda_f}{2}}\big[\cos{\frac{\pi\lambda_f}{2}}\langle J_{s_1}J_{s_2}J_{s_3}J_{s_4}\rangle_{\text{FF}}+i \sin{\frac{\pi\lambda_f}{2}}\langle J_{s_1}J_{s_2}J_{s_3}J_{s_4}\rangle_{\text{CB}}\big]\cr
		\end{align}
	where we have suppressed helicity indices. 	
		It is clear that for $\lambda_f=0$, we get the expression for FF correlator. For $\lambda_f=1$ we get the CB result. For $\langle J_sOOO\rangle$ and $\langle J_{s_1}J_{s_2}J_{s_3}O\rangle$ at $\lambda_f=1$ we need to appropriately redefine the correlator by absorbing the coupling constant dependent factor\footnote{In this paper we have done the analysis for the QF theory. It should be easy to generalize this for the QB theory. In the QB theory, the four point function of scalar operators $O_{\text{QB}}$ differs from the scalar four point function of the FB theory by some exchange diagrams in $AdS_4$ that come from $\phi^3$ vertices, see equation (1.2) of \cite{Turiaci:2018nua}. The HSE will generate similar contact diagrams for spinning correlators.}.
		\\As was discussed in \eqref{3ptbf}, in the case of  three-point correlators we can get an even stronger statement that just the FB or just the FF theory is enough to construct QF theory correlation function. However as of yet we don't have such a statement for four point functions \footnote{
		Analysing four point functions of spinning operators in momentum space is a difficult task and has not yet been done. It would be interesting to understand the homogeneous and non-homogeneous distinction for this case as well. It might give us a stronger result as in the case of the three point function as discussed in \eqref{3ptbf}.}. 
 		\begin{figure}[H]\label{thcir11a}
		\centering
		\def\svgwidth{4cm}
		\includegraphics[scale=0.8]{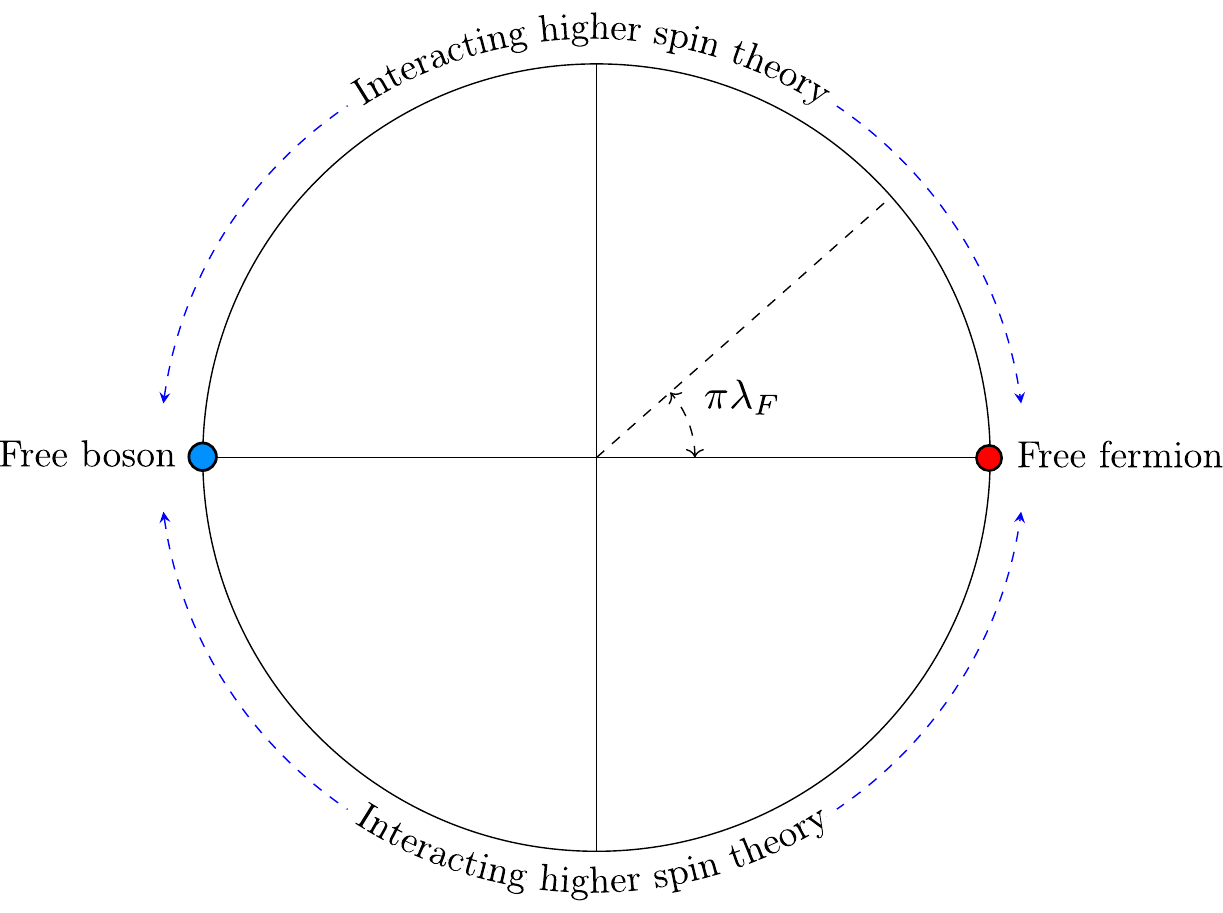}
		\caption{A chart visualizing the space of theories. Exactly conserved or weakly broken at large$-N$ higher spin theories lie on the circle of unit radius.}
		\label{chart}
	    \end{figure}
	 The above figure summarises our finding that correlation functions in CS matter theory are obtainable from the free theories with some anyonic phase factor \cite{Jain:2021whr}. For two and three point functions we can just start with the FB or FF answer and appropriately multiply with the anyonic phase factor to obtain the result in CS matter theory. For the four (and higher) point case, we do require both the FF and the FB results to get the result for the CS matter theory.
\section{Mapping Slightly broken HS correlators to free theory correlators}\label{Mapmth} 
In the previous section we saw that the results in the QF theory can be written in terms of the FF and FB theory results.
In this section we outline the methodology which maps correlation functions in SBHS theories to the free theories. In the next section we show how our methodology works explicitly.
\subsection{Method}
We make use of the SBHS equation following \cite{Maldacena:2012sf} to compute the result for correlation functions. Our method can be summarized by the following steps::\\

\noindent\textbf{\color{blue}Step 1:} Choose an appropriate charge operator and a seed correlator to write down the higher spin equation in the interacting theory, say the quasi fermionic theory.\\
\\
\textbf{\color{blue}Step 2:} Repeat the same for the free and critical theories.\\
\\
\textbf{\color{blue}Step 3:} Write down the ansatz for each correlator that appears in the slightly broken higher spin equation in the interacting theory.\\
\\
\textbf{\color{blue}Step 4:} Map the equations that are at the lowest $O(\tilde\lambda^0)$ and highest orders in the coupling to the free fermion and critical bosonic theories respectively. This helps us identify the contributions 
at the lowest and highest orders as the ones from the free and critical theories, respectively.\\
\\
\textbf{\color{blue}Step 5:} Write the pole equations which are obtained by expanding the HSE around $\tilde\lambda=\pm i$ to obtain the remaining unknowns in the ansatz 
\footnote{For certain correlators such as $\langle JJJJ\rangle$ it turns out that the pole equations are not sufficient and one has to resort to the higher spin equations at intermediate orders in the coupling to extract out the remaining unknowns.}.\\
\\
\textbf{\color{blue}Step 6:} Plug back the solution in the higher spin equation and map it to a linear combination of the equations in the free and critical theories \footnote{For the case of four-point functions, spinor-helicity variables are extremely useful at this step since the map between parity even and odd parts of the HSE is much more transparent in these variables. }.   
\\
The above map allows us to identify the unknowns in the ansatz of the interacting theory correlator purely in terms of the FF and CB theory results. 

\section{Example: 3 point functions}\label{3ptfunctions}
The aim of this section is to implement the methodology described in section \ref{Mapmth}. The case of two-point functions is straightforward and is dealt with in the appendix \ref{2pthse}. We illustrate our methodology for three point functions.
The simplest spinning correlator is of the kind $\langle J_s OO\rangle$. However, we know from \eqref{msexp} that this correlator does not have an odd contribution and is completely fixed by the free theory correlator. We start with the simplest nontrivial correlator $\langle JJO \rangle.$
	\subsection{$\langle JJO\rangle_\text{QF}$} \label{jjoqf}
		As discussed earlier in \eqref{3ptansmz}, $\langle J_{\alpha}J_{\beta}O \rangle $ in the QF theory  has an odd part. In our analysis we make use of higher spin equations and follow the steps  given at the end of section \ref{Mapmth}. 
		
		\textbf{\color{blue}Step 1:} We choose the charge operator and the seed correlator to be $Q_3$ and $\langle JOO \rangle$ respectively to write the following HSE in position space   \cite{Jain:2020puw} 
		
		\begin{align}
			\label{genjjohse}
			&\langle[Q_{\mu\nu},  J_{\alpha}(x_1)]O(x_2)O(x_3)\rangle_\text{QF}+\langle J_{\alpha}(x_1)[Q_{\mu\nu}, O(x_2)]O(x_3)\rangle_\text{QF}\cr
			&+\langle  J_{\alpha}(x_1)O(x_2)[Q_{\mu\nu},O(x_3)]\rangle_\text{QF}=\int_{x}^{} \langle \partial_\sigma J^\sigma_{\mu\nu}(x) J_{\alpha}(x_1)O(x_2)O(x_3)\rangle_\text{QF}.
		\end{align}
		We utilize the higher spin algebra \eqref{q4oqf} \footnote{We keep the coefficients in the algebra arbitrary since fixing them will not affect our computation.}
		and the current non conservation for $J_{\mu\nu\rho}$ \eqref{ncj4qf} in the HSE \eqref{genjjohse}. 
		We then perform an integration by parts and use the large $ N $ factorisation of the 5-point correlator that appears on the RHS. After a subsequent Fourier transform of the HSE we obtain the following HSE as an algebraic equation in terms of the correlators of the interacting theory 
		\begin{equation}\label{jjohse}
			\begin{aligned}
				{}&p_{1(\mu}\langle T_{\nu)\alpha}(p_1)O(p_2)O(p_3)\rangle_\text{QF} +p_{1\alpha}\langle T_{\mu\nu}(p_1)O(p_2)O(p_3) \rangle_\text{QF}\\
				&+\bigg(\epsilon_{(\mu a b}p_{2a}p_{2\nu)}\langle J_{\alpha} (p_1)J_{b}(p_2)O(p_3) \rangle_\text{QF}+\left\lbrace 2\leftrightarrow3\right\rbrace\bigg)\cr
				&=\bigg(\frac{\tilde{\lambda}}{1+\tilde{\lambda}^2}p_{2(\mu}p_2\langle J_{\alpha} (p_1)J_{\nu)}(p_2)O(p_3) \rangle_\text{QF}
				+\left\lbrace 2\leftrightarrow3\right\rbrace\bigg)
			\end{aligned}
		\end{equation} 
		where the notation $p_{1(\mu} T_{\nu)\alpha}$ denotes $\mu, \nu$ symmetrisation of $p_{1\mu} T_{\nu\alpha}$. We note that Fourier transforming $\eqref{jjohse}$ gets rid of the integral on the RHS of \eqref{genjjohse} and thus makes it easier to factorise the resulting 5-point function \cite{Jain:2020puw}.
		
		\textbf{\color{blue} Step 2:} We now write down the corresponding HSEs for the FF theory
		\begin{equation}\label{jjoff}
			\begin{aligned}
				{}&p_{1(\mu}\langle T_{\nu)\alpha}(p_1)O(p_2)O(p_3)\rangle_\text{FF} +p_{1\alpha}\langle T_{\mu\nu}(p_1)O(p_2)O(p_3) \rangle_\text{FF}\\
				&+\bigg(\epsilon_{(\mu a b}p_{2a}p_{2\nu)}\langle J_{\alpha} (p_1)J_{b}(p_2)O(p_3) \rangle_\text{FF}+\left\lbrace 2\leftrightarrow3\right\rbrace\bigg)=0
			\end{aligned}
		\end{equation} 
		and the CB theory,
		\begin{align}\label{jjocb}
			p_{1(\mu}\langle T_{\nu)\alpha}(p_1)O(p_2)O(p_3)\rangle_\text{CB} +c_2p_{1\alpha}\langle T_{\mu\nu}(p_1)O(p_2)O(p_3) \rangle_\text{CB}\cr
			=p_{2(\mu}p_2\langle J_{\alpha} (p_1)J_{\nu)}(p_2)O(p_3) \rangle_\text{CB}+\left\lbrace 2\leftrightarrow3\right\rbrace
		\end{align}

		\textbf{\color{blue}Step 3:} We consider the following ansatz for the correlators that appear in the HSE \eqref{jjohse} \cite{Maldacena:2012sf}
		\begin{equation}
			\begin{split}
				\langle T_{\nu\alpha}(p_1)O(p_2)O(p_3)\rangle_\text{QF}&=\tilde{N}(1+\tilde{\lambda}^2)\langle T_{\nu\alpha}(p_1)O(p_2)O(p_3)\rangle_{\text{FF}} \label{jjoansatz} \\
				\langle J_{\alpha} (p_1)J_{\beta}(p_2)O(p_3) \rangle_\text{QF}&=\tilde{N}\langle J_{\alpha} (p_1)J_{\beta}(p_2)O(p_3) \rangle_{Y_0}+\tilde{N}\tilde{\lambda}\langle J_{\alpha} (p_1)J_{\beta}(p_2)O(p_3) \rangle_\text{odd}
			\end{split}
		\end{equation}
		where $ \langle J_{\alpha} (p_1)J_{\beta}(p_2)O(p_3) \rangle_{Y_0} $ and $\langle J_{\alpha} (p_1)J_{\beta}(p_2)O(p_3) \rangle_\text{odd}$ are the unknown parts that we wish to find. The HSE \eqref{jjohse} can then be written at different orders in the coupling.
		
		\textbf{\color{blue}Step 4:} At $O(\tilde\lambda^0)$ of \eqref{jjohse} the HSE is identical to the FF theory equation  \eqref{jjoff} which gives 
		\begin{align}
			\langle J_{\alpha}(p_1)J_{\beta}(p_2)O(p_3) \rangle_{Y_0}=\langle J_{\alpha}(p_1)J_{\beta}(p_2)O(p_3) \rangle_{FF}. 
		\end{align}
		Similarly, the highest order equation, namely the one at $O(\tilde{\lambda}^2)$ is identical to the CB\footnote{This is because the CB theory is obtained in the limit $\tilde{\lambda}\rightarrow\infty$  of the quasi fermionic theory.} equation \eqref{jjocb}. Thus, we identify 
		\begin{align}\label{oddcb}
			\langle J_{\alpha} (p_1)J_{\beta}(p_2)O(p_3) \rangle_\text{odd}=\langle J_{\alpha} (p_1)J_{\beta}(p_2)O(p_3) \rangle_\text{CB}\cr
			\langle T_{\nu\alpha}(p_1)O(p_2)O(p_3)\rangle_{\text{FF}}=\langle T_{\nu\alpha}(p_1)O(p_2)O(p_3)\rangle_{\text{CB}}
		\end{align}
	
		\textbf{\color{blue}Step 5:}We now write the pole equation.  We expand \eqref{jjohse} around the pole $\tilde{\lambda}=\pm i$ to get the following pole equations
		\begin{align}\label{pejjo}
			\epsilon_{(\mu a b}p_{2a}p_{2\nu)}\langle J_{\alpha} (p_1)J_{b}(p_2)O(p_3) \rangle_\text{odd}+\left\lbrace 2\leftrightarrow3\right\rbrace
			=p_{2(\mu}p_2\langle J_{\alpha} (p_1)J_{\nu)}(p_2)O(p_3) \rangle_\text{FF}+\left\lbrace 2\leftrightarrow3\right\rbrace\cr
			\epsilon_{(\mu a b}p_{2a}p_{2\nu)}\langle J_{\alpha} (p_1)J_{b}(p_2)O(p_3) \rangle_\text{FF}+\left\lbrace 2\leftrightarrow3\right\rbrace= p_{2(\mu}p_2\langle J_{\alpha} (p_1)J_{\nu)}(p_2)O(p_3) \rangle_\text{odd}+\left\lbrace 2\leftrightarrow3\right\rbrace 
		\end{align}
		which helps us identify the unknown correlator $\langle J_{\alpha} (p_1)J_{\nu}(p_2)O(p_3) \rangle_\text{odd}$ in terms of the same correlator in free theory \cite{Jain:2021gwa,Gandhi:2021gwn}
		\begin{align}\label{expjjoodd}
			\langle J_{\alpha} (p_1)J_{\nu}(p_2)O(p_3) \rangle_\text{odd}=\frac{1}{p_2}\epsilon_{\nu a b}\,p_{2a}\,\langle J_{\alpha} (p_1)J_{b}(p_2)O(p_3) \rangle_\text{FF} 
		\end{align}
		The expression for $\langle JJO\rangle_\text{odd}$ obtained from \eqref{expjjoodd} is consistent with the results obtained using perturbative techniques in special kinematic regimes \cite{Aharony:2012nh,GurAri:2012is} and by solving conformal Ward identities in momentum space \cite{Jain:2021gwa,Gandhi:2021gwn}.
		\footnote{We note that \eqref{expjjoodd} is one of the solutions to \eqref{pejjo} where we ignore $\left\lbrace 2\leftrightarrow3\right\rbrace $ exchanges. We will adopt a similar strategy while computing  4-point functions where we again ignore such permutations. However as we shall see, pole equations are not sufficient to get the odd piece in case of certain 4-point functions and we will then have to make full use of the slightly broken HS equations and provide a consistent solution to the higher spin equation. }
		
		\textbf{\color{blue}Step 6:} We now use our results to map the SBHS equation to the free theory HSE. To do this, we use \eqref{expjjoodd} and substitute it back into \eqref{jjohse} and see that the remaining HSE maps to the free theory equation. Thus we see that  the solution for the odd piece that we obtained from the pole equation is consistent with the HSE at any order.
		
		This confirms the result obtained for $\langle J_{\alpha} (p_1)J_{\beta}(p_2)O(p_3) \rangle_\text{odd}$ in \eqref{pejjo}. Thus we have completely determined the 3-point spinning correlator  $\langle J_{\alpha} (p_1)J_{\beta}(p_2)O(p_3) \rangle_\text{odd}$ in the interacting theory purely in terms of free theory correlators i.e
		\begin{align}
			\langle J_{\alpha} (p_1)J_{\beta}(p_2)O(p_3) \rangle_\text{QF}&=\tilde{N}\langle J_{\alpha} (p_1)J_{\beta}(p_2)O(p_3) \rangle_\text{FF}+\tilde{N}\tilde{\lambda}\frac{\epsilon_{\beta a b}\,p_{2a}}{p_2}\langle J_{\alpha} (p_1)J_{b}(p_2)O(p_3) \rangle_\text{FF}
		\end{align}
Now in spinor-helicity we have $\langle\epsilon\cdot JJO\rangle\rightarrow \pm i\langle JJO\rangle$ depending on the helicity. Thus the final expression becomes,
\begin{align}
    \langle JJO\rangle_\text{QF}&=\tilde{N}(1+i\tilde{\lambda})\langle JJO\rangle_\text{FF}\cr
	&=\frac{iN(1-e^{-i\pi\lambda_{f}})}{\pi\lambda_{f}}\langle JJO\rangle_\text{FF}
\end{align}
We note that the expression of the correlator picks up an anyonic phase when we express the full correlator only in terms of the FF theory correlator.
\subsection{$\langle TTO\rangle_\text{QF}$} \label{ttoqf}
We now compute the correlator $\langle T_{\alpha\beta}T_{\mu\nu}O \rangle $ in the QF theory. The analysis here is very similar to $\langle JJO \rangle_\text{QF}$ obtained in the previous section. We make use of the higher spin algebra of $Q_4$ in \eqref{q4oqf} and \eqref{q4tqf} and the current non-conservation equation associated to  $J_4$ in \eqref{ncj4qf} to obtain the momentum space HSE in terms of the interacting theory correlators.

Now we assume $\langle TTO \rangle_\text{QF}$ has the following structure
\begin{align}
	\langle T_{\alpha\beta} (p_1)T_{\gamma\delta}(p_2)O(p_3) \rangle_\text{QF}=\tilde{N}\langle T_{\alpha\beta} (p_1)T_{\gamma\delta}(p_2)O(p_3) \rangle_{Y_0}+\tilde{N}\tilde{\lambda}\langle T_{\alpha\beta} (p_1)T_{\gamma\delta}(p_2)O(p_3) \rangle_\text{odd}
\end{align}
Repeating the steps as in the previous section, we obtain a set of algebraic equations at different orders in the coupling. The $O(\tilde\lambda^0)$ equation is the free theory equation after we identify  
\begin{align}\label{FF1tto}
	\langle T_{\alpha\beta} (p_1)T_{\gamma\delta}(p_2)O(p_3) \rangle_{Y_0}&=\langle T_{\alpha\beta} (p_1)T_{\gamma\delta}(p_2)O(p_3) \rangle_\text{FF}
\end{align}
The pole equations help us identify $ \langle TTO \rangle_\text{odd} $ in terms of the same correlator in the free theory
\begin{align}\label{od1tto}
	\langle T_{\alpha\beta} (p_1)T_{\gamma\delta}(p_2)O(p_3) \rangle_\text{odd}&=\epsilon_{\gamma a b}\frac{p_{2a}}{p_2}\langle T_{\alpha\beta} (p_1)T_{b \delta}(p_2)O(p_3) \rangle_\text{FF}
\end{align}
We then substitute this back \eqref{FF1tto}, \eqref{od1tto} in the SBHS equation and map it to the free theory equation. Therefore the interacting theory correlator  $\langle TTT \rangle_\text{QF}$ is given by
\begin{align}\label{ttofinal}
	\langle T_{\alpha\beta} (p_1)T_{\gamma\delta}(p_2)O(p_3)\rangle_\text{QF}&=\tilde{N}\langle T_{\alpha\beta} (p_1)T_{\gamma\delta}(p_2)O(p_3) \rangle_\text{FF}+\tilde{N}\tilde{\lambda}\epsilon_{\gamma a b}\frac{p_{2a}}{p_2}\langle T_{\alpha\beta} (p_1)T_{b \delta}(p_2)O(p_3) \rangle_\text{FF}
\end{align}
Now in spinor-helicity we have $\langle\epsilon\cdot TTO\rangle\rightarrow \pm i\langle TTO\rangle$. Thus the final expression becomes,
\begin{align}
    \langle TTO\rangle_\text{QF}&=\tilde{N}(1+i\tilde{\lambda})\langle TTO\rangle_\text{FF}\cr
		&=\frac{iN(1-e^{-i\pi\lambda_f})}{\pi\lambda_f}\langle TTO\rangle_\text{FF}
\end{align}
Yet again we see that the expression of the correlator picks up an anyonic phase.
\subsection{ $\langle TTT\rangle_\text{QF}$}\label{ttt}
	In this section we will make use of the HSE to obtain the odd part of $\langle TTT\rangle$ in the QF theory. As before we follow the steps presented at the end of section \ref{Mapmth}.
	
	\textbf{\color{blue}Step 1:} We choose the charge operator and the seed correlator to be $Q_4$ and $\langle OTT \rangle$ respectively to write the following HSE in position space
	\begin{align}
		\label{HSETTT}
		&	\langle[Q_{\mu\nu\rho}, O(x_1)]T_{\alpha\beta}(x_2)T_{\gamma\theta}(x_3)\rangle_\text{QF}+\langle O(x_1)[Q_{\mu\nu\rho},T_{\alpha\beta}(x_2)]T_{\gamma\theta}(x_3)]\rangle_\text{QF}\cr
		&\hspace{-.1cm}+\langle O(x_1)T_{\alpha\beta}(x_2)[Q_{\mu\nu\rho},T_{\gamma\theta}(x_3)]\rangle_\text{QF}=\int_{x}^{} \langle \partial_\sigma J^\sigma_{\mu\nu\rho}(x) O(x_1)T_{\alpha\beta}(x_2)T_{\gamma\theta}(x_3)\rangle_\text{QF}
	\end{align}
	 We make use of the higher spin algebra \eqref{q4oqf} and \eqref{q4tqf} along with the current non conservation \eqref{ncj4qf} to obtain the following HSE in momentum space
	\begin{align}
		\label{qfttthse}
		&p_{1(\mu}p_{1\nu}p_{1\rho)}\langle O(p_1)T_{\alpha\beta}(p_2)T_{\gamma\theta}(p_3)\rangle_{\text{QF}}+\epsilon_{(\mu a b}p_{1a}p_{1\nu}\langle T_{b\rho)}(p_1)T_{\alpha\beta}(p_2)T_{\gamma\theta}(p_3)\rangle_\text{QF}\cr
		&+\bigg(p_{2(\mu}p_{2\nu}p_{2\rho)}\langle O(p_1)T_{\alpha\beta}(p_2)T_{\gamma\theta}(p_3)\rangle_\text{QF}\cr
		&\hspace{1cm}+p_{2(\mu}p_{2\nu}p_{2\alpha}\langle O(p_1)T_{\rho)\beta}(p_2)T_{\gamma\theta}(p_3)\rangle_\text{QF}+ \left\lbrace 2\leftrightarrow3\right\rbrace \bigg)=\cr
		&\frac{\tilde{\lambda}}{1+\tilde{\lambda}^2}\Bigg[{p_{1(\mu}}p_1 \langle T_{\nu\rho)}(p_2)T_{\alpha\beta}(p_2)T_{\gamma\theta}(p_3)\rangle_\text{QF}+\bigg(p_{2(\mu}\langle T_{\nu\rho)}T_{\alpha\beta} \rangle_\text{QF} \langle O(p_1)O(p_2)T_{\gamma\theta}(p_3) \rangle_\text{QF}\cr
		&+ \left\lbrace 2 \leftrightarrow 3\right\rbrace\bigg)\Bigg]
	\end{align}	
	\textbf{\color{blue}Step 2:} We now write down the corresponding higher spin equation for the FF theory
	\begin{align}
		\label{tttff}
		&p_{1(\mu}p_{1\nu}p_{1\rho)}\langle O(p_1)T_{\alpha\beta}(p_2)T_{\gamma\theta}(p_3)\rangle_{\text{FF}}-\epsilon_{(\mu a b}p_{1a}p_{1\nu}\langle T_{b \rho)}(p_1)T_{\alpha\beta}(p_2)T_{\gamma\theta}(p_3)\rangle_\text{FF}\cr
		&+\left(p_{2(\mu}p_{2\nu}p_{2\rho)}\langle O(p_1)T_{\alpha\beta}(p_2)T_{\gamma\theta}(p_3)\rangle_{\text{FF}}\right.\cr
		&\left.\hspace{1cm}+p_{2(\mu}p_{2\nu}p_{2\alpha}\langle O(p_1)T_{\rho)\beta}(p_2)T_{\gamma\theta}(p_3)\rangle_{\text{FF}}+ \left\lbrace 2\leftrightarrow3\right\rbrace \right)=0
	\end{align}	
	and similarly for the CB theory,
	\begin{align}
		\label{tttcb}
		&p_{1(\mu}p_{1\nu}p_{1\rho)}\langle O(p_1)T_{\alpha\beta}(p_2)T_{\gamma\theta}(p_3)\rangle_{\text{CB}}+\left(p_{2(\mu}p_{2\nu}p_{2\rho)}\langle O(p_1)T_{\alpha\beta}(p_2)T_{\gamma\theta}(p_3)\rangle_{\text{CB}}\right.\cr
		&\left.\hspace{4cm}+p_{2(\mu}p_{2\nu}p_{2\alpha}\langle O(p_1)T_{\rho)\beta}(p_2)T_{\gamma\theta}(p_3)\rangle_{\text{CB}}+\left\lbrace 2\leftrightarrow3\right\rbrace \right)\cr
		&=\bigg[p_{1(\mu}p_1\langle T_{\nu \rho)}(p_1)T_{\alpha\beta}(p_2)T_{\gamma\theta}(p_3)\rangle_\text{CB}+\bigg(p_{2(\mu}\langle T_{\nu\rho)}T_{\alpha\beta} \rangle_\text{QF} \langle O(p_1)O(p_2)T_{\gamma\theta}(p_3) \rangle_\text{CB}\cr
		&+ \left\lbrace 2 \leftrightarrow 3\right\rbrace\bigg)\Bigg]
	\end{align}
	\textbf{\color{blue}Step 3:} We consider the following ansatz for the correlators that appear in the HSE \eqref{qfttthse} \cite{Maldacena:2012sf}
	%
	\begin{align}\label{pettt}
		\begin{split}
			\langle O(p_1)O(-p_1)\rangle_{\text{QF}}&=\tilde{N}(1+\tilde{\lambda}^2)\langle O(p_1)O(-p_1)\rangle_{\text{FF}}\\
			\langle O(p_1)T_{\alpha\beta}(p_2)T_{\gamma\theta}(p_3)\rangle_\text{QF}&=\tilde{N}\langle O(p_1)T_{\alpha\beta}(p_2)T_{\gamma\theta}(p_3)\rangle_\text{FF}+\tilde{N}\tilde{\lambda}\langle O(p_1)T_{\alpha\beta}(p_2)T_{\gamma\theta}(p_3)\rangle_\text{CB}\\
			\langle T_{\rho\sigma}(p_1)T_{\alpha\beta}(p_2)T_{\gamma\theta}(p_3)\rangle_\text{QF}&=\tilde{N}\frac{1}{1+\tilde\lambda^2}\left[\langle T_{\rho\sigma}(p_1)T_{\alpha\beta}(p_2)T_{\gamma\theta}(p_3)\rangle_{Y_0}\right.\cr
			&\hspace{-2cm}\left.+\tilde\lambda^2\langle T_{\rho\sigma}(p_1)T_{\alpha\beta}(p_2)T_{\gamma\theta}(p_3)\rangle_{Y_2}+\tilde\lambda\langle T_{\rho\sigma}(p_1)T_{\alpha\beta}(p_2)T_{\gamma\theta}(p_3)\rangle_{odd}\right]
		\end{split}
	\end{align}
	 Our goal is to determine the parity odd part of $\langle TTT\rangle$, viz. $\langle T_{ \rho\sigma}(p_1)T_{\alpha\beta}(p_2)T_{\gamma\theta}(p_3)\rangle_\text{odd}$ in terms of the free theory correlators. We can now write the HSE at various orders in the coupling constant. 
	
	\textbf{\color{blue}Step 4:} At $O(\tilde\lambda^0)$ of \eqref{qfttthse}, the HSE is identical to the FF HSE  \eqref{tttff} which gives
	\begin{align}
		\langle T_{\rho\sigma}(p_1)T_{\alpha\beta}(p_2)T_{\gamma\theta}(p_3)\rangle_{Y_0}=\langle T_{\rho\sigma}(p_1)T_{\alpha\beta}(p_2)T_{\gamma\theta}(p_3)\rangle_{\text{FF}}
	\end{align}
	Similarly, the highest order equation namely the $O(\tilde{\lambda}^3)$ is identical to the critical bosonic equation \eqref{tttcb} since the CB theory is the $\tilde{\lambda}\rightarrow\infty$ limit of the QF theory. This happens after we identify 
	\begin{align}\label{tttoddcb}
		\langle T_{\rho\sigma}(p_1)T_{\alpha\beta}(p_2)T_{\gamma\theta}(p_3)\rangle_{Y_2}=\langle T_{\rho\sigma}(p_1)T_{\alpha\beta}(p_2)T_{\gamma\theta}(p_3)\rangle_{\text{CB}}
	\end{align}
	Hence we get 2 of the 3 unknowns. 
	
	\textbf{\color{blue}Step 5:} To find the third unknown we expand \eqref{qfttthse} around the pole $\tilde{\lambda}=\pm i$ to get the following pole equations
	\begin{align}
		\label{p1eqn}
		{}&\hspace{-1.4cm}\epsilon_{(\mu a b}p_{1a}p_{1\nu}\langle T_{b \rho)}(p_1)T_{\alpha\beta}(p_2)T_{\gamma\theta}(p_3)\rangle_\text{odd}\cr
		&=p_{1(\mu}p_1 \bigg(\langle T_{\nu \rho)}(p_1)T_{\alpha\beta}(p_2)T_{\gamma\theta}(p_3)\rangle_\text{FB}-\langle T_{\nu \rho}(p_1)T_{\alpha\beta}(p_2)T_{\gamma\theta}(p_3)\rangle_\text{FF}\bigg)\cr
		& \hspace{-1.5cm}\epsilon_{(\mu a b}p_{1a}p_{1\nu}\bigg(\langle T_{b \rho)}(p_1)T_{\alpha\beta}(p_2)T_{\gamma\theta}(p_3)\rangle_\text{FB}-\langle T_{b \rho)}(p_1)T_{\alpha\beta}(p_2)T_{\gamma\theta}(p_3)\rangle_\text{FF}\bigg)\cr
		&=p_{1(\mu}p_1 \langle T_{\nu \rho)}(p_1)T_{\alpha\beta}(p_2)T_{\gamma\theta}(p_3)\rangle_\text{odd}
	\end{align}
	which helps us identify the unknown correlator $ \langle T_{\nu \rho}(p_1)T_{\alpha\beta}(p_2)T_{\gamma\theta}(p_3)\rangle_\text{odd} $ in terms of the same correlator in the free theories. Thus from \eqref{pettt}, after contracting with $\Pi_{\mu\nu}(p_1)$ and $ p_{1\rho} $ we get $\langle TTT\rangle_\text{odd}$ to be
		\begin{align}
			\label{p2eqn}
			&\langle T_{\mu \nu}(p_1)T_{\alpha\beta}(p_2)T_{\gamma\theta}(p_3)\rangle_\text{odd}\cr
			&=\frac{1}{p_1}\epsilon_{\mu a b}p_{1a}\bigg(\langle T_{b \nu}(p_1)T_{\alpha\beta}(p_2)T_{\gamma\theta}(p_3)\rangle_\text{FB}
			-\langle T_{b \rho)}(p_1)T_{\alpha\beta}(p_2)T_{\gamma\theta}(p_3)\rangle_\text{FF}\bigg)
		\end{align}
	
	\textbf{\color{blue}Step 6:} We now use our results to map the SBHS equation to the free theory HSE.  We use the expression for $\langle TTT\rangle_\text{odd}$ and substitute it back into the $O(\tilde{\lambda})$ and $O(\tilde{\lambda}^2)$ equations and see that they map to the free theory equations.
		Therefore we see that that our solution for $\langle TTT\rangle_{\text{odd}}$ obtained from \eqref{p2eqn} solves the entire higher spin equation and is also consistent with results obtained in \cite{Jain:2021gwa,Gandhi:2021gwn}. The same can also be obtained by solving conformal Ward identities in momentum space. Thus we have seen that in the QF theory the correlator $\langle TTT\rangle$ is given by 
		\begin{align}
			\langle TTT\rangle_\text{QF}&=\frac{\tilde{N}}{1+\tilde{\lambda}^2}\big(\langle TTT\rangle_\text{FF}+\tilde{\lambda}\langle TTT\rangle_\text{odd}+\tilde{\lambda}^2\langle TTT\rangle_\text{FB}\big)\cr
			&=\frac{\tilde{N}}{2}\left(\langle TTT \rangle_\text{FF+FB}+\frac{1-\tilde{\lambda}^2}{1+\tilde{\lambda}^2}\langle TTT \rangle_\text{FF-FB}+\frac{2\tilde{\lambda}}{1+\tilde{\lambda}^2} \langle\epsilon\cdot TTT \rangle_\text{FB-FF}\right)
		\end{align}
		Now in spinor-helicity variables we have $\langle\epsilon\cdot TTT\rangle\rightarrow i\langle TTT\rangle$ and thus
		\begin{align}
			\langle TTT\rangle_\text{QF}
			&=\frac{\tilde{N}}{2}\left(\langle TTT \rangle_\text{FF+FB}+\left(\frac{1+i\tilde{\lambda}}{1-i\tilde{\lambda}}\right)\langle TTT \rangle_\text{FF-FB}\right)\cr
			&=\frac{\tilde{N}}{2}\left(\langle TTT \rangle_\text{FF+FB}+e^{-i\pi\lambda_f}\langle TTT \rangle_\text{FF-FB}\right)
		\end{align}
		Thus we see the presence of an anyonic phase yet again in the expression for the correlator.
There is one more representation which makes the duality manifest
\begin{align}\label{tttsph}
			\langle TTT\rangle_\text{QF}
			&= e^{-i  \pi \frac{\lambda_f}{2}}\left( \cos\frac{\pi \lambda_f}{2} 	\langle TTT\rangle_\text{FF}+i\sin\frac{\pi \lambda_f}{2} 	\langle TTT\rangle_\text{FB}\right)
		\end{align}
Note that at $\lambda_f=0$ it gives the FF and at $\lambda_f=1,$ it gives the FB answer.		
		\subsection{$\langle J_{s_1}J_{s_2}T\rangle_{QF}$}
		In this section we use HSEs to constrain the  general 3-point correlator $\langle J_{s_1}J_{s_2}T\rangle$ in the QF theory.
		We choose the charge operator and the seed correlator to be $Q_4$ and $\langle J_{s_1}J_{s_2}O\rangle$  respectively. The relevant operator algebra is given by \cite{Maldacena:2012sf}
		\begin{align}\label{genqbalg}
			[Q_4,J_s]&=c_{s,s-2}\partial^5 J_{s-2}+c_{s,s}\partial^3 J_{s}+c_{s,s+2}\partial J_{s+2}\cr
			[Q_4,O]&=\partial^3O+\epsilon \cdot \partial T
		\end{align}
		%
		To solve the resulting HSEs we make use of the  following structure of the correlators,
		\begin{align}
			\langle OO\rangle_\text{QF}&=\tilde{N}(1+\tilde{\lambda}^2)\langle OO\rangle_\text{FF}
			\cr
			\langle J_{s_1}J_{s_2}O\rangle_\text{QF}&=\tilde{N}\langle J_{s_1}J_{s_2}O\rangle_\text{FF}+\tilde{N}\tilde{\lambda}\langle J_{s_1}J_{s_2}O\rangle_\text{odd}
			\cr
			\langle J_{s_1}J_{s_2}T\rangle_\text{QF}&=\tilde{N}\frac{1}{1+\tilde{\lambda}^2}\big(\langle J_{s_1}J_{s_2}T\rangle_{Y_0}+\tilde{\lambda}\langle J_{s_1}J_{s_2}T\rangle_\text{odd}+\tilde{\lambda}^2\langle J_{s_1}J_{s_2}T\rangle_{Y_2}\big)
		\end{align}
		where our goal is to obtain $\langle J_{s_1}J_{s_2}T\rangle_{Y_0}$, $\langle J_{s_1}J_{s_2}T\rangle_{Y_2}$ ,$\langle J_{s_1}J_{s_2}T\rangle_\text{odd}$ in terms of the free theory correlators. We use the free and critical theory equations to identify two unknowns
		\begin{align}
			\langle J_{s_1}J_{s_2}T\rangle_{Y_0}=\langle J_{s_1}J_{s_2}T\rangle_\text{FF}\cr
			\langle J_{s_1}J_{s_2}T\rangle_{Y_2}=\langle J_{s_1}J_{s_2}T\rangle_\text{CB}
		\end{align}
		The pole equations take the form
		\begin{align}\label{jjtgeneralpole}
			&\langle J_{s_1}J_{s_2}T\rangle_\text{FF} - \langle J_{s_1}J_{s_2}T\rangle_\text{CB}=\epsilon\cdot\langle J_{s_1}J_{s_2}T\rangle_\text{odd}
			\cr
			&\langle J_{s_1}J_{s_2}T\rangle_\text{odd} = \epsilon\cdot \bigg(\langle J_{s_1}J_{s_2}T\rangle_\text{FF}- 
			\langle J_{s_1}J_{s_2}T\rangle_\text{CB}\bigg)
		\end{align}
		where the dot indicates a contraction of the Levi-Civita tensor with one of the indices of the spin $s_1$ current in the correlator.
		From the second  equation in \eqref{jjtgeneralpole} we obtain the remaining unknown piece of $\langle J_{s_1}J_{s_2}T\rangle_\text{QF}$
		\begin{equation}
			\langle J_{s_1}J_{s_2}T\rangle_\text{odd}=\epsilon \cdot \bigg(\langle J_{s_1}J_{s_2}T\rangle_\text{FF}- 
			\langle J_{s_1}J_{s_2}T\rangle_\text{CB}\bigg)
		\end{equation}
		Therefore we have the full correlator in the QF theory given by
		\begin{equation}
		\langle J_{s_1}J_{s_2}T\rangle_\text{QF}=\tilde{N}\frac{1}{1+\tilde{\lambda}^2}\left(\langle J_{s_1}J_{s_2}T\rangle_{FF}+\tilde{\lambda}\epsilon \cdot \left(\langle J_{s_1}J_{s_2}T\rangle_\text{FF}- 
			\langle J_{s_1}J_{s_2}T\rangle_\text{CB}\right)+\tilde{\lambda}^2\langle J_{s_1}J_{s_2}T\rangle_{CB}\right)
			\end{equation}
		
		It can be easily checked that the higher spin equations at all orders are satisfied by the above expression for $\langle J_{s_1}J_{s_2}T\rangle_\text{odd}$.
		\subsection{$\langle J_{s_1}J_{s_2}J_{s_3}\rangle_{QF}$}
		In this section we use HSEs to constrain the general 3-point correlator $\langle J_{s_1}J_{s_2}J_{s_3}\rangle$ with $s_i>2$ in the QF theory. We look at the Ward identity corresponding to the non-conservation equation of the spin-4 current in the correlator $\langle J_{s_1}J_{s_2}J_{s_3}\rangle_\text{QF}$. We use the relevant operator algebra \eqref{genqbalg} for $J_s$. The current non-conservation does not contribute to the RHS after we perform a large $N$ factorisation of the resulting 5-point function. The momentum space HSE becomes
		\begin{align}\label{jjjgeneralhse}
			&c_{s_1,s_1-2}p_{1\mu}p_{1\nu}p_{1\rho}p_{1\alpha_1}p_{1\alpha_2}\langle J_{\alpha_3\alpha_4...\alpha_{s_1}}J_{\beta_1\beta_2...\beta_{s_2}}J_{\gamma_1\gamma_2...\gamma_{s_3}}\rangle_{QF}\cr
			&+c_{s_1,s_1}p_{1\mu}p_{1\nu}p_{1\rho}\langle J_{\alpha_1\alpha_2...\alpha_{s_1}}J_{\beta_1\beta_2...\beta_{s_2}}J_{\gamma_1\gamma_2...\gamma_{s_3}}\rangle_{QF}\cr
			&+c_{s_1,s_1+2}p_{1\mu}\langle J_{\nu\rho\alpha_1\alpha_2...\alpha_{s_1}}J_{\beta_1\beta_2...\beta_{s_2}}J_{\gamma_1\gamma_2...\gamma_{s_3}}\rangle_{QF}+\lbrace1\leftrightarrow2\rbrace+\lbrace1\leftrightarrow3\rbrace=0
		\end{align}
		The corresponding FF equation is 
		\begin{align}
			&c_{s_1,s_1-2}p_{1\mu}p_{1\nu}p_{1\rho}p_{1\alpha_1}p_{1\alpha_2}\langle J_{\alpha_3\alpha_4...\alpha_{s_1}}J_{\beta_1\beta_2...\beta_{s_2}}J_{\gamma_1\gamma_2...\gamma_{s_3}}\rangle_{FF}\cr
			&+c_{s_1,s_1}p_{1\mu}p_{1\nu}p_{1\rho}\langle J_{\alpha_1\alpha_2...\alpha_{s_1}}J_{\beta_1\beta_2...\beta_{s_2}}J_{\gamma_1\gamma_2...\gamma_{s_3}}\rangle_{FF}\cr
			&+c_{s_1,s_1+2}p_{1\mu}\langle J_{\nu\rho\alpha_1\alpha_2...\alpha_{s_1}}J_{\beta_1\beta_2...\beta_{s_2}}J_{\gamma_1\gamma_2...\gamma_{s_3}}\rangle_{FF}+\lbrace1\leftrightarrow2\rbrace+\lbrace1\leftrightarrow3\rbrace=0
		\end{align}
		We use the following ansatz for the general 3-point spinning correlator 
		\begin{align}
			\langle J_{s_1}J_{s_2}J_{s_3}\rangle_{QF}=\frac{\tilde{N}}{1+\tilde{{\lambda}}^2} \bigg(\langle J_{s_1}J_{s_2}J_{s_3}\rangle_{Y_0}
			+\tilde{\lambda}\langle J_{s_1}J_{s_2}J_{s_3}\rangle_{\text{odd}}
			+\tilde{\lambda}^2\langle J_{s_1}J_{s_2}J_{s_3}\rangle_{Y_2}\bigg)
		\end{align}
		where $\langle J_{s_1}J_{s_2}J_{s_3}\rangle_{Y_0}$, $\langle J_{s_1}J_{s_2}J_{s_3}\rangle_{\text{odd}}$ and $\langle J_{s_1}J_{s_2}J_{s_3}\rangle_{Y_2}$, are the unknowns that we wish to find.
		We substitute this back into \eqref{jjjgeneralhse} and see that the $O(\tilde\lambda^0)$ equation is like the FF equation after we identify 
		\begin{align}
			\langle J_{\alpha_3\alpha_4...\alpha_{s_1}}J_{\beta_1\beta_2...\beta_{s_2}}J_{\gamma_1\gamma_2...\gamma_{s_3}}\rangle_{Y_0}=\langle J_{\alpha_3\alpha_4...\alpha_{s_1}}J_{\beta_1\beta_2...\beta_{s_2}}J_{\gamma_1\gamma_2...\gamma_{s_3}}\rangle_{FF}
		\end{align}
		The highest order equation in $\tilde{\lambda}$ in \eqref{jjjgeneralhse} is like the CB equation. This can be seen after we use the CB algebra to write the CB higher spin equation and identify
		\begin{align}
			\langle J_{\nu\rho\alpha_1\alpha_2...\alpha_{s_1}}J_{\beta_1\beta_2...\beta_{s_2}}J_{\gamma_1\gamma_2...\gamma_{s_3}}\rangle_{Y_2}=\langle J_{\nu\rho\alpha_1\alpha_2...\alpha_{s_1}}J_{\beta_1\beta_2...\beta_{s_2}}J_{\gamma_1\gamma_2...\gamma_{s_3}}\rangle_{CB}
		\end{align}
		We now use the following ansatz for $ \langle J_{s_1}J_{s_2}J_{s_3} \rangle_{\text{odd}} $
		\begin{align}\label{jsansatz}
			p_{1\mu}p_1\langle J_{\alpha_1\alpha_2...\alpha_{s_1}}J_{\beta_1\beta_2...\beta_{s_2}}J_{\gamma_1\gamma_2...\gamma_{s_3}}\rangle_{\text{odd}}=\epsilon_{\nu a b}p_{1a}p_{1\mu}\bigg(\langle J_{b\alpha_2\alpha_3.....\alpha_{s_1}}J_{\beta_1\beta_2...\beta_{s_2}}J_{\gamma_1\gamma_2...\gamma_{s_3}}\rangle_\text{FF}\cr
			-\langle J_{b\alpha_2\alpha_3.....\alpha_{s_1}}J_{\beta_1\beta_2...\beta_{s_2}}J_{\gamma_1\gamma_2...\gamma_{s_3}}\rangle_\text{CB}\bigg)
		\end{align}
		and plug it back into the HSE \eqref{jjjgeneralhse}.
		The above ansatz helps map the $O(\tilde{\lambda})$ equation to the free and critical theory equations. Therefore our ansatz of $\langle J_{s_1}J_{s_2}J_{s_3}\rangle_{\text{odd}}$ \eqref{jsansatz} is consistent with the above set of equations. Hence we could compute $\langle J_{s_1}J_{s_2}J_{s_3}\rangle_{QF}$ in terms of free theory correlators as
		\begin{align}
			\langle J_{s_1}J_{s_2}J_{s_3}\rangle_{QF}=\frac{\tilde{N}}{1+\tilde{\lambda}^2}\bigg(\langle J_{s_1}J_{s_2}J_{s_3}\rangle_{FF}  +\tilde{\lambda}\epsilon \cdot \bigg(\langle J_{s_1}J_{s_2}J_{s_3}\rangle_{FF}-\langle J_{s_1}J_{s_2}J_{s_3}\rangle_{CB}\bigg) +\tilde{\lambda}^2\langle J_{s_1}J_{s_2}J_{s_3}\rangle_{CB}\bigg)
		\end{align}
	    which is consistent with the result obtained by solving the conformal Ward identities in momentum space and reproduces the same result as the one obtained via perturbative calculations which are in general quite difficult.
	    
		If we now  write the same in the helicity basis, the steps parallel the analysis done for $\langle TTT\rangle$ in \ref{ttt}  which gives
		\begin{align}
			\langle J_{s_1}J_{s_2}J_{s_3}\rangle_\text{QF}&=\frac{\tilde{N}}{2}\left(\langle J_{s_1}J_{s_2}J_{s_3}\rangle_\text{FF+FB}+e^{-i\pi\lambda_f}\langle J_{s_1}J_{s_2}J_{s_3}\rangle_\text{FF-FB}\right)
		\end{align}
		and thus the presence of an anyonic phase is a generic occurrence in three-point functions. Now, if the spins of the operators satisfy the triangle inequality in \eqref{tineq} then from the discussion in \ref{hnh} we can write this result in terms of homogeneous/non-homogeneous parts of just the FF theory or just the FB theory.
	
		\section{Example-II: 4-point functions}\label{4ptex} 
In the previous section we solved for 3-point functions using HSEs. In this section we look into 4-point functions comprising of operator insertions with arbitrary spins.
\subsection{$\langle TOOO\rangle_\text{QF}$}\label{tooohse1}
 To start with, we look at the simple example of $\langle TOOO\rangle_{\text{QF}}.$ The result of this part was obtained first in position space by \cite{Li:2019twz}, in momentum space in \cite{Jain:2020puw} and in Mellin space in \cite{Silva:2021ece}. Below we work in momentum space.


\textbf{\color{blue}Step 1:} We choose the charge operator and the seed correlator to be $Q_4$ and $\langle OOOO \rangle$ respectively. Then we make use of the higher spin algebra \eqref{q4oqf} and \eqref{q4tqf} along with the current non conservation \eqref{ncj4qf} to obtain the following HSE in momentum space \footnote{Strictly speaking, we should have the constants from the algebra appearing in each term of the LHS but since our aim is only to \textit{map} the SBHS HSE to the free theory HSEs, we do not need to fix these constants to their numerical values. All we had to do was fix their $\tilde{\lambda}$ dependence.}
\begin{align}
	\label{qfhsetooo}
	\begin{split}
		&p_{1\mu}p_{1\nu}p_{1\rho}\langle OOOO\rangle_{\text{QF}}+g_{(\mu\nu}p_{1\rho)}p_1^2\langle OOOO\rangle_{\text{QF}}+\epsilon_{ a b(\mu}p_{1\nu}p_{1a}\langle T_{b\rho)}OOO\rangle_{\text{QF}}\\
		&+\lbrace1\leftrightarrow2\rbrace+\lbrace1\leftrightarrow3\rbrace+\lbrace1\leftrightarrow4\rbrace=\tilde{\lambda}p_{1(\mu}p_1\langle T_{\nu\rho)}OOO\rangle_{\text{QF}}+\lbrace1\leftrightarrow2\rbrace+\lbrace1\leftrightarrow3\rbrace+\lbrace1\leftrightarrow4\rbrace
	\end{split}
\end{align}
\textbf{\color{blue}Step 2:} We now write down the corresponding higher spin equations for the free theory
\begin{align}
	\label{ffhsetooo}
	\begin{split}
		&p_{1\mu}p_{1\nu}p_{1\rho}\langle OOOO\rangle_{\text{FF}}+g_{(\mu\nu}p_{1\rho)}p_1^2\langle OOOO\rangle_{\text{FF}}+\epsilon_{ a b(\mu}p_{1\nu}p_{1a}\langle T_{b\rho)}OOO\rangle_{\text{FF}}\\
		&+\lbrace1\leftrightarrow2\rbrace+\lbrace1\leftrightarrow3\rbrace+\lbrace1\leftrightarrow4\rbrace=0
	\end{split}	
\end{align}
and similarly for the CB theory,
\begin{align}
	\label{cbhsetooo}
	\begin{split}
		&p_{1(\mu}p_{1\nu}p_{1\rho)}\langle OOOO\rangle_{\text{CB}}+g_{(\mu\nu}p_{1\rho)}p_1^2\langle OOOO\rangle_{\text{CB}}+\lbrace1\leftrightarrow2\rbrace+\lbrace1\leftrightarrow3\rbrace+\lbrace1\leftrightarrow4\rbrace\\
		&=p_{1(\mu}p_1\langle T_{\nu\rho)}OOO\rangle_{\text{CB}}+\lbrace1\leftrightarrow2\rbrace+\lbrace1\leftrightarrow3\rbrace+\lbrace1\leftrightarrow4\rbrace
	\end{split}
\end{align}
\textbf{\color{blue}Step 3:} We consider the following ansatz for the correlators that appear in the HSE \eqref{qfhsetooo} 
\begin{align}
	\begin{split}
	&\langle TOOO\rangle_{\text{QF}}=\tilde{N}(1+\tilde{\lambda}^2)\bigg( \langle TOOO\rangle_{\text{$Y_0$}}+\tilde{\lambda}\langle TOOO\rangle_{\text{$Y_1$}} \bigg)\cr
		&\langle OOOO\rangle_{\text{QF}}=\tilde{N}(1+\tilde{\lambda}^2)(\langle OOOO\rangle_{\text{FF}}+\tilde{\lambda}^2\langle OOOO\rangle_{\text{CB}})
	\end{split}
\end{align}
\textbf{\color{blue}Step 4:} At $O(\tilde\lambda^0)$ of \eqref{qfhsetooo} we obtain the HSE to be identical to the FF theory equation  \eqref{ffhsetooo} which gives
\begin{align}
	\label{TOOOsoln}
	\langle TOOO\rangle_{\text{$Y_0$}}=\langle TOOO\rangle_{\text{FF}}
\end{align}
Similarly, the highest order equation( $O(\tilde{\lambda}^4)$ ) is identical to the CB equation \eqref{cbhsetooo} since the CB theory is the $\tilde{\lambda}\rightarrow\infty$ limit of the quasi fermionic theory. This happens after we identify 
\begin{align}
	\langle TOOO\rangle_{\text{$Y_1$}}=\langle TOOO\rangle_{\text{CB}}
\end{align}\\
\textbf{\color{blue}Step 5:}
We expand the HSE \eqref{qfhsetooo} around the point $\tilde{\lambda}=\pm i$ and obtain the following pole equations
\begin{align}
	\begin{split}\label{tooopol1}
		&\epsilon_{ a b(\mu}p_{1\nu}p_{1a}\langle T_{b\rho)}OOO\rangle_{\text{FF}}+\lbrace1\leftrightarrow2\rbrace+\lbrace1\leftrightarrow3\rbrace+\lbrace1\leftrightarrow4\rbrace\\
		&=p_{1(\mu}p_1\langle T_{\nu\rho)}OOO\rangle_{\text{CB}}+\lbrace1\leftrightarrow2\rbrace+\lbrace1\leftrightarrow3\rbrace+\lbrace1\leftrightarrow4\rbrace
	\end{split}
\end{align}
and
\begin{align}\label{toofnla}
	\begin{split}
		&\epsilon_{ a b(\mu}p_{1\nu}p_{1a}\langle T_{b\rho)}OOO\rangle_{\text{CB}}+\lbrace1\leftrightarrow2\rbrace+\lbrace1\leftrightarrow3\rbrace+\lbrace1\leftrightarrow4\rbrace\\
		&=p_{1(\mu}p_1\langle T_{\nu\rho)}OOO\rangle_{\text{FF}}+\lbrace1\leftrightarrow2\rbrace+\lbrace1\leftrightarrow3\rbrace+\lbrace1\leftrightarrow4\rbrace
	\end{split}
\end{align}
One can check again that \eqref{toofnla} can be mapped to FF and CB equations. In momentum space it looks complicated to map the above equation to FF and CB equations as it requires some epsilon transforms. However, going to spinor helicity variables solves this problem as in spinor helicity variables the epsilon transform becomes trivial \footnote{
One possible solution to these pole equations is
\begin{align}\label{tooosol1}
	\epsilon_{ a b(\mu}p_{1\nu}p_{1a}\langle T_{b\nu)}OOO\rangle_{\text{FF}}=\langle T_{\mu\nu)}OOO\rangle_{\text{CB}}
\end{align}
We chose this solution such that equation \eqref{tooopol1} is satisfied individually for each permutation. A direct verification of this result requires a proper analysis of the contact terms which we leave for future works. However, we note that certain HSEs demand that this individual equivalence such as that of $\langle TTTO\rangle ($\ref{tttoqfsec}). It is easy to check that using naive bootstrap argument involving single trace operator only, one gets \eqref{tooosol1}.}\label{TOOOFFCBept}.\\
\textbf{\color{blue}Step 6:}
Considering different order equations of \eqref{qfhsetooo}, it can be shown directly in momentum space that
\begin{align}
    \text{HSE at O}(\tilde{\lambda}^2)&=\text{FF HSE+CB HSE}
\end{align}
The $O(\tilde{\lambda})$ equation which is the same as the $O(\tilde{\lambda^3})$ equation is the same as an epsilon transform of the difference between the FF and CB equations. To see this we move to spinor helicity variables where it is easily seen. Thus we get,
\begin{align}
 \text{HSE at O}(\tilde{\lambda})= \text{HSE at O}(\tilde{\lambda}^3)=\pm i(\text{FF HSE-CB HSE})
\end{align}
For more details on how the mapping is carried out, refer to \ref{toooextra}.
Thus we obtain the following form for $\langle TOOO\rangle$ in the QF theory in spinor helicity variables
\begin{equation}
	\begin{aligned}
		\langle TOOO\rangle_{\text{QF}}&=\tilde{N}(1+\tilde{\lambda}^2)\bigg( \langle TOOO\rangle_{\text{FF}}+\tilde{\lambda}\langle TOOO\rangle_{\text{CB}} \bigg)
	\end{aligned}
\end{equation}

Using the same procedure, we can also generalize the above result for correlators with arbitrary spin as follows in spinor helicity variables.
		\begin{align}
			\langle J_sOOO\rangle_\text{QF}=\tilde{N}(1+\tilde{\lambda}^2)(\langle J_sOOO\rangle_\text{FF}+\tilde{\lambda}\langle J_sOOO\rangle_\text{CB})
		\end{align}
One can check the consistency of the result using the following argument: If we consider the divergence of the correlator,
		\begin{align}
			\langle \partial\cdot J_sOOO\rangle_\text{QF}=(1+\tilde{\lambda}^2)(\langle \partial\cdot J_sOOO\rangle_\text{FF}+\tilde{\lambda}\langle \partial\cdot J_sOOO\rangle_\text{CB})
		\end{align}
		due to the nonconservation of $J_s$ the left hand side is nonzero while in the RHS the free theory contribution drops out but the CB term is nonzero due to its nonconservation which is exactly the same as the left hand side.

\subsection{$\langle JJOO\rangle_{\text{QF}}$}\label{jjoosec}
Let us now see how higher spin equations can be used to solve for the 4-point spinning correlator $\langle J_{\alpha}J_{\beta}OO\rangle$ in the QF  theory. 
\\
\\
\noindent \textbf{\color{blue}Step 1:} We choose the charge operator and the seed correlator to be $Q_3$ and $\langle JOOO \rangle$ respectively. Then we make use of the higher spin algebra \eqref{q3oqf} and \eqref{q3jqf} along with the current non conservation \eqref{ncj3qf} to obtain the following HSE in momentum space 
\begin{align}
\begin{split}\label{jjoohse}
&\frac{1}{1+\tilde{\lambda}^2}\epsilon_{\alpha a(\mu}p_{1\nu)}p_1^a\langle O(p_1)O(p_2)O(p_3)O(p_4)\rangle_{QF}+ p_{1(\mu}\langle T_{\nu)\alpha}(p_1)O(p_2)O(p_3)O(p_4)\rangle_{QF}\\
	&+ p_{1\alpha}\langle T_{\mu\nu}(p_1)O(p_2)O(p_3)O(p_4)\rangle_{QF}+ 2( \epsilon_{ a b(\mu}p_2^ap_{2\nu)}\langle J_{\alpha}(p_1)J^b(p_2)O(p_3)O(p_4)\rangle_{QF}+\lbrace 2\leftrightarrow3\rbrace+\lbrace 2\leftrightarrow4\rbrace)\\
		&=\frac{16\tilde{\lambda}}{(1+\tilde{\lambda}^2)}\bigg[ p_{1(\mu}\langle J_{\alpha}J_{\nu)}\rangle_{QF}\langle O(p_1)O(p_2)O(p_3)O(p_4)\rangle_{QF} \\
	&+( p_{2(\mu}\langle O(p_2)O(-p_2)\rangle_{QF} \langle J_{\alpha}(p_1)J_{\nu)}(p_2)O(p_3)O(p_4)\rangle_{QF} +\lbrace 2\leftrightarrow3\rbrace+\lbrace 2\leftrightarrow4\rbrace) \bigg]
\end{split}
\end{align}

\noindent \textbf{\color{blue}Step 2:} We now write down the corresponding higher spin equation for the free theory
\begin{align}
	\label{ffhsejjoo}
	\begin{split}
		&\epsilon_{\alpha a(\mu}p_{1\nu)}p_{1a}\langle O(p_1)O(p_2)O(p_3)O(p_4)\rangle_{\text{FF}}+p_{1(\mu}\langle T_{\nu)\alpha}(p_1)O(p_2)O(p_3)O(p_4)\rangle_{\text{FF}}\\
		&+p_{1\alpha}\langle T_{\mu\nu}(p_1)O(p_2)O(p_3)O(p_4)\rangle_{\text{FF}}+\,2\bigg(\epsilon_{ a b(\mu}p_{2a}p_{2\nu)}\langle J_{\alpha}(p_1)J_{b}(p_2)O(p_3)O(p_4)\rangle_{\text{FF}}\\
		&+\lbrace 2\leftrightarrow3\rbrace+\lbrace 2\leftrightarrow4\rbrace \bigg)=0
	\end{split}
\end{align}
and similarly we write down the higher spin equation for the CB theory.
\begin{align}
	\label{cbhsejjoo}
	\begin{split}
		&p_{1(\mu}\langle T_{\nu)\alpha}(p_1)O(p_2)O(p_3)O(p_4)\rangle_{\text{CB}}+p_{1\alpha}\langle T_{\mu\nu}(p_1)O(p_2)O(p_3)O(p_4)\rangle_{\text{CB}}\\
		&=\bigg[ 16 p_{1(\mu}\langle J_{\alpha}J_{\nu)}\rangle_{\text{FF}}(\langle O(p_1)O(p_2)O(p_3)O(p_4)\rangle_{\text{CB}}) \\
		&+\bigg( 2p_{2(\mu}p_2 (\langle J_{\alpha}(p_1)J_{\nu)}(p_2)O(p_3)O(p_4)\rangle_{\text{CB}})+\lbrace 2\leftrightarrow3\rbrace+\lbrace 2\leftrightarrow4\rbrace \bigg)\bigg]
	\end{split}
\end{align}
\noindent \textbf{\color{blue}Step 3:}  We consider the following ansatz for the correlators that appear in the HSE in \eqref{jjoohse} \cite{Kalloor:2019xjb,Gandhi:2021gwn}.
\begin{align}\label{jjooansatz}
	\langle J_{\alpha}J_{\beta}OO\rangle_{\text{QF}} &= \tilde{N}\langle J_{\alpha}J_{\beta}OO\rangle_{Y_0}+\tilde{N}\tilde{\lambda}\langle J_{\alpha}J_{\beta}OO\rangle_{\text{odd}}+\tilde{N}\tilde{\lambda}^2\langle J_{\alpha}J_{\beta}OO\rangle_{Y_2}
\end{align} 
\textbf{\color{blue}Step 4:} At $O(\tilde\lambda^0)$ of \eqref{jjoohse} the HSE is identical to the FF theory equation  \eqref{ffhsejjoo} which gives
\begin{align}\label{y0y2jjoo}
	\begin{split}
		&\langle JJOO\rangle_{Y_0}=\langle JJOO\rangle_{\text{FF}}
	\end{split}
\end{align}
Similarly, the highest order equation namely the $O(\tilde{\lambda}^4)$ is identical to the CB equation \eqref{cbhsejjoo} since the CB theory is the $\tilde{\lambda}\rightarrow\infty$ limit of the QF theory. This happens after we identify 
\begin{align}
	\begin{split}
		&\langle JJOO\rangle_{Y_2}=\langle JJOO\rangle_{\text{CB}}
	\end{split}
\end{align}
\textbf{\color{blue}Step 5:} We expand the HSE around the point $\tilde{\lambda}=\pm i$ and obtain the following pole equations
\begin{align}\label{pejjooa}
	&\epsilon_{ a b(\mu}p_{2a} p_{2\nu)}\bigg(\langle J_{\alpha}(p_1)J_b(p_2)O(p_3)O(p_4)\rangle_{\text{FF}}-\langle J_{\alpha}(p_1)J_b(p_2)O(p_3)O(p_4)\rangle_{\text{CB}}\bigg)+\lbrace 2\leftrightarrow3\rbrace+\lbrace 2\leftrightarrow4\rbrace\cr
	&\hspace{0cm}= p_2 p_{2(\mu}\langle J_{\alpha}(p_1)J_{\nu)}(p_2)O(p_3)O(p_4)\rangle_{\text{odd}}+\lbrace 2\leftrightarrow3\rbrace+\lbrace 2\leftrightarrow4\rbrace
\end{align}
and
\begin{align}\label{pejjoob}
	&\epsilon_{ a b(\mu}p_{2a} p_{2\nu)}\langle J_{\alpha}(p_1)J_b(p_2)O(p_3)O(p_4)\rangle_{\text{odd}}+\lbrace 2\leftrightarrow3\rbrace+\lbrace 2\leftrightarrow4\rbrace\cr
	&=-p_2 p_{2(\mu}\bigg(\langle J_{\alpha}(p_1)J_{\nu)}(p_2)O(p_3)O(p_4)\rangle_{\text{FF}}-\langle J_{\alpha}(p_1)J_{\nu)}(p_2)O(p_3)O(p_4)\rangle_{\text{CB}}\bigg)+\lbrace 2\leftrightarrow3\rbrace+\lbrace 2\leftrightarrow4\rbrace\cr
\end{align}
To solve this equation, as in the case of the three point function, we neglect the permutation first and then
we dot the first pole equation \eqref{pejjooa} with $p_{2\mu}$ and make use of the trivial transverse Ward identity for $\langle JJOO\rangle$ to obtain the following expression for the unknown $\langle JJOO\rangle_{\text{odd}}$\footnote{This result is also consistent with what was obtained in \cite{Kalloor:2019xjb}. We thank R.R. Kalloor and Trivko Kukolj for informing us of this.}
\begin{align}
	\label{polesoln}
	\langle J_{\alpha}J_{\nu}OO\rangle_{\text{odd}}=\frac{\epsilon_{\nu a b}p_{2a}}{p_2}(\langle J_{\alpha}J_{b}OO\rangle_{\text{FF}}-\langle J_{\alpha}J_{b}OO\rangle_{\text{CB}}).
\end{align}
\noindent \textbf{\color{blue}Step 6:} We now use our results to map slightly broken HSE to free theory HSE. We can show that, 
\begin{align}
    \text{HSE at O}(\tilde{\lambda})&=\text{CB HSE}\cr
    \text{HSE at O}(\tilde{\lambda}^2)&=\text{FF HSE}
\end{align}
For more details on how the mapping is carried out, refer to \ref{jjooextra}. The above analysis gives us all the unknowns that appear in the ansatz \eqref{jjooansatz} for the correlator purely in terms of free theory results. Thus, we can write the QF correlator as
\begin{align}
\langle J_{\alpha}J_{\nu}OO\rangle_{\text{QF}}=\tilde{N}\langle J_{\alpha}J_{\nu}OO\rangle_{\text{FF}}+\tilde{N}\tilde{\lambda}\frac{\epsilon_{\nu a b}p_{2a}}{p_2}(\langle J_{\alpha}J_{b}OO\rangle_{\text{FF}}-\langle J_{\alpha}J_{b}OO\rangle_{\text{CB}})+\tilde{N}\tilde{\lambda}^2\langle J_{\alpha}J_{\nu}OO\rangle_{\text{CB}}
	\end{align}

Now, to write the expression in spinor helicity we dot the above expression with the null polarization tensors $z_1^{\alpha}z_2^{\nu}$. In spinor helicity variables we have $\langle\epsilon\cdot JJOO\rangle\rightarrow i\langle JJOO\rangle$ and thus
\begin{align}\label{jjooqf}
	\begin{split}
		\langle JJOO\rangle_\text{QF}&=\tilde{N}\bigg[\langle JJOO\rangle_{\text{FF}}-i\tilde{\lambda}(\langle JJOO\rangle_{\text{FF}}-\langle JJOO\rangle_{\text{CB}})+\tilde{\lambda}^2\langle JJOO\rangle_{\text{CB}}\bigg]\\
		&=\frac{\tilde{N}}{2}\bigg((1+\tilde{\lambda}^2)\langle JJOO\rangle_{\text{FF+CB}}+(1-2i\tilde{\lambda}-\tilde{\lambda}^2)\langle JJOO\rangle_{\text{FF-CB}}\bigg)
	\end{split}
\end{align}
To get a more intuitive representation of this, we chose a different normalization and use \eqref{cplconst} to write the QF correlator as 
\begin{align}
		\langle JJOO\rangle_{\text{QF}}&=\frac{\tilde{N}}{2(1+\tilde{\lambda}^2)}\bigg[ (1+\tilde{\lambda}^2)\langle JJOO\rangle_{\text{FF+CB}}+(1-2i\tilde{\lambda}-\tilde{\lambda}^2)\langle JJOO\rangle_{\text{FF-CB}} \bigg]\cr
		&=\frac{\tilde{N}}{2}\bigg[ \langle JJOO\rangle_{\text{FF+CB}}-\frac{\tilde{\lambda}+i}{\tilde{\lambda}-i}\langle JJOO\rangle_{\text{FF-CB}}\bigg]\cr
	&=\frac{\tilde{N}}{2}\bigg[ \langle JJOO\rangle_{\text{FF+CB}}+e^{-i\pi \lambda_f}\langle JJOO\rangle_{\text{FF-CB}}\bigg]\cr
	&=\tilde{N}e^{-\frac{i\pi \lambda_f}{2}}\big[\cos{\frac{\pi\lambda_f}{2}}\langle  JJOO\rangle_\text{FF}+i\sin{\frac{\pi\lambda_f}{2}}\langle  JJOO\rangle_\text{CB}\big]
\end{align}

\subsection{$\langle TTTO\rangle$}\label{tttoqfsec}
\textbf{\color{blue}Step 1:~~}The charge operator and seed correlator that we choose are $Q_4$ and $\langle TTOO\rangle$ respectively.\\
After using \eqref{q4tqf},\eqref{q4oqf} and \eqref{ncj4qf}, the HSE in momentum space reads,
\vspace{-0.5cm}
\begin{align}\label{tttoqfhse}
\begin{split}
    &\bigg[p_{1\mu}p_{1\nu}p_{1\rho}\langle T_{\alpha\beta}(p_1)T_{\gamma\sigma}(p_2)O(p_3)O(p_4)\rangle+ p_{1\mu}p_{1\nu}p_{1\alpha}\langle T_{\rho\beta}(p_1)T_{\gamma\sigma}(p_2)O(p_3)O(p_4)\rangle\\& +p_{1\mu}p_{1\alpha}p_{1\beta}\langle T_{\nu\rho}(p_1)T_{\gamma\sigma}(p_2)O(p_3)O(p_4)\rangle+p_{1\alpha}\langle J_{4\mu\nu\rho\beta}(p_1)T_{\gamma\sigma}(p_2)O(p_3)O(p_4)\rangle\\&+p_{1\mu}\langle J_{4\nu\rho\alpha\beta}(p_1)T_{\gamma\sigma}(p_2)O(p_3)O(p_4)\rangle+\frac{1}{1+\tilde{\lambda}^2}\epsilon_{\nu\alpha a}p_{1a}p_{1\mu}p_{1\rho}p_{1\beta}\langle O(p_1)T_{\gamma\sigma}(p_2)O(p_3)O(p_4)\rangle\\&+\{1\leftrightarrow 2,(\alpha,\beta)\leftrightarrow(\gamma,\sigma)\}\bigg]\\&+\bigg[p_{3\mu}p_{3\nu}p_{3\rho}\langle T_{\alpha\beta}(p_1)T_{\gamma\sigma}(p_2)O(p_3)O(p_4)\rangle+\epsilon_{\mu a b}p_{3\nu}p_{3a}\langle T_{\alpha\beta}(p_1)T_{\gamma\sigma}(p_2)T_{\rho b}(p_3)O(p_4)\rangle+\{3\leftrightarrow 4\}\bigg]\\
    &=\frac{\tilde{\lambda}}{1+\tilde{\lambda}^2}\bigg[p_{3\mu}\langle O(-p_3)O(p_3)\rangle\langle T_{\alpha\beta}(p_1)T_{\gamma\sigma}(p_2)T_{\nu\rho}(p_3)O(p_4)\rangle+\{3\leftrightarrow 4\}\\
    &+p_{1\mu}\langle T_{\nu\rho}(-p_1)T_{\alpha\beta}(p_1)\rangle\langle O(p_1)T_{\gamma\sigma}(p_2)O(p_3)O(p_4)\rangle+\{(1\leftrightarrow 2),(\alpha,\beta)\leftrightarrow(\gamma,\sigma)\}\bigg]
    \end{split}
\end{align}
\textbf{\color{blue}Step 2:~~}The FF and CB equations are,
\begin{align}\label{ffhsettto}
     &\bigg[p_{1\mu}p_{1\nu}p_{1\rho}\langle T_{\alpha\beta}(p_1)T_{\gamma\sigma}(p_2)O(p_3)O(p_4)\rangle_{\text{FF}}+ p_{1\mu}p_{1\nu}p_{1\alpha}\langle T_{\rho\beta}(p_1)T_{\gamma\sigma}(p_2)O(p_3)O(p_4)\rangle_{\text{FF}}\notag\\& +p_{1\mu}p_{1\alpha}p_{1\beta}\langle T_{\nu\rho}(p_1)T_{\gamma\sigma}(p_2)O(p_3)O(p_4)\rangle_{\text{FF}}+p_{1\alpha}\langle J_{4\mu\nu\rho\beta}(p_1)T_{\gamma\sigma}(p_2)O(p_3)O(p_4)\rangle_{\text{FF}}\notag\\&+p_{1\mu}\langle J_{4\nu\rho\alpha\beta}(p_1)T_{\gamma\sigma}(p_2)O(p_3)O(p_4)\rangle_{\text{FF}}+\epsilon_{\nu\alpha a}p_{1a}p_{1\mu}p_{1\rho}p_{1\beta}\langle O(p_1)T_{\gamma\sigma}(p_2)O(p_3)O(p_4)\rangle_{\text{FF}}\notag\\&+\{1\leftrightarrow 2,(\alpha,\beta)\leftrightarrow(\gamma,\sigma)\}\bigg]\notag\\&+\bigg[p_{3\mu}p_{3\nu}p_{3\rho}\langle T_{\alpha\beta}(p_1)T_{\gamma\sigma}(p_2)O(p_3)O(p_4)\rangle_{\text{FF}}+\epsilon_{\mu a b}p_{3\nu}p_{3a}\langle T_{\alpha\beta}(p_1)T_{\gamma\sigma}(p_2)T_{\rho b}(p_3)O(p_4)\rangle_{\text{FF}}+\{3\leftrightarrow 4\}\bigg]=0
\end{align}
and
\begin{align}\label{cbhsettto}
     &\bigg[p_{1\mu}p_{1\nu}p_{1\rho}\langle T_{\alpha\beta}(p_1)T_{\gamma\sigma}(p_2)O(p_3)O(p_4)\rangle_{\text{CB}}+ p_{1\mu}p_{1\nu}p_{1\alpha}\langle T_{\rho\beta}(p_1)T_{\gamma\sigma}(p_2)O(p_3)O(p_4)\rangle_{\text{CB}}\notag\\& +p_{1\mu}p_{1\alpha}p_{1\beta}\langle T_{\nu\rho}(p_1)T_{\gamma\sigma}(p_2)O(p_3)O(p_4)\rangle_{\text{CB}}+p_{1\alpha}\langle J_{4\mu\nu\rho\beta}(p_1)T_{\gamma\sigma}(p_2)O(p_3)O(p_4)\rangle_{\text{CB}}\notag\\&+p_{1\mu}\langle J_{4\nu\rho\alpha\beta}(p_1)T_{\gamma\sigma}(p_2)O(p_3)O(p_4)\rangle_{\text{CB}}+\{1\leftrightarrow 2,(\alpha,\beta)\leftrightarrow(\gamma,\sigma)\}\bigg]\notag\\&+\bigg[p_{3\mu}p_{3\nu}p_{3\rho}\langle T_{\alpha\beta}(p_1)T_{\gamma\sigma}(p_2)O(p_3)O(p_4)\rangle_{\text{CB}}+\{3\leftrightarrow 4\}\bigg]\notag\\
    &-\bigg[p_{3\mu}\langle O(-p_3)O(p_3)\rangle_{\text{CB}}\langle T_{\alpha\beta}(p_1)T_{\gamma\sigma}(p_2)T_{\nu\rho}(p_3)O(p_4)\rangle_{\text{CB}}+\{3\leftrightarrow 4\}\notag\\
    &+p_{1\mu}\langle T_{\nu\rho}(-p_1)T_{\alpha\beta}(p_1)\rangle_{\text{CB}}\langle O(p_1)T_{\gamma\sigma}(p_2)O(p_3)O(p_4)\rangle_{\text{CB}}+\{(1\leftrightarrow 2),(\alpha,\beta)\leftrightarrow(\gamma,\sigma)\}\bigg]=0
\end{align}

\textbf{\color{blue}Step 3:~~}Our ansatz is,
\begin{align}\label{tttoansatz}
    \langle TTTO\rangle_{\text{QF}}=\frac{\tilde{N}}{1+\tilde{\lambda}^2}\bigg(\langle TTTO\rangle_{\text{Y0}}+\tilde{\lambda}\langle TTTO\rangle_{\text{Y1}}+\tilde{\lambda}^2\langle TTTO\rangle_{\text{Y2}}+\tilde{\lambda}^3\langle TTTO\rangle_{\text{Y3}}\bigg)
\end{align}
\textbf{\color{blue}Step 4:~~}
Comparing the lowest and highest order equations with the FF and CB equations give,
\begin{align}
    &\langle TTTO\rangle_{Y0}=\langle TTTO\rangle_{\text{FF}}\notag\\
    &\langle TTTO\rangle_{Y3}=\langle TTTO\rangle_{\text{CB}}
\end{align}
\textbf{\color{blue}Step 5:~~}
Expanding the HSE about $\tilde{\lambda}=i$, we obtain the pole equations,
\begin{align}\label{pettto}
    &p_3 p_{3\mu}\bigg(-\langle T_{\alpha\beta}(p_1)T_{\gamma\sigma}(p_2)T_{\nu\rho}(p_3)O(p_4)\rangle_{FF}+\langle T_{\alpha\beta}(p_1)T_{\gamma\sigma}(p_2)T_{\nu\rho}(p_3)O(p_4)\rangle_{\text{Y2}}+\{3\leftrightarrow 4\}\bigg)\notag\\&+\epsilon_{\nu a b}p_{3a}p_{3\mu}\bigg(-\langle T_{\alpha\beta}(p_1)T_{\gamma\sigma}(p_2)T_{\rho b}(p_3)O(p_4)\rangle_{CB}+\langle T_{\alpha\beta}(p_1)T_{\gamma\sigma}(p_2)T_{\rho b}(p_3)O(p_4)\rangle_{\text{Y1}}\bigg)=0\notag\\
    &p_3 p_{3\mu}\bigg(\langle T_{\alpha\beta}(p_1)T_{\gamma\sigma}(p_2)T_{\nu\rho}(p_3)O(p_4)\rangle_{CB}-\langle T_{\alpha\beta}(p_1)T_{\gamma\sigma}(p_2)T_{\nu\rho}(p_3)O(p_4)\rangle_{\text{Y1}}+\{3\leftrightarrow 4\}\bigg)\notag\\&+\epsilon_{\nu a b}p_{3a}p_{3\mu}\bigg(-\langle T_{\alpha\beta}(p_1)T_{\gamma\sigma}(p_2)T_{\rho b}(p_3)O(p_4)\rangle_{FF}+\langle T_{\alpha\beta}(p_1)T_{\gamma\sigma}(p_2)T_{\rho b}(p_3)O(p_4)\rangle_{\text{Y2}}\bigg)=0
\end{align}
One of the solutions of the pole equations is,
\begin{align}\label{petttosol}
    &\langle TTTO\rangle_{Y1}=\langle TTTO\rangle_{\text{CB}}\notag\\
    &\langle TTTO\rangle_{Y2}=\langle TTTO\rangle_{\text{FF}}
\end{align}
\textbf{\color{blue}Step 6:~~}Plugging these into the HSE yields in momentum space,
\begin{align}
    &\text{HSE at~}O(\tilde{\lambda}^2)=\text{FF HSE }+\text{CB HSE}\notag\\
    &\text{HSE at~}O(\tilde{\lambda}^3)=\text{HSE at~}O(\tilde{\lambda})
\end{align}
Further, in spinor helicity variables one can show that the $O(\tilde{\lambda})$ maps to the epsilon transform of the difference of the FF and CB HSEs \footnote{Here the mapping requires that $\langle TOOO\rangle_{\text{CB}}=\epsilon\cdot \langle TOOO\rangle_{\text{FF}}$ as we discussed in the footnote \ref{TOOOFFCBept}.}.
Thus we have,
\begin{align}\label{tttoqf}
    \langle TTTO\rangle=\tilde{N}\bigg(\langle TTTO\rangle_{\text{FF}}+\tilde{\lambda}\langle TTTO\rangle_{\text{CB}}\bigg)
\end{align}
For more details on how the mapping is done, please refer to \ref{tttoextra}.
\subsection{$\langle JJJJ\rangle_{\text{QF}}$}\label{jjjjsec}
We now deal with the case of $\langle JJJJ\rangle_{\text{QF}}$ which has 4 spinning operator insertions and we will see that its analysis is quite involved compared to the previous examples. In this case, as we will see, the pole equations by themselves are insufficient to solve for the QF correlator and thus we have to look at the different order equations in order to arrive at a solution. 
\\
\noindent\textbf{\color{blue}Step 1:} We choose the charge operator and the seed correlator to be $Q_3$ and $\langle JJJO \rangle$ respectively. Then we make use of the higher spin algebra ((B.22) and (B.23)) along with the current non conservation (B.15) to obtain the following HSE in momentum space 
\begin{align}\label{jjjjqfhse}
	\begin{split}
		&\bigg[ \epsilon_{\alpha a(\mu}p_{1\nu)}p_{1a}\langle O(p_1)J_{\beta}(p_2)J_{\gamma}(p_3)O(p_4)\rangle_{\text{QF}}+p_{1(\mu}\langle T_{\nu)\alpha}(p_1)J_{\beta}(p_2)J_{\gamma}(p_3)O(p_4)\rangle_{\text{QF}}\\&+p_{1\alpha}\langle T_{\mu\nu}(p_1)J_{\beta}(p_2)J_{\gamma}(p_3)O(p_4)\rangle_{\text{QF}}+\lbrace 1\leftrightarrow2,\alpha\leftrightarrow\beta\rbrace+\lbrace1\leftrightarrow3,\alpha\leftrightarrow\gamma\rbrace\bigg]\\
		&+2\epsilon_{ a b(\mu}p_{4a}p_{4\nu)}\langle J_{\alpha}(p_1)J_{\beta}(p_2)J_{\gamma}(p_3)J_b(p_4)\rangle_{\text{QF}}=\frac{-16i\tilde{\lambda}}{1+\tilde{\lambda}^2}\bigg[ \bigg( p_{1(\mu}\langle J_{\alpha}J_{\nu)}\rangle_{\text{QF}}\langle O(p_1)J_{\beta}(p_2)J_{\gamma}(p_3)O(p_4)\rangle_{\text{QF}}\\&+\lbrace 1\leftrightarrow2,\alpha\leftrightarrow\beta\rbrace+\lbrace1\leftrightarrow3,\alpha\leftrightarrow\gamma\rbrace \bigg)\bigg]+p_{4(\mu}\langle O(p_4)O(-p_4)\rangle\langle J_{\alpha}(p_1)J_{\beta}(p_2)J_{\gamma}(p_3)J_{\nu)}(p_4)\rangle_{\text{QF}}
	\end{split}
\end{align}
\textbf{\color{blue}Step 2:} We now write down the corresponding HSE for the FF theory
\begin{align}\label{jjjjffhse}
	\begin{split}
		&\bigg[ \epsilon_{\alpha a(\mu}p_{1\nu)}p_{1a}\langle O(p_1)J_{\beta}(p_2)J_{\gamma}(p_3)O(p_4)\rangle_{\text{FF}}+p_{1(\mu}\langle T_{\nu)\alpha}(p_1)J_{\beta}(p_2)J_{\gamma}(p_3)O(p_4)\rangle_{\text{FF}}\\&+p_{1\alpha}\langle T_{\mu\nu}(p_1)J_{\beta}(p_2)J_{\gamma}(p_3)O(p_4)\rangle_{\text{FF}}+(1\leftrightarrow2,\alpha\leftrightarrow\beta)+(1\leftrightarrow3,\alpha\leftrightarrow\gamma)\bigg]\\
		&+2\epsilon_{ a b(\mu}p_{4a}p_{4\nu)}\langle J_{\alpha}(p_1)J_{\beta}(p_2)J_{\gamma}(p_3)J_b(p_4)\rangle_{\text{FF}}=0
	\end{split}
\end{align}
and similarly for the CB theory.
\small
\begin{align}\label{jjjjcbhse}
	\begin{split}
		&\bigg[p_{1(\mu}\langle T_{\nu)\alpha}(p_1)J_{\beta}(p_2)J_{\gamma}(p_3)O(p_4)\rangle_{\text{CB}}+p_{1\alpha}\langle T_{\mu\nu}(p_1)J_{\beta}(p_2)J_{\gamma}(p_3)O(p_4)\rangle_{\text{CB}}+(1\leftrightarrow2,\alpha\leftrightarrow\beta)+(1\leftrightarrow3,\alpha\leftrightarrow\gamma)\bigg]\\
		&=-16i\bigg[ p_{1(\mu}\langle J_{\alpha}(p_1)J_{\nu)}(-p_1)\rangle_{\text{FF}}\langle O(p_1)J_{\beta}(p_2)J_{\gamma}(p_3)O(p_4)\rangle_{\text{CB}}+(1\leftrightarrow2,\alpha\leftrightarrow\beta)+(1\leftrightarrow3,\alpha\leftrightarrow\gamma)\bigg]\\
		&+2ip_{4(\mu}p_4\langle J_{\alpha}(p_1)J_{\beta}(p_2)J_{\gamma}(p_3)J_{\nu)}(p_4)\rangle_{\text{CB}}
	\end{split}
\end{align}
\normalsize
\textbf{\color{blue}Step 3:} We consider the following ansatz for the correlator $\langle JJJJ\rangle_{\text{QF}}$ in momentum space(momentum and spin labels suppressed for clarity) expanded in orders of the coupling as in \cite{Kalloor:2019xjb}
\begin{align}
	\begin{split}
		\langle JJJJ\rangle=\frac{\tilde{N}}{(1+\tilde{\lambda}^2)^2}\bigg( \langle JJJJ\rangle_{Y_0}+\tilde{\lambda}\langle JJJJ\rangle_{Y_1}+\tilde{\lambda}^2\langle JJJJ\rangle_{Y_2}+\tilde{\lambda}^3\langle JJJJ\rangle_{Y_3}+\tilde{\lambda}^4\langle JJJJ\rangle_{Y_4} \bigg)
	\end{split}
\end{align} 
\textbf{\color{blue}Step 4:} We substitute the above ansatz into the HSE and at $O(\tilde\lambda^0)$ of \eqref{jjjjqfhse} the HSE is that of the FF theory \eqref{jjjjffhse} which gives
\begin{align}
	\langle JJJJ\rangle_{Y_0}=\langle JJJJ\rangle_{\text{FF}}
\end{align}
Similarly, the highest order equation namely the $O(\tilde{\lambda}^5)$ equation is that of the CB \eqref{jjjjcbhse} since the CB theory is the $\tilde{\lambda}\rightarrow\infty$ limit of the QF theory. Thus, we identify
\begin{align}
	\langle JJJJ\rangle_{Y_4}=\langle JJJJ\rangle_{\text{CB}}
\end{align}
\textbf{\color{blue}Step 5:} We expand the HSE  around the point $\tilde{\lambda}=\pm i$ and obtain the following pole equations
\begin{align}
	\begin{split}
		\epsilon_{ a b(\mu}p_{4a}p_{4\nu)}\bigg(\langle J_{\alpha}J_{\beta}J_{\gamma}J_b\rangle_{Y_0}-&\langle J_{\alpha}J_{\beta}J_{\gamma}J_b\rangle_{Y_2}+\langle J_{\alpha}J_{\beta}J_{\gamma}J_b\rangle_{Y_4}\bigg)=\\
		    &2ip_{4(\mu}p_4\bigg(\langle J_{\alpha}J_{\beta}J_{\gamma}J_{\nu)}\rangle_{Y_3}-\langle J_{\alpha}J_{\beta}J_{\gamma}J_{\nu}\rangle_{Y_1}\bigg)
	\end{split}
\end{align}
and
\begin{align}
	\begin{split}
		\epsilon_{ a b(\mu}p_{4a}p_{4\nu)}\bigg(\langle J_{\alpha}J_{\beta}J_{\gamma}J_b\rangle_{Y_1}&-\langle J_{\alpha}J_{\beta}J_{\gamma}J_b\rangle_{Y_3}\bigg)=\\
		    &2ip_{4(\mu}p_4\bigg(\langle J_{\alpha}J_{\beta}J_{\gamma}J_{\nu)}\rangle_{Y_0}-\langle J_{\alpha}J_{\beta}J_{\gamma}J_{\nu}\rangle_{Y_2}+\langle J_{\alpha}J_{\beta}J_{\gamma}J_{\nu}\rangle_{Y_4}\bigg)
	\end{split}
\end{align}
A possible solution to the above pole equations is
\begin{align}\label{jjjjordsol}
	\langle JJJJ\rangle_{Y_1}&=\langle JJJJ\rangle_{Y_3}\cr
	\langle JJJJ\rangle_{Y_2}&=\langle JJJJ\rangle_{Y_0}+\langle JJJJ\rangle_{Y_4}
\end{align}
In \cite{Kalloor:2019xjb} it was noticed by direct computation in specific  kinematic regime that these relations holds between various components. See Appendix B of \cite{Kalloor:2019xjb} for more details.
As alluded to earlier, the $\langle JJJJ\rangle$ pole equations are insufficient to solve for the entire correlator. The remaining expression in $\langle JJJJ\rangle_{\text{QF}}$ that is still to be determined is $\langle JJJJ\rangle_{Y_1}$. To get the expression of $Y_1$ in spinor helicity, consider the combination $O(\tilde{\lambda}^3)-2O(\tilde{\lambda})$ and dot it with the null polarization tensors $z_{1\alpha}z_{2\beta}z_{3\gamma}z_{2\mu}z_{2\nu}$ and then convert the equation to spinor helicity variables. Then the expression for $Y_1$ that we get is the following
\begin{align}
    \langle JJJJ\rangle_{Y_1}=\epsilon\cdot(\langle JJJJ\rangle_{Y_0}-\langle JJJJ\rangle_{Y_4}).
\end{align}
\\
\noindent \textbf{\color{blue}Step 6:} We now use our results to map the SBHS equation to the free theory HSEs\footnote{$\langle J_\nu(-p_1)J_\alpha(p_1)\rangle\langle O(p_1)J_\beta(p_2)J_\gamma(p_3)O(p_4)\rangle_{\text{odd}}+\epsilon_{a\alpha\nu}p_{1a}\langle O(p_1)J_\beta(p_2)J_\gamma(p_3)O(p_4)\rangle_{\text{CB-FF}}$ is left over while performing the mapping. However, we note that $\langle O(p_1)J_\beta(p_2)J_\gamma(p_3)O(p_4)\rangle_{\text{odd}}$ is given by an epsilon transform of the difference between the corresponding CB and FF correlators as in \eqref{polesoln} and $\epsilon_{a\alpha\nu}p_{1a}\langle O(p_1)J_\beta(p_2)J_\gamma(p_3)O(p_4)\rangle_{\text{CB-FF}}$ is also similar to the epsilon transformation of the difference between the CB and FF correlators. Thus, we believe that a more careful analysis of these terms should get rid of this leftover expression and render the mapping exact. Further evidence is provided by the fact that this issue does not appear in the general case $\langle J_{s_1}J_{s_2}J_{s_3}J_{s_4}\rangle$ as we discuss below. In order to understand this, we believe that we need a proper understanding of the structure of spinning 4 point correlators which we will come back to in a future work. } We can show that in spinor helicity variables, 
\begin{align}
    \text{HSE at O}(\tilde{\lambda})=&\text{CB HSE}\cr
    \text{HSE at O}(\tilde{\lambda}^2)=&2~\text{FF HSE}\cr
    \text{HSE at O}(\tilde{\lambda}^3)=&2~\text{CB HSE}\cr 
    \text{HSE at O}(\tilde{\lambda}^4)=&\text{FF HSE}
\end{align}
For more details on how the mapping is carried out, refer to \ref{jjjjextra}. Using the answers obtained at different orders in coupling \eqref{jjjjordsol} we get the full correlator to be
\begin{align}\label{jjjjqf}
		\langle JJJJ\rangle_{\text{QF}}&=\frac{\tilde{N}}{(1+\tilde\lambda^2)}\bigg[\langle JJJJ\rangle_\text{FF}+\tilde\lambda \langle\epsilon \cdot JJJJ\rangle_\text{FF-CB}+\tilde\lambda^2\langle JJJJ\rangle_\text{CB}\bigg]\cr
\end{align}
To get a more intuitive form, we can move to spinor-helicity variables by dotting with null polarisation tensors, which gives us $\langle JJJJ\rangle_{Y_1}$ to be  
\begin{align}
	\langle JJJJ\rangle_{Y_1}=i\bigg(\langle JJJJ\rangle_{Y_0}-\langle JJJJ\rangle_{Y_4}\bigg)
\end{align}
Thus, using the solutions of the pole equation\eqref{jjjjordsol} and the expression for $Y_1$ that we just obtained, we can write the full QF correlator in spinor-helicity variables as
\begin{align}
	\langle JJJJ\rangle_\text{QF}&=\frac{\tilde{N}}{(1+\tilde{\lambda}^2)^2}\bigg( \langle JJJJ\rangle_\text{FF}+i\tilde{\lambda}\langle JJJJ\rangle_\text{FF$-$CB}+\tilde{\lambda}^2\langle JJJJ\rangle_\text{FF+CB}\cr
	&+i\tilde{\lambda}^3\langle JJJJ\rangle_\text{FF$-$CB}+\tilde{\lambda}^4\langle JJJJ\rangle_\text{CB} \bigg)\cr
	&=\frac{\tilde{N}}{2}\bigg[ \langle JJJJ\rangle_\text{FF+CB}+\frac{\tilde{\lambda}+i}{\tilde{\lambda}-i}\langle JJJJ\rangle_\text{FF$-$CB} \bigg]\cr
		&=\frac{\tilde{N}}{2}\bigg[ \langle JJJJ\rangle_\text{FF+CB}+e^{-i\pi\lambda_f}\langle JJJJ\rangle_\text{FF$-$CB} \bigg]\cr
		&=\tilde{N}e^{-\frac{i\pi \lambda_f}{2}}\big[\cos{\frac{\pi\lambda_f}{2}}\langle JJJJ\rangle_{\text{FF}}+i \sin{\frac{\pi\lambda_f}{2}}\langle JJJJ\rangle_{\text{CB}}\big]
\end{align}
which yet again, exhibits the characteristic anyonic behaviour.

\subsection{$\langle J_{s_1}J_{s_2}J_{s_3}J_{s_4}\rangle_{\text{QF}}$}
Now we proceed to the general case of spinning correlators, $\langle J_{s_1}J_{s_2}J_{s_3}J_{s_4}\rangle_{\text{QF}}$ for $s_i>2$. For this analysis we choose to work with the action of the spin-4 current. A similar analysis can also be done with the spin-3 current.

\textbf{\color{blue}Step 1:} To write the HSE for general 3-point spinning correlator with $s_i>2$, consider the action of $Q_{\mu\nu\rho}$ on the correlator $\langle J_{s_1}J_{s_2}J_{s_3}J_{s_4}\rangle_{\text{QF}}$ which gives us
\begin{equation}
	\begin{aligned}
		Q_{\mu\nu\rho}&\langle J_{s_1}~J_{s_2}~J_{s_3}~J_{s_4}\rangle_{\text{QF}}=c_{s_1,s_1-2}~~p_{1\mu}p_{1\nu}p_{1\rho}p_{1(\alpha_1}p_{1\alpha_2}\langle J_{s_1-2}~J_{s_2}~J_{s_3}~J_{s_4}\rangle_{\text{QF}}\\
		&+c_{s_1,s_1}~~p_{1\mu}p_{1\nu}p_{1\rho}\langle J_{s_1}~J_{s_2}~J_{s_3}~J_{s_4}\rangle_{\text{QF}}+c_{s_1,s_1+2}~~p_{1(\mu}\langle J_{s_1+2}~J_{s_2}~J_{s_3}~J_{s_4}\rangle_{\text{QF}}\\
		&+(1\leftrightarrow2)+(1\leftrightarrow3)+(1\leftrightarrow4)=0
	\end{aligned}
\end{equation}
We have set the RHS for this HSE to zero because in the decomposition of six-point correlator $\langle O~T~J_{s_1}~J_{s_2}~J_{s_3}~J_{s_4}\rangle$ coming from the current divergence, the only possible decompositions at the large N limit are those involving multiplication of two 3-point functions and no possible decomposition into a 2-point function multiplied by a 4-point function. Also, we neglect the 3-point contributions while considering the 4-point HSE. Thus, the above HSE is valid upto 3-point functions.

\textbf{\color{blue}Step 2:} We now write down the corresponding HSEs for the free theory,
\begin{equation}
	\begin{aligned}
		&c_{s_1,s_1-2}~p_{1\mu}p_{1\nu}p_{1\rho}p_{\alpha_1}p_{1\alpha_2}\langle J_{s_1-2}~J_{s_2}~J_{s_3}~J_{s_4}\rangle_{\text{FF}}+c_{s_1,s_1}p_{1\mu}p_{1\nu}p_{1\rho}\langle J_{s_1}~J_{s_2}~J_{s_3}~J_{s_4}\rangle_{\text{FF}}\\
		&+c_{s_1,s_1+2}p_{1\mu}\langle J_{s_1+2}~J_{s_2}~J_{s_3}~J_{s_4}\rangle_{\text{FF}}+(1\leftrightarrow2)+(1\leftrightarrow3)+(1\leftrightarrow4)=0
	\end{aligned}
\end{equation}
and for the CB theory,
\begin{equation}
	\begin{aligned}
		&c_{s_1,s_1-2}~~p_{1\mu}p_{1\nu}p_{1\rho}p_{\alpha_1}p_{1\alpha_2}\langle J_{s_1-2}~J_{s_2}~J_{s_3}~J_{s_4}\rangle_{\text{CB}}+c_{s_1,s_1}p_{1\mu}p_{1\nu}p_{1\rho}\langle J_{s_1}~J_{s_2}~J_{s_3}~J_{s_4}\rangle_{\text{CB}}\\
		&+c_{s_1,s_1+2}~~p_{1\mu}\langle J_{s_1+2}~J_{s_2}~J_{s_3}~J_{s_4}\rangle_{\text{CB}}+(1\leftrightarrow2)+(1\leftrightarrow3)+(1\leftrightarrow4)=0
	\end{aligned}
\end{equation}

\noindent \textbf{\color{blue}Step 3:} We consider the following ansatz for  the general spinning correlator \cite{Kalloor:2019xjb}
\begin{equation}
	\begin{aligned}
		&\langle J_{s_1}J_{s_2}J_{s_3}J_{s_4}\rangle_{\text{QF}}=\frac{\tilde{N}}{(1+\tilde{\lambda}^2)^2}\bigg( \langle J_{s_1}J_{s_2}J_{s_3}J_{s_4}\rangle_{Y_0}+\tilde{\lambda}\langle J_{s_1}J_{s_2}J_{s_3}J_{s_4}\rangle_{Y_1}+\tilde{\lambda}^2\langle J_{s_1}J_{s_2}J_{s_3}J_{s_4}\rangle_{Y_2}\\
		&+\tilde{\lambda}^3\langle J_{s_1}J_{s_2}J_{s_3}J_{s_4}\rangle_{Y_3}+\tilde{\lambda}^4\langle J_{s_1}J_{s_2}J_{s_3}J_{s_4}\rangle_{Y_4} \bigg)
	\end{aligned}
\end{equation}
we get the equation
\begin{equation}
	\begin{aligned}\label{4ptgenhse}
		&c_{s_1,s_1-2}\frac{p_{1\mu}p_{1\nu}p_{1\rho}p_{1(\alpha_1}p_{1\alpha_2}}{(1+\tilde{\lambda}^2)^2}\bigg(\langle J_{s_1-2}J_{s_2}J_{s_3}J_{s_4}\rangle_{Y_0}+\tilde{\lambda}\langle J_{s_1-2}J_{s_2}J_{s_3}J_{s_4}\rangle_{Y_1}+\tilde{\lambda}^2\langle J_{s_1-2}J_{s_2}J_{s_3}J_{s_4}\rangle_{Y_2}\\
		&+\tilde{\lambda}^3\langle J_{s_1-2}J_{s_2}J_{s_3}J_{s_4}\rangle_{Y_3}+\tilde{\lambda}^4\langle J_{s_1-2}J_{s_2}J_{s_3}J_{s_4}\rangle_{Y_4}\bigg)\\
		&+c_{s_1,s_1}\frac{p_{1\mu}p_{1\nu}p_{1\rho}}{(1+\tilde{\lambda}^2)^2}\bigg(\langle J_{s_1}J_{s_2}J_{s_3}J_{s_4}\rangle_{Y_0}+\tilde{\lambda}\langle J_{s_1}J_{s_2}J_{s_3}J_{s_4}\rangle_{Y_1}+\tilde{\lambda}^2\langle J_{s_1}J_{s_2}J_{s_3}J_{s_4}\rangle_{Y_2}+\tilde{\lambda}^3\langle J_{s_1}J_{s_2}J_{s_3}J_{s_4}\rangle_{Y_3}\\
		&+\tilde{\lambda}^4\langle J_{s_1}J_{s_2}J_{s_3}J_{s_4}\rangle_{Y_4}\bigg)\\
		&+c_{s_1,s_1+2}\frac{p_{1(\mu}}{(1+\tilde{\lambda}^2)^2}\bigg(\langle J_{s_1+2}J_{s_2}J_{s_3}J_{s_4}\rangle_{Y_0}+\tilde{\lambda}\langle J_{s_1+2}J_{s_2}J_{s_3}J_{s_4}\rangle_{Y_1}+\tilde{\lambda}^2\langle J_{s_1+2}J_{s_2}J_{s_3}J_{s_4}\rangle_{Y_2}\\
		&+\tilde{\lambda}^3\langle J_{s_1+2}J_{s_2}J_{s_3}J_{s_4}\rangle_{Y_3}+\tilde{\lambda}^4\langle J_{s_1+2}J_{s_2}J_{s_3}J_{s_4}\rangle_{Y_4}\bigg)+(1\leftrightarrow2)+(1\leftrightarrow3)+(1\leftrightarrow4)=0
	\end{aligned}
\end{equation}

\noindent\textbf{\color{blue}Step 4:} Now as before we look at the lowest and the highest order equations of the HSE and note that they match the FF and CB equations respectively after we identify
\begin{align}
	\langle J_{s_1}J_{s_2}J_{s_3}J_{s_4}\rangle_{Y_0}=\langle J_{s_1}J_{s_2}J_{s_3}J_{s_4}\rangle_{\text{FF}},\cr
	\langle J_{s_1}J_{s_2}J_{s_3}J_{s_4}\rangle_{Y_4}=\langle J_{s_1}J_{s_2}J_{s_3}J_{s_4}\rangle_{\text{CB}}
\end{align}
\textbf{\color{blue}Step 5:} From the pole equations we get the following relations among the unknowns
\begin{align}
	&\langle J_{s_1}J_{s_2}J_{s_3}J_{s_4}\rangle_{Y_2}=\langle J_{s_1}J_{s_2}J_{s_3}J_{s_4}\rangle_{Y_0}+\langle J_{s_1}J_{s_2}J_{s_3}J_{s_4}\rangle_{Y_4}\\
	&\langle J_{s_1}J_{s_2}J_{s_3}J_{s_4}\rangle_{Y_3}=\langle J_{s_1}J_{s_2}J_{s_3}J_{s_4}\rangle_{Y_1}
\end{align}
\textbf{\color{blue}Step 6:}We now use our results to map the slightly broken HSE to the free theory HSEs. We have already mapped the lowest and highest order equations in $\tilde\lambda$ and now we consider the intermediate order equations

\noindent $O(\tilde{\lambda}$):
\begin{equation}
	\begin{aligned}
		&c_{s_1,s_1-2}p_{1\mu}p_{1\nu}p_{1\rho}p_{\alpha_1}p_{1\alpha_2}\langle J_{s_1-2}J_{s_2}J_{s_3}J_{s_4}\rangle_{Y_1}+c_{s_1,s_1}p_{1\mu}p_{1\nu}p_{1\rho}\langle J_{s_1}J_{s_2}J_{s_3}J_{s_4}\rangle_{Y_1}\cr
		&+c_{s_1,s_1+2}p_{1\mu}\langle J_{s_1+2}J_{s_2}J_{s_3}J_{s_4}\rangle_{\text{FF}}+(1\leftrightarrow2)+(1\leftrightarrow3)+(1\leftrightarrow4)=0
	\end{aligned}
\end{equation}
$O(\tilde{\lambda}^2$):
\begin{equation}
	\begin{aligned}
		&c_{s_1,s_1-2}p_{1\mu}p_{1\nu}p_{1\rho}p_{\alpha_1}p_{1\alpha_2}\langle J_{s_1-2}J_{s_2}J_{s_3}J_{s_4}\rangle_{Y_2}+c_{s_1,s_1}p_{1\mu}p_{1\nu}p_{1\rho}\langle J_{s_1}J_{s_2}J_{s_3}J_{s_4}\rangle_{Y_2}\\
		&+c_{s_1,s_1+2}p_{1\mu}\langle J_{s_1+2}J_{s_2}J_{s_3}J_{s_4}\rangle_{Y_2}+(1\leftrightarrow2)+(1\leftrightarrow3)+(1\leftrightarrow4)=0
	\end{aligned}
\end{equation}
$O(\tilde{\lambda}^3$):
\begin{equation}
	\begin{aligned}
		&c_{s_1,s_1-2}p_{1\mu}p_{1\nu}p_{1\rho}p_{\alpha_1}p_{1\alpha_2}\langle J_{s_1-2}J_{s_2}J_{s_3}J_{s_4}\rangle_{Y_3}+c_{s_1,s_1}p_{1\mu}p_{1\nu}p_{1\rho}\langle J_{s_1}J_{s_2}J_{s_3}J_{s_4}\rangle_{Y_3}\\
		&+c_{s_1,s_1+2}p_{1\mu}\langle J_{s_1+2}J_{s_2}J_{s_3}J_{s_4}\rangle_{Y_3}+(1\leftrightarrow2)+(1\leftrightarrow3)+(1\leftrightarrow4)=0
	\end{aligned}
\end{equation}
Generalizing the results for $\langle JJJJ\rangle_{QF}$, we propose
\small
\begin{align}\label{genordsol}
	&\langle J_{s_1}J_{s_2}J_{s_3}J_{s_4}\rangle_{Y_2}=\langle J_{s_1}J_{s_2}J_{s_3}J_{s_4}\rangle_{\text{FF}}+\langle J_{s_1}J_{s_2}J_{s_3}J_{s_4}\rangle_{\text{CB}}\cr
	&\langle J_{s_1}J_{s_2}J_{s_3}J_{s_4}\rangle_{Y_3}=\langle J_{s_1}J_{s_2}J_{s_3}J_{s_4}\rangle_{Y_1}=\epsilon_{\eta_1 a b}p_{4a}(\langle J_{s_1}J_{s_2}J_{s_3}J_{s_4}\rangle_{\text{FF}}-\langle J_{s_1}J_{s_2}J_{s_3}J_{s_4}\rangle_{\text{CB}})
\end{align}
\normalsize
Plugging these expressions into the different order equations, we can map the equations to the free and critical theory higher spin equations as 
\begin{align}
	&\text{HSE at O}(\tilde{\lambda})=\text{HSE at O}(\tilde{\lambda}^3)=\text{FF HSE}-\text{CB HSE}\\
	&\text{HSE at O}(\tilde{\lambda}^2)=\text{FF HSE}+\text{CB HSE}
\end{align}

Finally, inputting \eqref{genordsol} in the expression for the QF correlator we get it to be
\begin{align}
		\langle J_{s_1}J_{s_2}J_{s_3}J_{s_4}\rangle_{\text{QF}}&=\frac{\tilde{N}}{(1+\tilde\lambda^2)}\bigg[\langle J_{s_1}J_{s_2}J_{s_3}J_{s_4}\rangle_\text{FF}+\tilde\lambda \langle\epsilon \cdot J_{s_1}J_{s_2}J_{s_3}J_{s_4}\rangle_\text{FF-CB}+\tilde\lambda^2\langle J_{s_1}J_{s_2}J_{s_3}J_{s_4}\rangle_\text{CB}\bigg]\cr
\end{align}
and converting it into spinor helicity, we get,
\begin{align}
	\begin{split}
		&\langle J_{s_1}J_{s_2}J_{s_3}J_{s_4}\rangle_{\text{QF}}=\frac{\tilde{N}}{(1+\tilde{\lambda}^2)^2}\bigg( \langle J_{s_1}J_{s_2}J_{s_3}J_{s_4}\rangle_{\text{FF}}-i\tilde{\lambda}\langle J_{s_1}J_{s_2}J_{s_3}J_{s_4}\rangle_{\text{FF$-$CB}}\\
		&+\tilde{\lambda}^2\langle J_{s_1}J_{s_2}J_{s_3}J_{s_4}\rangle_{\text{FF+CB}}-i\tilde{\lambda}^3\langle J_{s_1}J_{s_2}J_{s_3}J_{s_4}\rangle_{\text{FF$-$CB}}+\tilde{\lambda}^4\langle J_{s_1}J_{s_2}J_{s_3}J_{s_4}\rangle_{\text{CB}} \bigg)
	\end{split}
\end{align}
Now using \eqref{cplconst} we finally have
\begin{align}\label{4ptgensph}
    \begin{split}
		\langle J_{s_1}J_{s_2}J_{s_3}J_{s_4}\rangle_{\text{QF}}&=\frac{\tilde{N}}{2}\bigg[ \langle J_{s_1}J_{s_2}J_{s_3}J_{s_4}\rangle_{\text{FF+CB}}+e^{-i\pi\lambda_{f}}\langle J_{s_1}J_{s_2}J_{s_3}J_{s_4}\rangle_{\text{FF-CB}} \bigg]
	\end{split}
\end{align}
	
Thus we conclude that the correlation functions in a SBSH theory can be computed using the free theory correlators. We would like to point out once again that the solution found in this paper is one particular solution to the HSE. However we also saw how the HSEs for different correlators are interconnected. For example, if we want to solve the HSE for $\langle JJJJ\rangle$ we need information about $\langle JJTO \rangle$ and $\langle JJOO \rangle.$ To solve for $\langle JJTO\rangle$ we need information about $\langle JJOO \rangle$ and $\langle J_3 J OO \rangle$ and so on. Thus, we see that we need to solve an interconnected set of HSEs. We suspect that finding out any other solution which would simultaneously satisfy all such constraints would really be a difficult job.\\

For an extension of the above analysis to the five point case, please refer to appendix \ref{5ptappendix}.
\section{Ward Takahashi Identity of QF theory in terms of free theory}\label{WtFr1}
In this section we show that the non-conservation Ward Takahashi Identity can be obtained from the Ward Takahashi Identity of conserved currents for the boson and fermion. To see this concretely consider the Ward Takahashi identity for the correlator $\langle J_{s_1}J_{s_2}J_{s_3}\rangle_{\text{QF}}$. To analyse this, let's write the correlator in terms of the free theory and parity odd correlators:
\begin{align}
\langle J_{s_1}J_{s_2}J_{s_3}\rangle_{\text{QF}}=\frac{\tilde{N}}{1+\tilde{\lambda}^2}\bigg(\langle J_{s_1}J_{s_2}J_{s_3}\rangle_{\text{FF}}+\tilde{\lambda}^2\langle J_{s_1}J_{s_2}J_{s_3}\rangle_{\text{FB}}+\tilde{\lambda}\langle J_{s_1}J_{s_2}J_{s_3}\rangle_{\text{odd}}\bigg)
\end{align} 
Now consider the Ward Takahashi identity
\begin{align}\label{wtqf11}
\langle \partial. J_{s_1}J_{s_2}J_{s_3}\rangle_{\text{QF}}=\frac{1}{1+\tilde{\lambda}^2}\bigg(\langle \partial\cdot J_{s_1}J_{s_2}J_{s_3}\rangle_{\text{FF}}+\tilde{\lambda}^2\langle \partial\cdot J_{s_1}J_{s_2}J_{s_3}\rangle_{\text{FB}}+\tilde{\lambda}\langle \partial\cdot J_{s_1}J_{s_2}J_{s_3}\rangle_{\text{odd}}\bigg)
\end{align}
From the pole equation of $\langle J_{s_1}J_{s_2}J_{s_3}\rangle$\eqref{jsansatz}, we know that parity odd correlator is
\begin{align}\label{jjjansatz}
\langle J_{s_1}J_{s_2}J_{s_3}\rangle_{odd}=\epsilon\cdot(\langle J_{s_1}J_{s_2}J_{s_3}\rangle_{FF}-\langle J_{s_1}J_{s_2}J_{s_3}\rangle_{FB})
\end{align}
Inputting \eqref{jjjansatz} in \eqref{wtqf11}, we get
\begin{align}
\begin{split}\label{wtqf}
\langle &\partial. J_{s_1}J_{s_2}J_{s_3}\rangle_{\text{QF}}=\\
	&\frac{1}{1+\tilde{\lambda}^2}\bigg(\langle \partial\cdot J_{s_1}J_{s_2}J_{s_3}\rangle_{\text{FF}}+\tilde{\lambda}^2\langle \partial\cdot J_{s_1}J_{s_2}J_{s_3}\rangle_{\text{FB}}+\tilde{\lambda}~\epsilon\cdot(\langle \partial\cdot J_{s_1}J_{s_2}J_{s_3}\rangle_{\text{FF}}-\langle \partial\cdot J_{s_1}J_{s_2}J_{s_3}\rangle_{\text{FB}})\bigg)
\end{split}
\end{align}
In the LHS, we get two contributions, one from the Ward Takahashi identity of the conservation and the other from the Ward Takahashi identity of the non-conservation.
\begin{align}
\begin{split}
\langle \partial\cdot J_{s_1}J_{s_2}J_{s_3}\rangle_{\textbf{c}}+\langle \partial\cdot J_{s_1}J_{s_2}J_{s_3}\rangle_{\textbf{nc}}&=\frac{1}{1+\tilde{\lambda}^2}\bigg(\langle \partial\cdot J_{s_1}J_{s_2}J_{s_3}\rangle_{\text{FF}}+\tilde{\lambda}^2\langle \partial\cdot J_{s_1}J_{s_2}J_{s_3}\rangle_{\text{FB}}\\
	&+\tilde{\lambda}~\epsilon\cdot (\langle \partial\cdot J_{s_1}J_{s_2}J_{s_3}\rangle_{\text{FF}}-\langle \partial\cdot J_{s_1}J_{s_2}J_{s_3}\rangle_{\text{FB}})\bigg)
\end{split}
\end{align}
Now, the Ward Takahashi contribution from the conservation in the LHS comes from the free fermion and free boson terms. Thus, we can write
\begin{align}
\begin{split}
&\frac{1}{1+\tilde{\lambda}^2}\bigg(\langle \partial\cdot J_{s_1}J_{s_2}J_{s_3}\rangle_{\text{FF}}+\tilde{\lambda}^2\langle \partial\cdot J_{s_1}J_{s_2}J_{s_3}\rangle_{\text{FB}}\bigg)+\langle \partial\cdot J_{s_1}J_{s_2}J_{s_3}\rangle_{\textbf{nc}}=\\
	&\frac{1}{1+\tilde{\lambda}^2}\bigg(\langle \partial\cdot J_{s_1}J_{s_2}J_{s_3}\rangle_{\text{FF}}+\tilde{\lambda}^2\langle \partial\cdot J_{s_1}J_{s_2}J_{s_3}\rangle_{\text{FB}}+\tilde{\lambda}~\epsilon\cdot(\langle \partial\cdot J_{s_1}J_{s_2}J_{s_3}\rangle_{\text{FF}}-\langle \partial\cdot J_{s_1}J_{s_2}J_{s_3}\rangle_{\text{FB}})\bigg)
\end{split}
\end{align}
Thus, the FF and FB contributions cancel out from both sides and we are left with a expression for the Ward Takahashi identity from the non-conservation as
\begin{align}
\begin{split}\label{wtnoncons}
\langle \partial\cdot J_{s_1}J_{s_2}J_{s_3}\rangle_{\textbf{nc}}=\frac{1}{1+\tilde{\lambda}^2}\tilde{\lambda}~ \epsilon\cdot(\langle \partial\cdot J_{s_1}J_{s_2}J_{s_3}\rangle_{FF}-\langle \partial\cdot J_{s_1}J_{s_2}J_{s_3}\rangle_{FB})
\end{split}
\end{align}
Now, we can have two different cases. For the case of outside the triangle, $s_1>s_2+s_3$, the Ward Takahashi identity for free fermion and free boson are not equal and the the Ward Takahashi identity for the correlator has contribution from non-conservation. But, for the case of inside the triangle, i.e, $s_1\le s_2+s_3$, the Ward Takahashi identity for the FF and FB are identical and hence looking at \eqref{wtnoncons}, we can conclude that the contribution to the Ward Takahashi identity fom the non-conservation in this case is vanishing. It is consistent with the fact that inside the triangle , the parity odd part of $\langle J_{s_1}J_{s_2}J_{s_3}\rangle$ is homogeneous. It can also be explicitly checked that the non-conservation part does not contribute to the Ward Takahashi identity in this case.

Finally, returning to \eqref{wtqf}
\begin{align}
\text{WT}_{\text{QF}}=\frac{1}{1+\tilde{\lambda}^2}\bigg( \text{WT}_{\text{FF}}+\tilde{\lambda}^2\text{WT}_{\text{FB}}+\tilde{\lambda}~\epsilon\cdot (\text{WT}_{\text{FF}}-\text{WT}_{\text{FB}}) \bigg)
\end{align}
Now, to get an interesting perspective, we go to spinor helicity variables where the epsilon transform of the correlator just becomes $i$ times the correlator. We get
\begin{align}
&\text{WT}_{\text{QF}}=\frac{1}{1+\tilde{\lambda}^2}\bigg( \text{WT}_{\text{FF}}+\tilde{\lambda}^2\text{WT}_{\text{FB}}-\tilde{\lambda}~i (\text{WT}_{\text{FF}}-\text{WT}_{\text{FB}}) \bigg)\\
	&=\frac{1}{1+\tilde{\lambda}^2}\bigg( \text{WT}_{\text{FF}}(1-i\tilde{\lambda})+i\tilde{\lambda}(1-i\tilde{\lambda})\text{WT}_{\text{FB}}) \bigg)
\end{align}
This can be modified and written as 
\begin{align}
\text{WT}_{\text{QF}}=\frac{1}{2}\bigg( (\text{WT}_{\text{FF}}+\text{WT}_{\text{FB}})-\frac{\tilde{\lambda}+i}{\tilde{\lambda}-i}(\text{WT}_{\text{FF}}-\text{WT}_{\text{FB}}) \bigg)
\end{align}
Using the definition, $\tilde{\lambda}=\tan{\frac{\pi\lambda_f}{2}}$ we get
\begin{align}
\text{WT}_{\text{QF}}&=\frac{1}{2}\bigg( (\text{WT}_{\text{FF}}+\text{WT}_{\text{FB}})+e^{-i\pi\lambda_f}(\text{WT}_{\text{FF}}-\text{WT}_{\text{FB}}) \bigg)\cr
&=\frac{1}{2}\bigg( \left(1+e^{-i\pi\lambda_f}\right)\text{WT}_{\text{FF}}+\left(1-e^{-i\pi\lambda_f}\right)\text{WT}_{\text{FB}} \bigg)\cr
&=e^{- i \pi \frac{\lambda_f}{2}}\bigg( \cos\frac{\pi \lambda_f}{2}\text{WT}_{\text{FF}}+i \sin\frac{\pi \lambda_f}{2}\text{WT}_{\text{FB}} \bigg)
\end{align}
This gives us an interesting interpretation that the QF correlator can be expressed in terms of the free theory correlators with modified Ward Takahashi identity. As is clear from the previous equation, at $\lambda_f=0$ we get the WT of the FF whereas for $\lambda_f=1$ we get the WT identity for the FB. For the case when we are inside the triangle $s_i\le s_j+s_k$ we have $WT_{FF} = WT_{FB}$ which gives
\begin{equation}
 \text{WT}_{\text{QF}}= e^{- i \pi \lambda_f}  ~ WT_{FF}.
\end{equation}
A similar calculation can be repeated for higher point functions.

This mapping of WT identities for SBHS theories to the free theories suggests that the non conservation should be accounted for by doing simple modification to the HS algebra for the conserved currents. In Appendix \ref{SBHSinTermsofFree}, we elaborate on this topic.

\section{Discussion}\label{dis}
In this paper, we develop a methodology to systematically solve SBHS equations for spinning correlation functions. Our solution involves mapping the SBHS theory correlation functions to the free theory correlation functions. We demonstrate our procedure first with three point functions which reproduce known results. We then apply our methodology to four point functions of spinning correlators. We show that the four point functions take on a remarkably simple form and  can be mapped to the free theory correlation functions. Our analysis in this paper demonstrates the usefulness of momentum  space or spinor helicity variables to deal with slightly broken HSEs. Our main strategy was to map the slightly broken HSE to exactly conserved HSEs which in turn  maps the interacting correlation functions in terms of the free theory correlators\footnote{We should emphasize that slightly broken HS equation can have more solutions. We checked some more possibilities but they all lead to inconsistencies. At the level of three point functions, the answers are unique as can be verified directly by using conformal symmetry \cite{Jain:2021vrv}.}. We also showed that one can rewrite the SBHS algebra in terms of the exactly conserved HS algebra with an appropriate redefinition of currents.  

There are a number of interesting followups on our work. 
\subsection*{Perturbative calculation}
In this paper we have made use of SBHS symmetry to compute the correlation functions. It would be interesting to understand the same result from perturbative calculation in CS matter theory. In general, even for two and three point functions,
the perturbative calculation is very complicated and is only performed for a few simple cases that too in special kinemetic regions \cite{Aharony:2012nh,GurAri:2012is}. For four point functions see, \cite{Bedhotiya:2015uga,Turiaci:2018nua,Yacoby:2018yvy,Inbasekar:2019wdw,Kalloor:2019xjb}. However our results suggests that there should be a better way of formulating the perturbative calculation. For example, if we consider a two point function, the full answer has both parity even and parity odd results. The parity odd results appear only in the odd loop calculation.  The parity odd results are just an epsilon transform of the parity even free theory result and since any even loop gives the free theory result, there should be a direct way to see this happening at the perturbative level.  We display our results in terms of perturbative calculations in a fermion coupled to CS gauge field by the following set of diagrams.
\begin{figure}[H]
		\centering
		\def\svgwidth{4cm}
		\includegraphics[scale=1]{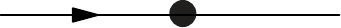}
		\caption{We denote the effective propagator with a black dot. This black dot contains all possible quantum corrections.}
		\label{dprop}
\end{figure}
\begin{figure}[H]
		\centering
		\def\svgwidth{4cm}
		\includegraphics[scale=0.8]{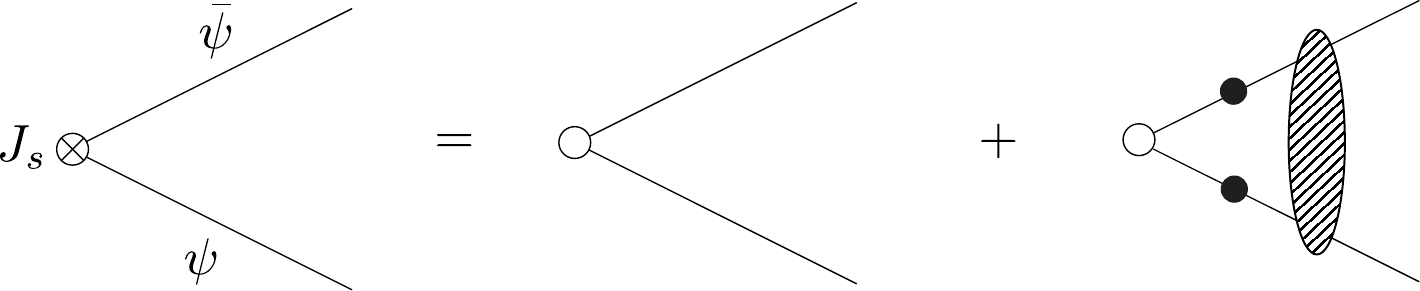}
		\caption{The vertex in interacting theory}
		\label{dvertex}
\end{figure}
\begin{figure}[H]
		\centering
		\def\svgwidth{7cm}
		\includegraphics[scale=0.8]{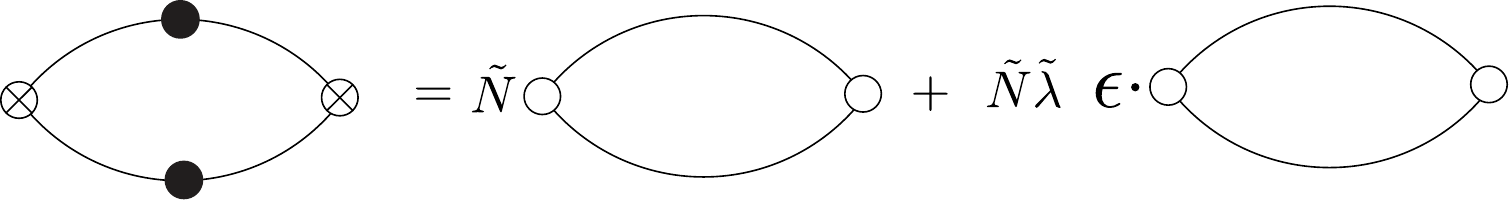}
		\caption{The left hand side of the above figure represents a fully quantum corrected two point function. The right hand side expresses the same as a free theory correlation function and its epsilon transform.}
		\label{d2ptqf}
\end{figure}
\begin{figure}[H]
		\centering
		\def\svgwidth{7cm}
		\includegraphics[scale=0.6]{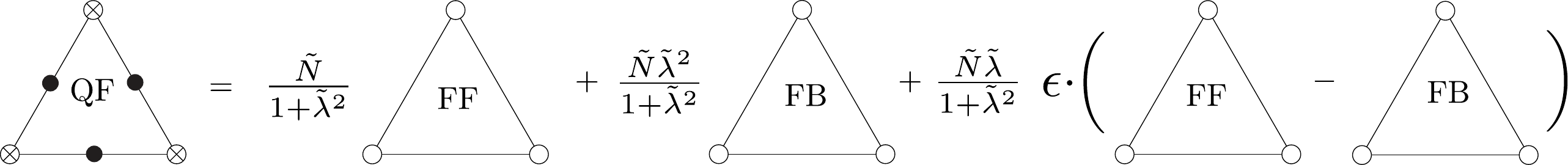}
    	\caption{The left hand side of the above figure represents a fully quantum corrected three point function in the quasi-fermionic theory. The right side expresses the same as a combination of the free fermionic and the free bosonic correlation functions and their epsilon transform.}
		\label{d3ptqf}
\end{figure}

\subsection*{Hilbert space interpretation}
In \cite{Minwalla:2022sef}, it was demonstrated that the thermal partition function
of CS matter theories at large N  can be given a simple   Hilbert space interpretation. They showed that the partition function can be understood in terms of an associated ungauged theory Hilbert space subject to a Gauss' law constraint in CS matter theory. This Hilbert space interpretation is at the level of single and multiparticle fock space and in some sense, the high energy sector. On the other hand currents can be interpreted as bound states and hence can be understood to appear in the low energy sector of Hilbert space. The fact that correlation functions of these currents can be understood from the free theories and the fact that they show anyonic behaviour indicates that for the low energy sector as well there should be some interesting Hilbert space interpretation \footnote{We thank S. Minwalla for a discussion on a discussion related to this point.}.

\subsection*{Bosonization, Anyon and Anyonic current}
In 2D bosonization, there is a direct map between currents of the fermionic and bosonic theories. In three dimensions, there may not exist a direct map as in this case bosonization is realised through anyonic behaviour. However results in this paper indicate that one might be able to define effective anyonic currents which interpolate between fermionic and bosonic currents. One should as well be able to use these anyonic currents to define some effective Wick contraction to get the correlation functions.
\subsection*{CFT bootstrap in momentum or spinor-helicity variables}
Three point function of conserved currents  can have three structures
\begin{equation}
  \langle J_{s_1} J_{s_2} J_{s_3}\rangle= n_F\langle J_{s_1} J_{s_2} J_{s_3}\rangle_{FF}+ n_B\langle J_{s_1} J_{s_2} J_{s_3}\rangle_{FB}  +n_{Odd} \langle J_{s_1} J_{s_2} J_{s_3}\rangle_{odd}.
\end{equation}
However solving in spinor helicity and momentum space breaks up things naturally in the following way: 
\begin{equation}\label{nhhid}
    \langle J_{s_1} J_{s_2} J_{s_3}\rangle= \left(n_F+ n_B\right)\langle J_{s_1} J_{s_2} J_{s_3}\rangle_{nh}+  \left(n_F- n_B\right)\langle J_{s_1} J_{s_2} J_{s_3}\rangle_{h,even}+ n_{odd}\langle J_{s_1} J_{s_2} J_{s_3}\rangle_{h,odd}
\end{equation}
where the "nh" peice saturates the WT identity and the coefficient of it is fixed by two point function, whereas the homogeneous piece depends on the OPE coeffcient and contains both parity even and parity odd contributions. 
Further, using the fact that in spinor helicity variables the homogeneous odd and the homogeneous even parts are related trivially which leads to 
\begin{equation}\label{arbotCFT3pt}
    \langle J_{s_1} J_{s_2} J_{s_3}\rangle=c_{2pt}\langle J_{s_1} J_{s_2} J_{s_3}\rangle_{nh}+ c_se^{i\theta}\langle J_{s_1} J_{s_2} J_{s_3}\rangle_{h}.
\end{equation}
Now, if we compute the free theory correlators, the homogeneous and non homogeneous parts can be isolated just from say, the free bosonic theory correlator. This gives us a powerful result that the full three point function can just be obtained from just the free bosonic theory or just from the free fermionic theory. 

We would like to repeat the same for four point functions, that is write them in the form, 
\begin{equation}
    \langle J_{s_1} J_{s_2} J_{s_3} J_{s_4}\rangle= \langle J_{s_1} J_{s_2} J_{s_3} J_{s_4}\rangle_{nh}+ \langle J_{s_1} J_{s_2} J_{s_3} J_{s_4}\rangle_{h}. 
\end{equation}
We believe separation of homogeneous and non-homogeneous parts in the four point function would lead to important insights into the structure of four point functions as it did in the case of three point functions. We believe this will lead to an even more beautiful structure in CS matter theory. 

It would also be very interesting to develop bootstrapping ideas directly in momentum space or in spinor helicity variables. See \cite{Caron-Huot:2021kjy} for some progress towards this. A naive analysis is presented below, see also \cite{Gandhi:2021gwn}.

\subsection*{$\langle TOOO\rangle_{\text{QF}}$}
We can use conformal block decomposition to write 
\begin{align}
\begin{split}
\langle TOOO\rangle_{\text{QF}}&=\sum_{s}\langle TOJ_{s}\rangle_{\text{QF}}\langle J_{s}OO\rangle_{\text{QF}}\\
	&=\sum_{s}\bigg( \langle TOJ_{s}\rangle_{\text{FF}}+\tilde{\lambda}\langle TOJ\rangle_{\text{CB}} \bigg)\bigg( (1+\tilde{\lambda}^2)\langle J_{s}OO\rangle_{\text{FF}} \bigg)\\
	&=(1+\tilde{\lambda}^2)\sum_{s}\bigg( \langle TOJ_{s}\rangle_{\text{FF}}\langle J_{s}OO\rangle_{\text{FF}}+\tilde{\lambda}\langle TOJ_{s}\rangle_{\text{CB}}\langle J_{s}OO\rangle_{\text{CB}} \bigg)\\
	&=(1+\tilde{\lambda}^2)\bigg( \langle TOOO\rangle_{\text{FF}}+\tilde{\lambda}\langle TOOO\rangle_{\text{CB}} \bigg)
\end{split}	
\end{align}
Going from the second to the third line we have used the fact that $\langle J_sOO\rangle_{\text{FF}}=\langle J_sO
O\rangle_{\text{CB}}$
\subsection*{$\langle JJOO\rangle_{\text{QF}}$}
Doing the same process for $\langle JJOO\rangle_{QF}$, we can write
\begin{align}
\begin{split}
&\langle JJOO\rangle_{\text{QF}}=\sum_{s}\langle JJJ_{s}\rangle_{\text{QF}}\langle J_{s}OO\rangle_{\text{QF}}\\
	&=\sum_{s}\bigg(\frac{1}{(1+\tilde{\lambda}^2)}(\langle JJJ_{s}\rangle_{\text{FF}}+\tilde{\lambda}~\epsilon\cdot(\langle JJJ_{s}\rangle_{\text{FF}}-\langle JJJ_{s}\rangle_{\text{CB}})+\tilde{\lambda}^2\langle JJJ_{s}\rangle_{\text{CB}})\bigg)\bigg((1+\tilde{\lambda}^2)\langle J_{s}OO\rangle_{\text{FF}}\bigg)\\
	&=\sum_{s}\langle JJJ_{s}\rangle_{\text{FF}}\langle J_{s}OO\rangle_{\text{FF}}+\tilde{\lambda}~\epsilon\cdot\langle JJJ_{s}\rangle_{\text{FF}}\langle J_{s}OO\rangle_{\text{FF}}-\epsilon\cdot\langle JJJ_{s}\rangle_{\text{CB}}\langle J_{s}OO\rangle_{\text{CB}}\\
	&+\tilde{\lambda}^2\langle JJJ_{s}\rangle_{\text{CB}}\langle J_{s}OO\rangle_{\text{CB}}=\langle JJOO\rangle_{\text{FF}}+\tilde{\lambda}~\epsilon\cdot(\langle JJOO\rangle_{\text{FF}}-\langle JJOO\rangle_{\text{CB}})+\tilde{\lambda}^2\langle JJOO\rangle_{\text{CB}}
\end{split}
\end{align}
Here again, while going from the second to the third line we have used $\langle J_sOO\rangle_{FF}=\langle J_sOO\rangle_{CB}$.
Even though the above analysis is very naive, it gives the same result as we obtained in \eqref{4ptans121}. We have not taken care of the double trace contributions. However, as was explained in the case of scalar four point functions \cite{Turiaci:2018nua}, these do not contribute. A naive analysis for spinning correlators also suggest the same. It would be very interesting to put these calculations on a firmer footing. Naive analyses for more general cases are discussed in appendix \ref{naivebtst}.

We believe that since spinor-helicity variables give interesting insights into the CFT correlation functions even at the level of three point functions, it would also reveal some interesting structures at the level of the four point functions which have not been understood as of yet in position or Mellin variables. An other attractive feature of the momentum space analysis is its connection to S-matrix bootstrap programme. At the level of three point functions, a detailed relation of flat space amplitudes and CFT correlators appeared in \cite{Jain:2022ujj}. It would be interesting to understand this connection at the level of the four point function better and thus making a connection to the S-matrix bootstrap.

\subsection*{Supersymmetric extension}
For the case of three point functions it was observed  \cite{Nizami:2013tpa,Aharony:2019mbc,Buchbinder:2021qlb} that conserved currents have two structures, one parity even and one parity odd. In spinor-helicity variables, this fact can easily be shown, see \cite{Jain:2021whr}. This just follows from plugging $n_B=n_F$ in \eqref{nhhid}. This implies that there will  be one non-homogeneous and only one homogeneous odd contribution. In the context of four point functions, it should lead to very simple structures.  It would be interesting to understand this both in terms of the higher spin equation as well as the conformal bootstrap perspective in momentum space. See \cite{Binder:2021cif} for a calculation of four point correlation function for the stress tensor multiplet for ${\mathcal{N}}=6$ susy SBHS theory. Like non-susy  case considered in this paper, \cite{Binder:2021cif}  also showed that there is no contact term contributing to four point function using super symmetric localization.  

\subsection*{Correlation function under relevant or marginal deformation and at finite tempearture}
In this paper we have considered CS matter theories at the conformal fixed point. Three point and four functions results are fully expressible in terms of free theory results. It would be interesting to check if similar conclusions hold true even away from the conformal fixed point. It would also be interesting to understand if at finite temperature similar conclusions can be reached. To start with we can just concentrate on turning on a mass term \cite{Jain:2020puw}. In this regard we consider a concrete example of determining the two point function $\langle J_\mu J_\nu\rangle$ of conserved higher spin operators in the free massive bosonic theory where we have deformed the free bosonic theory by introducing the mass term. We start by acting $Q_{\mu\nu}$ on $\langle J_\alpha(x_1)O(x_2)\rangle$ in position space. 
\begin{align}
\langle \left[ Q_{\mu\nu},J_{\alpha}(x_1) \right] O(x_2)\rangle_{\text{FMB}}+\langle  J_{\alpha}(x_1) \left[ Q_{\mu\nu},O(x_1) \right]\rangle_{\text{FMB}}=0
\end{align}
Here $\langle ...\rangle_{\text{FMB}}$ denotes a correlator in free massive bosonic theory. The algebra in the massive case is the same as in the case of free theory \cite{Jain:2020puw}. Using the algebra\eqref{q3oqb} we get the equation
\begin{align}
&\langle\partial_\mu\partial_\nu\partial_\alpha O(x_1)O(x_2)\rangle_{\text{FMB}}+g_{\mu\nu}\langle\partial_\alpha\square O(x_1)O(x_2)\rangle_{\text{FMB}}+	\langle\partial_\mu T_{\nu\alpha}(x_1)O(x_2)\rangle_{\text{FMB}}\\
	&+\langle\partial_\alpha T_{\mu\nu}(x_1)O(x_2)\rangle_{\text{FMB}}+\langle J_\alpha(x_1)\partial_\mu J_\nu(x_2)\rangle_{\text{FMB}}=0
\end{align}
One important difference between the massless case and the massive one is that for the latter,  2-point functions of different spins may not be zero. Converting this to momentum space we get the equation
\begin{align}
\begin{split}\label{masshse}
&p_{1\mu}p_{1\nu}p_{1\alpha}\langle O(p_1)O(-p_1)\rangle_{\text{FMB}}+g_{\mu\nu}p_{1\alpha}p_1^2\langle O(p_1)O(-p_1)\rangle_{\text{FMB}}\\
	&+p_{1\mu}\langle T_{\nu\alpha}(p_1)O(-p_1)\rangle_{\text{FMB}}+p_{1\alpha}\langle T_{\mu\nu}(p_1)O(-p_1)\rangle_{\text{FMB}}+p_{2\mu}\langle J_\alpha(p_1) J_\nu(-p_1)\rangle_{\text{FMB}}=0
\end{split}
\end{align}
Now, we can compute $\langle O(p_1)O(-p_1)\rangle_{\text{FMB}}$ and $\langle T_{\mu\nu}(p_1)O(-p_1)\rangle_{\text{FMB}}$ explicitly. Inputting the explicit expression it can be checked that this equation is indeed satisfied \cite{Jain:2020puw}. It would be an interesting future goal to generalise this for the case of slightly broken higher spin theories and to check that the correlators in the slightly broken theories can be written in terms of the free bosonic, critical fermion correlators and their epsilon transforms.

\subsection*{Derivation of dual of CS-matter theory}
The free bosonic or free fermionic theory in three dimensions are dual to Vasiliev Type A or Type B theory. CS matter theory which is a parity violating theory is dual to Vasiliev theory with parity violating $\theta$ term. Given the fact that we are able to derive the correlation functions in the parity violating CS matter theory in terms of free theory, this indicates that there might be a way to write down these parity violating theta deformed theories in terms of Vasiliev-A or Vasileiv-B theories. Recently, in \cite{Aharony:2020omh}, it was shown how to rewrite the path integral over free theory vector models in d dimensions in terms of path integrals of fields in one higher dimension in AdS space. Since CS matter theory results can be mapped to free theory results, it might turn out that this rewriting of the path integral can be extended for the case of CS matter theory and show how the $\theta$ term arises. One possible difficulty \footnote{We thank Ofer Aharony  for a comment regarding this point.} is the fact that position space correlation functions when expressed in terms of free theory results becomes complicated. Also, even though we could express the three point function in terms of just the free fermion theory, the four point function requires both free fermion and crtical bosonic results \footnote{ Hopefully understanding the four point function in spinor helicity variables and their decomposition in terms of homogeneous and non-homogeneous terms can give us a much stronger result.}.

\subsection*{Contact diagram in AdS or double trace contribution}
As mentioned above, we could show that the full slightly broken HSEs can be mapped to the FF HSE, CB HSE their combinations or as we defined earlier, an epsilon transformed version of these. In this analysis we did not require any contact diagrams. As mentioned in the main-text, the result reported in the paper solves the SBHS equations but there could be multiple solutions.
In principle one can add contact diagrams to our results \cite{Turiaci:2018nua,Silva:2021ece,Chowdhury:2019kaq}. We suspect these contact diagrams would violate the HSEs. However, let us emphasize again that we could solve SBHS equations without requiring contact diagrams. 

For scalar four point function in \cite{Turiaci:2018nua} it was shown explicitly that contact terms are not required. For $\langle JJJJ\rangle$  in \cite{Kalloor:2019xjb} it was argued that no contact term is required. Our results indicate that, we may not require AdS contact diagrams and hence no double trace contribution os required. The full results are just obtained from single trace contribution. From AdS perspective this would imply existance of 
sub-AdS locality in the higher spin gauge theories \footnote{See discussion section in \cite{Turiaci:2018nua}.}.

\subsection*{Generating functional for correlation function in CS matter theory}
Given the simplicity of the results presented in this paper, it would be interesting to find out a generating functional which reproduces all the results. See for three point function for free theories \cite{Giombi:2011rz},
for four point function in free theories \cite{Didenko:2013bj}, see also  \cite{Gerasimenko:2021sxj} for recent abstract discussions on interacting theories. It would be interesting to find explicit expression of correlation function for free theories in momentum space for four point functions.
\subsection*{Finite N bootstrap}
Reproducing the results presented in this paper from the position space or Melin space bootstrap perspective would be very interesting. Also generalizing the results by taking subleading order in $\mathcal{O}(\frac{1}{N})$ corrections would be an interesting thing to pursue. In \cite{Aharony:2018npf} it was shown that to obtain CFT data at $\frac{1}{N^2}$ we need to know
spinning four-point functions $\langle J_{s_1} J_{s_2} J_{s_3} J_{s_4} \rangle $. It would be nice to revisit analytic bootstrap program in  \cite{Aharony:2018npf}  in light of the results presented in this paper.
\section*{Acknowledgements}
The work of S.J  is supported by the Ramanujan Fellowship. We thank R.R. John, G. Mandal, A. Mehta, S. Minwalla, N. Prabhakar for discussions. SJ would like to thank DTP, TIFR for providing excellent hospitality during the course of the work.
We thank O. Aharony, A. Nizami, S. Giombi, J.A. Silva, Z. Li, A. Zhiboedov for valuable comments on an earlier version of the draft.
We acknowledge our debt to the people of India for their steady support of research in basic sciences.

\newpage	
		\appendix
		\section*{Appendix}
		\section{CS matter theory}\label{qbtheory}
	In the main text we have defined the QF theory. Let us define the QB theory here. The QB theory is two different theories which are dual to each other. These are a boson coupled to the CS gauge field and a critical fermion coupled to the CS gauge field. Here we briefly discuss a boson coupled to the CS gauge field. More details can be found in \cite{Aharony:2018pjn}.
		\subsection{Bosonic theory coupled to CS field}
	The bosonic theory coupled to the Chern-Simons gauge field has the following action
	\begin{align}
		S=\int d^3x\left[D_\mu\bar\phi D^\mu\phi+i\epsilon^{\mu\nu\rho}\frac{\kappa_F}{4\pi}\text{Tr}(A_\mu\partial_\nu A_\rho-\frac{2i}{3}A_\mu A_\nu A_\rho)+\frac{(2\pi)^2}{\kappa_B^2}(x_6^B+1)(\bar\phi\phi)^3\right]
	\end{align}
	The scalar primary operator has conformal dimension $\Delta=1+\mathcal O\left(\frac 1N\right)$ and is parity even. The spin-1 and spin-2 conserved currents have dimensions 2 and 3 respectively. The theory also has an infinite tower of weakly broken higher spin currents $J_s$ with spin $s>2$ and conformal dimension $\Delta=s+1+\mathcal O\left(\frac 1N\right)$. In this text we will refer to this as the quasi bosonic(QB) theory.
	
	\subsection{Relationship between parameters}\label{params1}
	The coupling constant in the CS matter theories is defined as follows,
	\begin{equation}
		\lambda_b=\frac{N_b}{\kappa_b},\quad\lambda_f=\frac{N_f}{\kappa_f}
	\end{equation}
we now introduce a few other useful variables which will help simplify our expressions. \cite{Aharony:2012nh}\cite{GurAri:2012is}
\begin{align}
	&\tilde{N}=2N_b\frac{\sin\pi\lambda_b}{\pi\lambda_b}=2N_f\frac{\sin\pi\lambda_f}{\pi\lambda_f}\cr
	&\tilde{\lambda}_{qb}=\tan(\frac{\pi\lambda_b}{2})=\cot(\frac{\pi\lambda_f}{2})\cr
	&\tilde{\lambda}_{qf}=\cot(\frac{\pi\lambda_b}{2})=\tan(\frac{\pi\lambda_f}{2})
\end{align} 
		\section{Higher Spin Current Algebra}\label{HScurrentAlgebraAppendix}
		In theories possessing higher spin symmetry, there exist an infinite tower of conserved currents such that their associated charge $Q_s$ kills any correlator. i.e.
		
		\begin{equation}
			\sum_i \langle J_{s_1}...[Q_s,J_{s_i}]..J_{s_n}\rangle=0
		\end{equation}
		
		The conserved charges have an associated algebra with the current operators. We can constrain this algebra from covariance, dimensionality, parity, spin considerations etc. Throughout this text we will be mainly focusing on two charges, namely $Q_3$ and $Q_4$. For the FB theory we have a scalar operator $O$ with dimension $\Delta=1$ and an infinite tower of exactly conserved currents $J_s$ with integer spin $s$ and dimension $\Delta=s+1$. The algebra of charges $Q_3$ and $Q_4$ with  $O$ and $J_1$ is given by :

		\begin{align}
			[Q_3,O]&=[Q_{\mu\nu},O]=c_1\partial_{(\mu}J_{\nu)}
			\\
			[Q_3,J]&=[Q_{\mu\nu},J_{\rho}]=c_2\partial_{\mu}\partial_{\nu}\partial_{\rho}O+c_3g_{\mu\nu}\partial_{\rho}\Box O+c_4\partial_{(\mu}T_{\nu)\rho}+c_5\partial_{\rho}T_{\mu\nu}
			\\
			\cr
			[Q_4,O]&=[Q_{\mu\nu\rho},O]=d_1\partial_\mu\partial_\nu\partial_\rho O +d_2g_{(\mu\nu}\partial_{\rho)}\square O+d_3\partial_{(\mu}T_{\nu\rho)}
			\\
			[Q_4,J]&=[Q_{\mu\nu\rho},J_{\alpha}]=d_4\,\partial_{(\mu}\partial_{\nu}\partial_{\rho)}J_{\alpha}+d_5\,\partial_{\alpha}\partial_{(\mu}\partial_{\nu}J_{\rho)}+d_6g_{(\mu\nu}\partial_{\rho)}\Box J_{\alpha}\cr
			&\hspace{2.5cm}+d_7g_{\mu\nu}\partial_{\alpha}\Box J_{\rho}+d_8g_{\alpha(\mu}\partial_{\nu}\Box J_{\rho)}+d_9\partial_{\alpha}J_{(\mu\nu\rho)}
		\end{align}
		The several constants $\{c_i,d_j\}$ can be fixed by using two-point and three-point functions. 
		Similarly the FF theory in three dimensions has a pseudo-scalar operator $O$ with dimension $\Delta=2$ and an infinite tower of exactly conserved currents $J_s$ with integer spin $s$ and dimension $\Delta=s+1$.
		The algebra of charges $Q_3$ and $Q_4$ with  $O$ and $J_1$ is as follows :
		\begin{align}
			[Q_{\mu\nu},O]&=\epsilon_{\mu ab}\,\partial^a \partial_{\nu} J^b\label{q3oqb}
			\\
			[Q_{\mu\nu},J_{\rho}]&=\epsilon_{\sigma\rho(\mu}\partial_{\nu)}\partial_{\sigma}O + i\partial_{(\mu}T_{\nu)\rho}+\partial_{\rho}T_{\mu\nu}\label{q3jqb}
			\\
			[Q_{\mu\nu},T_{\alpha\beta}]&=p_{\mu}p_{\nu}p_{\alpha}J_{\beta}+p_{\mu}p_{\beta}p_{\alpha}J_{\nu}+g_{\mu\nu}p_{\alpha}p_1^2J_{\beta}+p_{\mu}J_{\nu\alpha\beta}\label{q3tqb}\\
			\cr
			[Q_{\mu\nu\rho},O]&=\partial_{\mu}\partial_{\nu}\partial_{\rho}O+g_{\mu\nu}\partial_{\rho}\Box O+\epsilon_{\mu a b}\partial_{\nu}\partial_{a}T_{b \rho}
			\\
			[Q_{\mu\nu\rho},J_{\alpha}]&=a\,\partial_{(\mu}\partial_{\nu}\partial_{\rho)}J_{\alpha}+b\,\partial_{\alpha}\partial_{(\mu}\partial_{\nu}J_{\rho)}+\partial_{\alpha}J_{(\mu\nu\rho)}\cr
			&\hspace{0.3cm}+g_{(\mu\nu}\partial_{\rho)}\Box J_{\alpha}+g_{\mu\nu}\partial_{\alpha}\Box J_{\rho}+cg_{\alpha(\mu}\partial_{\nu}\Box J_{\rho)}
			\\
			[Q_{\mu\nu\rho},T_{\alpha\beta}]&=b_1\partial_{\mu}\partial_{\nu}\partial_{\rho}T_{\alpha\beta}+b_2\partial_{\mu}\partial_{\nu}\partial_{\alpha}T_{\rho\beta}+b_3\partial_{\mu}J_{\nu\rho\alpha\beta}+b_4\partial_{\alpha}J_{\mu\nu\rho\beta}\label{q4tff}
		\end{align}
		In the case of theories with slightly broken higher spin symmetry, the non-conservation equation gets a contribution proportional to the coupling strength $\tilde\lambda$ and becomes,
		\begin{equation}
			\sum_i \langle J_{s_1}...[Q_s,J_{s_i}]..J_{s_n}\rangle=\int \langle \partial\cdot J_sJ_{s_1}...J_{s_n}\rangle
		\end{equation}
		The non-conservation for the currents $J_3$ and $J_4$ for the QB theory is given by \cite{Maldacena:2012sf,Giombi:2016zwa},
		\begin{align}
			\partial\cdot J_3&=\partial_\sigma J^\sigma_{\mu\nu}(x)=\epsilon_{\mu a b}\big( f_0 J_a(x)\partial_{b}\partial_{\nu}O(x)+b_1 \partial_{b}J_a(x)\partial_{\nu}O(x)\cr
			&\hspace{3cm}+f_1 \partial_{\nu}J_a(x)\partial_{b}O(x)+b_2O(x)\partial_{b}\partial_{\nu}J_a(x)\big)\label{ncj3qb}
			\\
			\partial\cdot J_4&=\partial_\sigma J^\sigma_{\mu\nu\rho}(x)=\epsilon_{\mu a b}\big( f_0 T_{a\nu}(x)\partial_{b}\partial_{\rho}O(x)+b_1 \partial_{b}T_{a\nu}(x)\partial_{\rho}O(x)\cr
			&\hspace{3.2cm}+f_1 \partial_{\rho}T_{a\nu}(x)\partial_{b}O(x)+b_2O(x) \partial_{b}\partial_{\rho}T_{a\nu}(x)\big)\label{ncj4qb}
		\end{align}
		and in the case of QF theories, the non-conservation for $J_3$ and $J_4$ are given by
		
		\begin{align}
			\partial_\sigma J^\sigma_{\mu\nu}&=a\partial_{(\mu} J_{\nu)} O+bJ_{(\nu}\partial_{\mu)}O\label{ncj3qf}
			\\
			\partial_\sigma J^\sigma_{\mu\nu\rho}&=r_0\partial_{\mu}O(x)T_{\nu\rho}(x)+\epsilon_{\mu a b}\big( a_0 J_a(x)\partial_{\rho}\partial_{\nu}J_{b}(x)+b_1 J_a(x)\partial_{b}\partial_{\nu}J_{\rho}(x)+e_0 J_{\nu}(x)\partial_{a}\partial_{\rho}J_{b}(x)\big)\label{ncj4qf}
		\end{align}
		here the several factors $\{f_i,b_j\}$ are no longer constants but pick up a $\tilde{\lambda}$ dependence. The precise values are given as follows,
		\begin{align}\label{const}
			&a=\frac{3}{5\tilde{N}}\frac{\tilde{\lambda}}{1+\tilde{\lambda}^2},\quad b=-\frac{2}{5\tilde{N}}\frac{\tilde{\lambda}}{1+\tilde{\lambda}^2},\quad\cr
			&r_0=\frac{480}{7\tilde{N}}\frac{\tilde{\lambda}}{1+\tilde{\lambda}^2},\quad b_1=-\frac{512}{3\tilde{N}}\frac{\tilde{\lambda}}{1+\tilde{\lambda}^2},\quad b_2=\frac{128}{3\tilde{N}}\frac{\tilde{\lambda}}{1+\tilde{\lambda}^2}\cr
			&f_0=\frac{64}{\tilde{N}}\frac{\tilde{\lambda}}{1+\tilde{\lambda}^2},\quad f_1=-\frac{128}{3\tilde{N}}\frac{\tilde{\lambda}}{1+\tilde{\lambda}^2}
		\end{align}
		The algebra for the slightly broken HS theory is also slightly modified where the coefficients may or may not pick up a dependence of the coupling constant. The modified algebra for each theory is given by,
		\\
		\textbf{Quasi-bosonic theory}
		\begin{align}
			[Q_3,O]&=[Q_{\mu\nu},O]=c_1\partial_{(\mu}J_{\nu)}
			\\
			[Q_3,J]&=[Q_{\mu\nu},J_{\rho}]=c_2\partial_{\mu}\partial_{\nu}\partial_{\rho}O+c_3g_{\mu\nu}\partial_{\rho}\Box O+c_4\partial_{(\mu}T_{\nu)\rho}+c_5\partial_{\rho}T_{\mu\nu}
			\\
			\cr
			[Q_4,O]&=[Q_{\mu\nu\rho},O]=d_1\partial_\mu\partial_\nu\partial_\rho O +d_2g_{(\mu\nu}\partial_{\rho)}\square O+d_3\partial_{(\mu}T_{\nu\rho)}
			\\
			[Q_4,J]&=[Q_{\mu\nu\rho},J_{\alpha}]=d_4\,\partial_{(\mu}\partial_{\nu}\partial_{\rho)}J_{\alpha}+d_5\,\partial_{\alpha}\partial_{(\mu}\partial_{\nu}J_{\rho)}+d_6g_{(\mu\nu}\partial_{\rho)}\Box J_{\alpha}\cr
			&\hspace{2.5cm}+d_7g_{\mu\nu}\partial_{\alpha}\Box J_{\rho}+d_8g_{\alpha(\mu}\partial_{\nu}\Box J_{\rho)}+d_9\partial_{\alpha}J_{(\mu\nu\rho)}
		\end{align}
		
		\textbf{Quasi-fermionic theory}
		\begin{align}
			[Q_{\mu\nu},O]&=c_1 \epsilon_{\mu ab}\,\partial^a \partial_{\nu} J^b\label{q3oqf}
			\\
			[Q_{\mu\nu},J_{\rho}]&=\frac{1}{1+\tilde{\lambda}^2}c_2 \epsilon_{\sigma\rho(\mu}\partial_{\nu)}\partial_{\sigma}O + c_3\partial_{(\mu}T_{\nu)\rho}+c_4\partial_{\rho}T_{\mu\nu}\label{q3jqf}\\
			\cr
			[Q_{\mu\nu\rho},O]&=a_1\partial_{\mu}\partial_{\nu}\partial_{\rho}O+a_2g_{\mu\nu}\partial_{\rho}\Box O+a_3\epsilon_{\mu a b}\partial_{\nu}\partial_{a}T_{b \rho}\label{q4oqf}
			\\
			[Q_{\mu\nu\rho},J_{\alpha}]&=b_1\,\partial_{(\mu}\partial_{\nu}\partial_{\rho)}J_{\alpha}+b_2\,\partial_{\alpha}\partial_{(\mu}\partial_{\nu}J_{\rho)}+b_3\partial_{\alpha}J_{(\mu\nu\rho)}\cr
			&\hspace{0.3cm}+g_{(\mu\nu}\partial_{\rho)}\Box J_{\alpha}+g_{\mu\nu}\partial_{\alpha}\Box J_{\rho}+cg_{\alpha(\mu}\partial_{\nu}\Box J_{\rho)}\label{q4jqf}
			\\
			[Q_{\mu\nu\rho},T_{\alpha\beta}]&=b_1\partial_{\mu}\partial_{\nu}\partial_{\rho}T_{\alpha\beta}+b_2\partial_{\mu}\partial_{\nu}\partial_{\alpha}T_{\rho\beta}+b_3\partial_{\mu}J_{\nu\rho\alpha\beta}+b_4\partial_{\alpha}J_{\mu\nu\rho\beta}\label{q4tqf}
		\end{align}
			the precise values of all the $\{a_i,b_j,c_k\}$ sometimes cannot be determined but  it is usually not required. The only information we know about these coefficients is if they are modified for SBHS theories. Our aim in this paper has been to map the correlation function in SBHS theories to the free theories which does not require explicit knowledge of these coefficients \footnote{ One can 
			fix some of these coefficients usinf two and three point functions and some of them are given by 
		\begin{align}\label{const}
			&c_1=1,\quad c_2+c_3=2i,\quad\cr
			&a_1=1,\quad a_2=1\cr.
		\end{align}We would like to again point out that for our analysis we do not require explicit values of there coefficients unless they explicitly depend on coupling constants.}.
		If we now move to lightcone coordinates, these coeffcients are explicitly known and can be found in \cite{Maldacena:2012sf,Jain:2020puw}.

\section{Spinor helicity variables}\label{sphident}
Spinor-helicity variables are a useful tool to represent scattering amplitudes and correlation functions in a compact manner. Here we will give a short introduction of the basic definitions and list the identities which we used throughout the text. For a more detailed discussion see \cite{Maldacena:2011nz}.
\\
\\
We begin by expressing the 4-momentum as,
\begin{equation}
	p_{\alpha\dot\alpha}=\lambda_\alpha\tilde\lambda_{\dot\alpha}
\end{equation}
We define barred spinors using,
\begin{equation}
	\bar\lambda_\alpha=\tilde\lambda_{\dot\alpha}\tau^{\dot\alpha}_\alpha\quad\text{where\quad}\tau^{\dot\alpha}_\alpha=-\epsilon^{\dot\alpha\dot\beta}\mathbb{I}_{\dot\beta\alpha}
\end{equation}
The 3-momentum can be expressed as
\begin{equation}
	p^i=\frac{1}{2}(\sigma^i)^\alpha_{\hphantom{\alpha}\beta}\lambda_\alpha\bar{\lambda}^\beta
\end{equation}
Using the above relations we can express the momentum derivatives in terms of spinorial derivatives and the special conformal generator then becomes
\begin{equation}
	K^\kappa=2\sum_{i=1}^{n}(\sigma^\kappa)_\alpha^{\hphantom{\alpha}\beta}\frac{\partial^2}{\partial \lambda_{i\alpha}\partial \lambda_i^\beta}
\end{equation}

We also define the inner product brackets as follows,
\begin{equation}
	\langle ij \rangle=\epsilon^{\alpha\beta}\lambda_{i\alpha}\lambda_{j\beta}\enspace,\;\langle i\bar j \rangle=\epsilon^{\alpha\beta}\lambda_{i\alpha}\bar\lambda_{j\beta}
\end{equation}

Finally we define the transverse polarization vectors as,
\begin{equation}
	z_{\alpha\beta}^-=\frac{\lambda_\alpha\lambda_{\beta}}{2p}\quad\quad	z_{\alpha\beta}^+=\frac{\bar\lambda_\alpha\bar\lambda_{\beta}}{2p}
\end{equation}
Given a momentum space expression we dot it with the polarization tensors to convert it into spinor-helicity notation. For example
\begin{equation}
	z_i\cdot p_j=-\frac{\langle ij \rangle \langle i\bar j \rangle}{2p_i}\quad\quad z_i\cdot z_j=-\frac{\langle ij \rangle^2}{4p_ip_j}
\end{equation}
When we have three momenta, the spinor brackets satisfy the following relations: \cite{Baumann:2020dch}
\begin{align}
	\langle ba\rangle\langle \bar a\bar c\rangle&=E\langle b\bar c\rangle\cr
	\langle ba\rangle\langle \bar ac\rangle&=(E-2k_c)\langle bc\rangle\cr
	\langle \bar b a\rangle\langle \bar a\bar c\rangle&=(E-2k_b)\langle\bar b\bar c\rangle\cr
	\langle\bar b a\rangle\langle\bar a c\rangle&=(E-2k_b-2k_c)\langle\bar b c\rangle\cr
	\langle ba\rangle\langle \bar ab\rangle+\langle bc\rangle\langle \bar cb\rangle&=0\cr
	\langle ab\rangle\langle \bar a\bar b\rangle&=E(E-2k_c)\cr
	\langle\bar ba\rangle\langle \bar ab\rangle&=(E-2k_a)(E-2k_b)=k_c^2-(k_a-k_b)^2
\end{align}
Similarly, when we have four momenta we have the following identities 
\begin{align}\label{sphidn4}
	\langle ba\rangle\langle \bar a\bar c\rangle+\langle bd\rangle\langle \bar d\bar c\rangle&=E\langle b\bar c\rangle\cr
	\langle ba\rangle\langle \bar ac\rangle+\langle bd\rangle\langle \bar dc\rangle&=(E-2k_c)\langle bc\rangle\cr
	\langle \bar b a\rangle\langle \bar a\bar c\rangle+\langle \bar b d\rangle\langle \bar d\bar c\rangle&=(E-2k_b)\langle\bar b\bar c\rangle\cr
	\langle\bar b a\rangle\langle\bar a c\rangle+\langle\bar b d\rangle\langle\bar d c\rangle&=(E-2k_b-2k_c)\langle\bar b c\rangle\cr
	\langle ba\rangle\langle \bar ab\rangle+\langle bc\rangle\langle \bar cb\rangle+\langle bd\rangle\langle \bar db\rangle&=0\cr
	\langle ba\rangle\langle \bar a\bar b\rangle-\langle cd\rangle\langle \bar d\bar c\rangle&=E(2k_c+2k_d-E)\cr
	\langle\bar ba\rangle\langle \bar ab\rangle-\langle\bar cd\rangle\langle \bar dc\rangle&=(k_c-k_b)^2-(k_a-k_b)^2
\end{align}
In our computation we also made use of the expression of dot products of momenta with each other and with null transverse polarization vectors $z_i$ which satisfy the null condition $z_i^2=0$ and the transverse condition $z_i\cdot p_i=0$. The dot products take the following form 
%

\noindent\underline{\textbf{Parity Even:}}
\begin{align}\label{sphid}
	2p_a\cdot p_b=-\langle ab\rangle\langle\bar a\bar b\rangle,\quad 2p_a\cdot z_b^+=\frac{\langle a\bar b\rangle\langle \bar b\bar a\rangle}{2p_b},\quad 2p_a\cdot z_b^-=\frac{\langle \bar a b\rangle\langle  ba\rangle}{2p_b}\cr
	2z_a^+\cdot z_b^+=-\frac{\langle \bar a\bar b\rangle^2}{4p_ap_b},\quad 2z_a^-\cdot z_b^-=-\frac{\langle  a b\rangle^2}{4p_ap_b},\quad 2z_a^+\cdot z_b^-=-\frac{\langle  \bar a b\rangle^2}{4p_ap_b}
\end{align}
\underline{\textbf{Parity Odd:}}
\begin{align}
	\epsilon_{p_ip_jz^+_k}&=\frac{-1}{p_k}\big(\langle i\bar k\rangle\langle \bar k\bar j\rangle+p_j\langle i\bar k\rangle\langle \bar k\bar i\rangle+p_i\langle j\bar k\rangle\langle \bar k\bar j\rangle\big)\cr
	\epsilon_{p_iz^+_jz^+_k}&=\frac{-1}{p_jp_k}(\langle i\bar k\rangle\langle \bar k\bar j\rangle\langle j\bar i\rangle-p_i\langle \bar j\bar k\rangle^2)\cr
	\epsilon_{p_iz^-_jz^+_k}&=\frac{-1}{p_jp_k}(-\langle i\bar k\rangle\langle j\bar k\rangle\langle j\bar i\rangle-p_i\langle j\bar k\rangle^2)\cr
\end{align}
\section{Epsilon transforms in spinor-helicity variables}\label{epstrans}
We denote the epsilon transform of a current $J_s$ as follows
\begin{equation}
	 \langle \epsilon\cdot J_s^{(\mu_1\cdots \mu_s)}(p)\cdots\rangle:=\epsilon^{\sigma\alpha(\mu_1}\frac{p^{\sigma}}{p}\langle J_s^{\mu_2\cdots\mu_s)\alpha}(p)\cdots\rangle.
\end{equation}
We illustrate the epsilon transform by an example.
Consider the expression of $\langle JJO\rangle_\text{odd}$ in the quasi-bosonic theory
\begin{align}
	\langle J_{\alpha} (p_1)J_{\nu}(p_2)O(p_3)\rangle_\text{odd}=\frac{\epsilon_{\alpha a b}\,p_{1a}}{p_1}\langle J_{b} (p_1)J_{\nu}(p_2)O(p_3) \rangle_\text{FB}
\end{align}
The explicit expression of $\langle JJO\rangle_\text{FB}$ is given by,
\begin{align}
	\langle J_{\alpha} (p_1)J_{\nu}(p_2)O(p_3) \rangle_\text{FB}=\pi_{\mu a}(p_1)\pi_{\nu b}(p_2)\big(A(p_i)p_{2a}p_{1b}+B(p_i)\delta_{ab}\big)
\end{align}
If we dot it with the polarization tensors $z^-_{1\mu},z^-_{2\nu}$ on both sides we will get 
\begin{align}
	\langle J_{-}J_{-}O(p_3) \rangle_\text{FB}&=\frac{A \langle 1,2\rangle ^2 \left\langle 1,\bar{2}\right\rangle  \left\langle 2,\bar{1}\right\rangle }{8 p_1 p_2}+\frac{B \langle 1,2\rangle ^2}{2 p_1 p_2}\cr
	\langle J_{-}J_{-}O(p_3) \rangle_\text{odd}&=i\frac{A \langle 1,2\rangle ^2 \left\langle 1,\bar{2}\right\rangle  \left\langle 2,\bar{1}\right\rangle }{8 p_1 p_2}+i\frac{B \langle 1,2\rangle ^2}{2 p_1 p_2}
\end{align}
hence we have $\langle \epsilon\cdot JJO\rangle_\text{FB}\rightarrow i\langle JJO\rangle_\text{FB}$. Similarly if we dot with $z^+_{1\mu},z^+_{2\nu}$ on both sides we will get 
\begin{align}
	\langle J_{-}J_{-}O(p_3) \rangle_\text{FB}&=\frac{A \left\langle 1,\bar{2}\right\rangle  \left\langle 2,\bar{1}\right\rangle  \left\langle \bar{1},\bar{2}\right\rangle ^2}{16 p_1 p_2}+\frac{B \left\langle
   \bar{1},\bar{2}\right\rangle ^2}{8 p_1 p_2}\cr
	\langle J_{-}J_{-}O(p_3) \rangle_\text{odd}&=-i\frac{A \left\langle 1,\bar{2}\right\rangle  \left\langle 2,\bar{1}\right\rangle  \left\langle \bar{1},\bar{2}\right\rangle ^2}{16 p_1 p_2}-i\frac{B \left\langle
   \bar{1},\bar{2}\right\rangle ^2}{8 p_1 p_2}
\end{align}
so we notice $\langle \epsilon\cdot JJO\rangle_\text{FB}\rightarrow -i\langle JJO\rangle_\text{FB}$. Thus we notice that in spinor-heilcity variables $\langle \epsilon\cdot JJO\rangle_\text{FB}\rightarrow\pm i\langle JJO\rangle_\text{FB}$
\\
\\
It turns out that this observation is not limited to just $\langle JJO\rangle$ and in-fact for any correlator we have $\langle\epsilon\cdot J_{s_1}J_{s_2}J_{s_3}\rangle\rightarrow \pm i\langle J_{s_1}J_{s_2}J_{s_3}\rangle$. The analogous result for four-point functions holds true as well.
\section{Two-point functions}\label{2pthse}
In this section, we use the SBHS equation to calculate the two point functions. 
	 We present a simple example and demonstrate how higher spin symmetry can be employed to calculate correlation functions.
	
	\subsection{$\langle JJ\rangle_{\text{QF}}$}
	Consider the action of $Q_3$ on the correlator $\langle J_{\alpha}O \rangle $ in the QF theory,
	\begin{equation}
		[Q_{\mu\nu}, \langle J_{\alpha}(x_1)O(x_2)\rangle_\text{QF}]=\int_{x} \langle \partial_\sigma J^\sigma_{(\mu\nu)}(x) J_{\alpha}(x_1)O(x_2)\rangle_\text{QF}
	\end{equation}
	Using \eqref{q3oqf} and \eqref{q3jqf} we expand the LHS as follows
	\begin{align}
	&[Q_{\mu\nu}, \langle J_{\alpha}(x_1)O(x_2)\rangle_\text{QF}]=\langle	[Q_{\mu\nu},  J_{\alpha}(x_1)]O(x_2)\rangle_\text{QF}+\langle J_{\alpha}(x_1)[Q_{\mu\nu}, O(x_2) ]\rangle_\text{QF}\cr
	&=c_2 \epsilon_{\sigma\rho(\mu}\partial_{1\nu)}\partial_{\sigma}\langle O(x_1)O(x_2)\rangle_\text{QF} + c_3\partial_{(1\mu}\langle T_{\nu)\rho}(x_1)O(x_2)\rangle_\text{QF}+c_4\partial_{1\rho}\langle T_{\mu\nu}(x_1)O(x_2)\rangle_\text{QF}\cr
	&+c_1 \epsilon_{\mu ab}\,\partial^{1a} \partial_{1\nu}\langle J^b(x_1)O(x_2)\rangle_\text{QF}
	\end{align}
	Now using the fact that $\langle J_s O\rangle$ is zero, the LHS simplifies to just two terms
	\begin{equation}
		c_2 \epsilon_{\sigma\rho(\mu}\partial_{1\nu)}\partial_{1\sigma}\langle O(x_1)O(x_2)\rangle_\text{QF}
		+c_1 \epsilon_{\mu ab}\,\partial^{1a} \partial_{1\nu}\langle J_{\alpha}(x_1)J^b(x_2)\rangle_\text{QF}
	\end{equation}
	Now, using the non-conservation of $J_3$ from \eqref{ncj3qf} we can write the RHS as,
	\begin{align}
		\int_{x}^{} \langle \partial_\sigma J^\sigma_{(\mu\nu)}(x) J_{\alpha}(x_1)O(x_2)\rangle_\text{QF}&=\int_{x}^{} \langle (a\partial_{(\mu} J_{\nu)}(x) O(x)+bJ_{(\nu}(x)\partial_{\mu)}O(x)) J_{\alpha}(x_1)O(x_2)\rangle_\text{QF}\cr
		&=\int_{x}^{} (a-b)\langle \partial_{(\mu} J_{\nu)}(x) O(x) J_{\alpha}(x_1)O(x_2)\rangle_\text{QF}\cr
		&=\int_{x}^{} (a-b)\langle \partial_{(\mu} J_{\nu)}(x)J_{\alpha}(x_1) \rangle_\text{QF} \langle O(x)O(x_2)\rangle_\text{QF}
	\end{align}
where in the second line we used integration by parts and in the third line we wrote the only nonzero large-$\tilde{N}$ contribution.
	
Now, to get rid of the integral and to make the overall HSE simpler we write it in momentum space as follows
	\begin{align}\label{jjqfhse}
		c_2 \epsilon_{\sigma\rho(\mu}p_{1\nu)}p_{1\sigma}\langle O(p_1)O(-p_1)\rangle_\text{QF}
		+c_1 \epsilon_{\mu ab}p^{1a} p_{1\nu}\langle J_{\alpha}(p_1)J^b(-p_1)\rangle_\text{QF}\cr
		=16i\frac{\tilde{\lambda}}{1+\tilde{\lambda}^2}p_{1(\mu}\langle  J_{\nu)}(-p_1)J_{\alpha}(p_1) \rangle_\text{QF} \langle O(p_1)O(-p_1)\rangle_\text{QF}
	\end{align}
	where we used the values for $a,b$ from \eqref{const}. Thus, we have an expression for the unknown correlator $\langle JJ\rangle_\text{QF}$ and now we will try to solve for it. The scalar 2-point function in the QF theory is given by,
	\begin{equation}\label{ooqf}
		\langle O(p)O(-p)\rangle_\text{QF}=\frac{\tilde{N}}{2}(1+\tilde{\lambda}^2)\langle O(p)O(-p)\rangle_\text{FF}=-\frac{\tilde{N}}{2}(1+\tilde{\lambda}^2)\frac{p}{8}
	\end{equation}
	while the $\langle JJ\rangle_\text{QF}$ correlator decomposes as follows,
	\begin{align} \label{jjqf}
		\langle J_{\mu}(p)J_\nu(-p)\rangle_\text{QF}&=\frac{\tilde{N}}{2}[\langle J_{\mu}(p)J_\nu(-p)\rangle_\text{even}+\langle J_{\mu}(p)J_\nu(-p)\rangle_\text{odd}]\cr
		&=\frac{\tilde{N}}{2}\left[\frac{p_\mu p_\nu-p^2g_{\mu\nu}}{16 p}+\tilde c_{JJ} \epsilon_{\mu\nu a}p^a \right]
	\end{align}
	
	where the even part of the correlator is just the FF correlator and for the odd part we make a guess knowing that the term should be parity odd and have the correct dimensions which leaves only one possibility. Substituting \eqref{ooqf} and \eqref{jjqf} back in \eqref{jjqfhse} and equating the parity odd tensor structures we get 
	\begin{equation}
		\tilde{c}_{JJ}=\frac{\tilde{\lambda}}{16}
	\end{equation}
	 
	 Thus, we see that the above expression for $\langle JJ\rangle_\text{QF}$ matches with what we had earlier. Now, using the definition of $\tilde{N}$ and $\tilde{\lambda}$ from \eqref{params} let us rewrite the expression in a different way,
	 \begin{align}
	 \langle J_{\mu}(p)J_\nu(-p)\rangle_\text{QF}&=\frac{\tilde{N}}{2}\left[\frac{p_\mu p_\nu-p^2g_{\mu\nu}}{16 p}+\frac{\tilde{\lambda}}{16} \epsilon_{\mu\nu a}p^a \right]\cr
	 	&=N_f\frac{\sin\pi\lambda_f}{\pi\lambda_f}\left[\frac{p_\mu p_\nu-p^2g_{\mu\nu}}{16 p}+\frac{\tan(\frac{\pi\lambda_f}{2})}{16} \epsilon_{\mu\nu a}p^a \right]
	 \end{align}
 If we now dot with the polarization tensors $z_{1\mu},z_{2\nu}$ and convert to spinor-helicity variables, we will get
 \begin{align}
 	z_{1\mu}z_{2\nu}\langle J_{\mu}(p)J_\nu(-p)\rangle_\text{QF}&=N_f\frac{\sin\pi\lambda_f}{\pi\lambda_f}\left[\frac{-pz_1\cdot z_2}{16 p}+\frac{\tan(\frac{\pi\lambda_f}{2})}{16} \epsilon_{z_1z_2 p}\right]\cr
 	&=-\frac{iNe^{i\pi\lambda_{f}}\langle12\rangle^2}{32\pi\lambda_{f}p}
 \end{align}

Thus we see that in spinor-helicity variables, the two-point function turns out to be just the free theory expression but with an overall phase factor. 
This procedure can be generalised for arbitrary spin-s conserved currents.

\section{Naive conformal bootstrapping and conformal block decomposition}\label{naivebtst}
In this section we present a naive conformal block decomposition of the four point function. In the discussion section we presented a naive analysis for $\langle TOOO\rangle_{\text{QF}}$ and $\langle JJOO\rangle_{\text{QF}}.$ In the following we consider two more cases of interest.

\subsection{$\langle JJTO\rangle_{\text{QF}}$}
\begin{align}
\begin{split}
&\langle JJTO\rangle_{\text{QF}}=\sum_{s}\langle JJJ_{s}\rangle_{\text{QF}}\langle J_{s}TO\rangle_{\text{QF}}\\
	&=\sum_s\bigg(\frac{1}{1+\tilde{\lambda}^2}(\langle JJJ_{s}\rangle_{\text{FF}}+\tilde{\lambda}~\epsilon\cdot(\langle JJJ_{s}\rangle_{\text{FF}}-\langle JJJ_{s}\rangle_{\text{CB}})+\tilde{\lambda}^2\langle JJJ_{s}\rangle_{\text{CB}})\bigg)\bigg(\langle J_{s}TO\rangle_{\text{FF}}+\tilde{\lambda}\langle J_{s}TO\rangle_{\text{CB}}\bigg)\\
	&=\frac{1}{1+\tilde{\lambda}^2}\sum_s\bigg(\langle JJJ_s\rangle_{\text{FF}}\langle J_{s}TO\rangle_{\text{FF}}
+\tilde{\lambda}\langle JJJ_s\rangle_{\text{FF}}\langle J_{s}TO\rangle_{\text{CB}}
+\tilde{\lambda}(\epsilon\cdot\langle JJJ_s\rangle_{\text{FF}})\langle J_{s}TO\rangle_{\text{FF}}\\
	&+\tilde{\lambda}^2(\epsilon\cdot\langle JJJ_s\rangle_{\text{FF}})\langle J_{s}TO\rangle_{\text{CB}}
-\tilde{\lambda}(\epsilon\cdot\langle JJJ_s\rangle_{\text{CB}})\langle J_{s}TO\rangle_{\text{FF}}
-\tilde{\lambda}^2(\epsilon\cdot\langle JJJ_s\rangle_{\text{CB}})\langle J_{s}TO\rangle_{\text{CB}}\\
&+\tilde{\lambda}^2\langle JJJ_s\rangle_{\text{CB}}\langle J_{s}TO\rangle_{\text{FF}}
+\tilde{\lambda}^3\langle JJJ_s\rangle_{\text{CB}}\langle J_{s}TO\rangle_{\text{CB}}\bigg)\label{jjtocfblock1}
\end{split}
\end{align}
Now, we can use 
\begin{align}
&\langle J_sTO\rangle_{\text{FF}}=\epsilon\cdot\langle J_sTO\rangle_{\text{CB}}~~ \text{and}~~ \langle J_sTO\rangle_{\text{CB}}=-\epsilon\cdot\langle J_sTO\rangle_{\text{FF}}
\end{align}
If we input this into \eqref{jjtocfblock1}, we get
\begin{align}
\begin{split}
&\frac{1}{1+\tilde{\lambda}^2}\sum_s\bigg(\langle JJJ_s\rangle_{\text{FF}}\langle J_{s}TO\rangle_{\text{FF}}
-\tilde{\lambda}\langle JJJ_s\rangle_{\text{FF}}(\epsilon\cdot\langle J_{s}TO\rangle_{\text{FF}})
+\tilde{\lambda}(\epsilon\cdot\langle JJJ_s\rangle_{\text{FF}})\langle J_{s}TO\rangle_{\text{FF}}\\
	&-\tilde{\lambda}^2(\epsilon\cdot\langle JJJ_s\rangle_{\text{FF}})(\epsilon\cdot\langle J_{s}TO\rangle_{\text{FF}})
-\tilde{\lambda}(\epsilon\cdot\langle JJJ_s\rangle_{\text{CB}})(\epsilon\cdot\langle J_{s}TO\rangle_{\text{CB}})
-\tilde{\lambda}^2(\epsilon\cdot\langle JJJ_s\rangle_{\text{CB}})\langle J_{s}TO\rangle_{\text{CB}}\\
&+\tilde{\lambda}^2\langle JJJ_s\rangle_{\text{CB}}(\epsilon\cdot\langle J_{s}TO\rangle_{\text{CB}})
+\tilde{\lambda}^3\langle JJJ_s\rangle_{\text{CB}}\langle J_{s}TO\rangle_{\text{CB}}\bigg)
\end{split}
\end{align}
Schematically, we write $(\epsilon\cdot\langle JJJ_s\rangle)(\epsilon\cdot\langle J_{s}TO\rangle)$ as $\pm\langle JJJ_s\rangle\langle J_sTO\rangle$. Thus the above equation becomes
\begin{equation}
\begin{aligned}
&\frac{1}{1+\tilde{\lambda}^2}\sum_s\bigg(\langle JJJ_s\rangle_{\text{FF}}\langle J_{s}TO\rangle_{\text{FF}}
-\tilde{\lambda}\langle JJJ_s\rangle_{\text{FF}}(\epsilon\cdot\langle J_{s}TO\rangle_{\text{FF}})
+\tilde{\lambda}(\epsilon\cdot\langle JJJ_s\rangle_{\text{FF}})\langle J_{s}TO\rangle_{\text{FF}}\\
	&\hspace{1cm}\mp\tilde{\lambda}^2(\langle JJJ_s\rangle_{\text{FF}})(\langle J_{s}TO\rangle_{\text{FF}})
\mp\tilde{\lambda}(\langle JJJ_s\rangle_{\text{CB}})(\langle J_{s}TO\rangle_{\text{CB}})
-\tilde{\lambda}^2(\epsilon\cdot\langle JJJ_s\rangle_{\text{CB}})\langle J_{s}TO\rangle_{\text{CB}}\\
&\hspace{1cm}+\tilde{\lambda}^2\langle JJJ_s\rangle_{\text{CB}}(\epsilon\cdot\langle J_{s}TO\rangle_{\text{CB}})
+\tilde{\lambda}^3\langle JJJ_s\rangle_{\text{CB}}\langle J_{s}TO\rangle_{\text{CB}}\bigg)\\
			&=\frac{1}{1+\tilde{\lambda^2}}\bigg(\langle JJTO\rangle_{\text{FF}}
-\tilde{\lambda}~\epsilon\cdot\langle JJTO\rangle_{\text{FF}}
+\tilde{\lambda}~\epsilon\cdot\langle JJTO\rangle_{\text{FF}}\mp\tilde{\lambda}^2\langle JJTO\rangle_{\text{FF}}\\
	&\hspace{1cm}\mp\tilde{\lambda}\langle JJTO\rangle_{\text{CB}}
-\tilde{\lambda}^2~\epsilon\cdot\langle JJTO\rangle_{\text{CB}}+\tilde{\lambda}^2~\epsilon\cdot\langle JJTO\rangle_{\text{CB}}
+\tilde{\lambda}^3\langle JJTO\rangle_{\text{CB}}\bigg)\\
			&=\frac{1}{1+\tilde{\lambda^2}}\bigg(\langle JJTO\rangle_{\text{FF}}
\mp\tilde{\lambda}^2\langle JJTO\rangle_{\text{FF}}\mp\tilde{\lambda}\langle JJTO\rangle_{\text{CB}}
+\tilde{\lambda}^3\langle JJTO\rangle_{\text{CB}}\bigg)
\end{aligned}
\end{equation}
Comparing to the ansatz that we had for $\langle JJTO\rangle_{\text{QF}}$\footnote{We have done a schematic analysis here. Hence, it is possible that there might be extra terms like $\tilde{\lambda}\epsilon\cdot\langle JJTO\rangle$ in the expression. But, if we include these terms, the pole equation of $\langle JJTO\rangle_{\text{QF}}$ doesn't get satisfied. Thus, using the higher spin equations we neglect these terms. There could also be some contact terms as well as contributions from products of 3-point correlators. There could also be contact diagram contributions from AdS which have to satisfy higher spin equations by themselves. Studying them will be interesting and we plan to return to this problem in the future.}, we can identify that
\begin{equation}
\langle JJTO\rangle_{\text{Y2}}=\mp\langle JJTO\rangle_{\text{FF}}~~~\langle JJTO\rangle_{\text{Y1}}=\mp\langle JJTO\rangle_{\text{CB}}
\end{equation}
However, only the identifications with the plus signs satisfy the pole equations for $\langle JJTO\rangle_{\text{QF}}$. Thus, we have
\begin{equation}
\langle JJTO\rangle_{\text{Y2}}=\langle JJTO\rangle_{\text{FF}}~~~\langle JJTO\rangle_{\text{Y1}}=\langle JJTO\rangle_{\text{CB}}
\end{equation}
This is analogous to $\langle TTTO\rangle_{\text{QF}}$ (\ref{tttoqf}).
\subsection{$\langle JJJJ\rangle_{\text{QF}}$}
\begin{align}
\begin{split}
&\langle JJJJ\rangle_{\text{QF}}=\sum_s\langle JJJ_s\rangle_{\text{QF}}\langle J_sJJ\rangle_{\text{QF}}\\
			&=\sum_s\frac{1}{1+\tilde{\lambda}^2}\bigg(\langle JJJ_s\rangle_{\text{FF}}+\tilde{\lambda}~\epsilon\cdot(\langle JJJ_s\rangle_{\text{FF}}-\langle JJJ_s\rangle_{\text{CB}})+\tilde{\lambda}^2\langle JJJ_s\rangle_{\text{CB}}\bigg)\\
	&\times\bigg(\langle J_sJJ\rangle_{\text{FF}}+\tilde{\lambda}~\epsilon\cdot(\langle J_sJJ\rangle_{\text{FF}}-\langle J_sJJ\rangle_{\text{CB}})+\tilde{\lambda}^2\langle J_sJJ\rangle_{\text{CB}}\bigg)\\
	&=\frac{1}{1+\tilde{\lambda}}\bigg[\bigg( \langle JJJJ\rangle_{\text{FF}}+2\tilde{\lambda}~\epsilon\cdot\langle JJJJ\rangle_{\text{FF}}-2\tilde{\lambda}^3~\epsilon\cdot\langle JJJJ\rangle_{\text{CB}}+\tilde{\lambda}^4\langle JJJJ\rangle_{\text{CB}}\\
	&+\tilde{\lambda}^2\langle JJJJ\rangle_{\text{FF}}+\tilde{\lambda}^2\langle JJJJ\rangle_{\text{CB}} \bigg)\\
	&+\sum_s \bigg(-\tilde{\lambda}\langle JJJ_s\rangle_{\text{FF}}(\epsilon\cdot\langle J_sJJ\rangle_{\text{CB}})+\tilde{\lambda}^2\langle JJJ_s\rangle_{\text{FF}}\langle J_sJJ\rangle_{\text{CB}}-\tilde{\lambda}^2(\epsilon.\langle JJJ_s\rangle_{\text{FF}})(\epsilon\cdot\langle J_sJJ\rangle_{\text{CB}})\\
		&+\tilde{\lambda}^3(\epsilon\cdot\langle JJJ_s\rangle_{\text{FF}})\langle J_sJJ\rangle_{\text{CB}}-\tilde{\lambda}(\epsilon\cdot\langle JJJ_s\rangle_{\text{CB}})\langle J_sJJ\rangle_{\text{FF}}-\tilde{\lambda}^2(\epsilon\cdot\langle JJJ_s\rangle_{\text{CB}})(\epsilon\cdot\langle J_sJJ\rangle_{\text{FF}}) \\
		&+\tilde{\lambda}^2\langle JJJ_s\rangle_{\text{CB}}\langle J_sJJ\rangle_{\text{FF}}+\tilde{\lambda}^3\langle JJJ_s\rangle_{\text{CB}}(\epsilon\cdot\langle J_sJJ\rangle_{\text{FF}})\bigg) \bigg]
\end{split}
\end{align}
Now, we cannot move forward with the terms inside the summation. This is because in this case we do not have a direct epsilon transform relation between $\langle J_sJJ\rangle_{\text{FF}}$ and $\langle J_sJJ\rangle_{\text{CB}}$. In case of $\langle JJTO\rangle_{\text{QF}}$, we could proceed because we could write $\langle J_sTO\rangle_{\text{FF}}$ and $\langle J_sTO\rangle_{\text{CB}}$ as epsilon transforms of each other, but this is not possible in the analysis of $\langle JJJJ\rangle$. Thus, the conformal block argument of $\langle JJJJ\rangle$ does not help us infer much. However, using the higher spin equation, we could solve for $\langle JJJJ\rangle_{\text{QF}}$.

\section{Details of four-point HSEs}\label{4PointDetails}
\subsection{$\langle TOOO\rangle_{\text{QF}}$}\label{toooextra}
We discussed the case of $\langle TOOO \rangle$ in section \ref{tooohse1}. It turns out that in this case the HSE allows for a stronger result as follows.

We shall now explicitly show the mapping of the SBHS equations to the free theory HSEs. At O($\tilde{\lambda}$) of \eqref{qfhsetooo} we have
\begin{align}
	\begin{split}
		&\epsilon_{ a b(\mu}p_{1\nu}p_{1a}\langle T_{b\rho)}(p_1)O(p_2)O(p_3)O(p_4)\rangle_{\text{$Y_1$}}+
	+\lbrace1\leftrightarrow2\rbrace+\lbrace1\leftrightarrow3\rbrace+\lbrace1\leftrightarrow4\rbrace\\
		&=-12p_{1(\mu}p_1\langle T_{\nu\rho)}(p_1)O(p_2)O(p_3)O(p_4)\rangle_{\text{$Y_0$}}+\lbrace1\leftrightarrow2\rbrace+\lbrace1\leftrightarrow3\rbrace+\lbrace1\leftrightarrow4\rbrace
	\end{split}
\end{align}
At O($\tilde{\lambda}^2$) we have
\small
\begin{align}
	\begin{split}
		&p_{1\mu}p_{1\nu}p_{1\rho}\langle O(p_1)O(p_2)O(p_3)O(p_4)\rangle_{\text{FF+CB}}+g_{(\mu\nu}p_{1\rho)}p_1^2\langle O(p_1)O(p_2)O(p_3)O(p_4)\rangle_{\text{FF+CB}}\\&+\epsilon_{ a b(\mu}p_{1\nu}p_{1a}\langle T_{b\rho)}(p_1)O(p_2)O(p_3)O(p_4)\rangle_{\text{$Y_0$}}\\
		&+\lbrace1\leftrightarrow2\rbrace+\lbrace1\leftrightarrow3\rbrace+\lbrace1\leftrightarrow4\rbrace=-12p_{1(\mu}p_1\langle T_{\nu\rho)}(p_1)O(p_2)O(p_3)O(p_4)\rangle_{\text{$Y_1$}}+\lbrace1\leftrightarrow2\rbrace+\lbrace1\leftrightarrow3\rbrace+\lbrace1\leftrightarrow4\rbrace
	\end{split}
\end{align}
\normalsize
At O($\tilde{\lambda}^3$) the equation takes the form
\begin{align}
		&\epsilon_{ a b(\mu}p_{1\nu}p_{1a}\langle T_{b\rho)}(p_1)O(p_2)O(p_3)O(p_4)\rangle_{\text{$Y_1$}}+\lbrace1\leftrightarrow2\rbrace+\lbrace1\leftrightarrow3\rbrace+\lbrace1\leftrightarrow4\rbrace\\
		&=-12p_{1(\mu}p_1\langle T_{\nu\rho)}(p_1)O(p_2)O(p_3)O(p_4)\rangle_{\text{$Y_0$}}+\lbrace1\leftrightarrow2\rbrace+\lbrace1\leftrightarrow3\rbrace+\lbrace1\leftrightarrow4\rbrace
\end{align}
It is easy to see directly in momentum space that the O($\tilde{\lambda}^2$) equation maps to the simple linear combination of the lowest  \eqref{ffhsetooo} and highest order equations  \eqref{cbhsetooo}, namely O($\tilde{\lambda}^0$)+O($\tilde{\lambda}^4$).To show that the O($\tilde{\lambda}$) equation maps to the combination O($\tilde{\lambda}^0$)-O($\tilde{\lambda}^4$) we move to spinor helicity variables where it is easily verified. That is,
\begin{align}
    \text{HSE at O}(\tilde{\lambda})=&\text{FF HSE}-\text{CB HSE}\cr
    \text{HSE at O}(\tilde{\lambda}^2)=&\text{FF HSE}+\text{CB HSE}
\end{align}
Thus. we map the entire HSE to the free and critical theory HSEs. Let us note that we have taken $\langle OOOO \rangle_{FF} =\langle OOOO \rangle_{CB}$ which can be checked explicitly \cite{Turiaci:2018nua}.

\subsection{$\langle JJOO\rangle_{\text{QF}}$}\label{jjooextra}
In \ref{jjoosec} we found that the correlator $\langle JJOO\rangle_{\text{QF}}$ is given by the expression in \eqref{jjooqf}. Using this expression and plugging it back into \eqref{jjoohse} we can show that the resulting HSE can be mapped to a combination of the free theory HSEs. As done in the main text, we expand the HSE in \eqref{jjoohse} at each order in the coupling constant $\tilde{\lambda}$. At $O(\tilde{\lambda})$ in \eqref{jjoohse} we have
\begin{align}
	\begin{split}\label{o1jjoo}
		&p_{1(\mu}\langle T_{\nu)\alpha}(p_1)O(p_2)O(p_3)O(p_4)\rangle_{\text{CB}}+p_{1\alpha}\langle T_{\mu\nu}(p_1)O(p_2)O(p_3)O(p_4)\rangle_{\text{CB}}+\\
		&2\bigg(\epsilon_{ a b(\mu}p_{2a}p_{2\nu)}\langle J_{\alpha}(p_1)J_b(p_2)O(p_3)O(p_4)\rangle_{\text{odd}}+\lbrace 2\leftrightarrow3\rbrace+\lbrace 2\leftrightarrow4\rbrace \bigg)\\
		&=-16\bigg[ p_{1(\mu}\langle J_{\alpha}J_{\nu)}\rangle_{\text{FF}}\langle O(p_1)O(p_2)O(p_3)O(p_4)\rangle_{\text{FF}}\\
		&+ \frac{p_{2(\mu}p_2}{8} (\langle J_{\alpha}(p_1)J_{\nu)}(p_2)O(p_3)O(p_4)\rangle_{Y_0})+\lbrace 2\leftrightarrow3\rbrace+\lbrace 2\leftrightarrow4\rbrace\bigg]
	\end{split}
\end{align}	
Now, if we plug the solutions that we got from the pole equation \eqref{polesoln} into the O($\lambda$) equation\eqref{o1jjoo}, we get
\begin{align}
\begin{split}\label{o1jjoosim}
		&p_{1\mu}\langle T_{\nu\alpha}(p_1)O(p_2)O(p_3)O(p_4)\rangle_{\text{CB}}+p_{1\alpha}\langle T_{\mu\nu}(p_1)O(p_2)O(p_3)O(p_4)\rangle_{\text{CB}}+(2\leftrightarrow3)+(2\leftrightarrow4)\\
	&=-2ip_2p_{2\mu}\langle J_{\alpha}(p_1)J_{\nu}(p_2)O(p_3)O(p_4)\rangle_{\text{CB}}-16i\langle J_{\alpha}(p_1)J_{\nu}(-p_1)\rangle_{FF}\langle O(p_1)O(p_2)O(p_3)O(p_4)\rangle_{\text{FF}}\\
	&+(2\leftrightarrow3)+(2\leftrightarrow4)
\end{split}
\end{align}
which is exactly the CB HSE. To get \eqref{o1jjoosim} from \eqref{o1jjoo}, we note that in the LHS of \eqref{o1jjoo}, we used various tensor identities involving $\epsilon_{\mu\nu\rho}$ which turns the  $\langle JJOO\rangle_{\text{odd}}$ term proportional to  $\langle JJOO\rangle_{\text{FF}}-\langle JJOO\rangle_{\text{CB}}$  using \eqref{polesoln}. Also using \eqref{y0y2jjoo} we observe that the free fermion terms cancel with the corresponding FF terms of $\langle JJOO\rangle$ in the RHS and we are left with \eqref{o1jjoosim}.

At the next order, that is at $O(\tilde{\lambda}^2)$ we have
\begin{align}
	\begin{split}
		&\epsilon_{\alpha a(\mu}p_{1\nu)}p_{1a}\langle O(p_1)O(p_2)O(p_3)O(p_4)\rangle_{\text{CB}}+p_{1(\mu}\langle T_{\nu)\alpha}(p_1)O(p_2)O(p_3)O(p_4)\rangle_{\text{FF}}\\
		&+p_{1\alpha}\langle T_{\mu\nu}(p_1)O(p_2)O(p_3)O(p_4)\rangle_{\text{FF}}+2\bigg(\epsilon_{ a b(\mu}p_{2a}p_{2\nu)}\langle J_{\alpha}(p_1)J_b(p_2)O(p_3)O(p_4)\rangle_{Y_2}\\
		&+\lbrace 2\leftrightarrow3\rbrace+\lbrace 2\leftrightarrow4\rbrace\bigg)\\
		&=-16\bigg[\frac{p_{2(\mu}p_2}{8} \langle J_{\alpha}(p_1)J_{\nu)}(p_2)O(p_3)O(p_4)\rangle_{\text{odd}}+\lbrace 2\leftrightarrow3\rbrace+\lbrace 2\leftrightarrow4\rbrace \bigg]
	\end{split}
\end{align}
 Now we plug in the solution of the pole equation \eqref{polesoln}into the O($\tilde{\lambda}^2$) equation which yields,
\begin{align}
&\epsilon_{(\mu|\alpha a}p_{1|\nu)}p_1^a\langle O(p_1)O(p_2)O(p_3)O(p_4)\rangle_{\text{FF}}+p_{1(\mu|}\langle T_{|\nu)\alpha}(p_1)O(p_2)O(p_3)O(p_4)\rangle_{\text{FF}}\cr
	&p_{1\alpha}\langle T_{\mu\nu}(p_1)O(p_2)O(p_3)O(p_4)\rangle_{\text{FF}}+2\epsilon_{(\mu| a b}p_2^a p_{2|\nu)}(\langle J_{\alpha}(p_1)J^b(p_2)O(p_3)O(p_4)\rangle_{\text{FF}})\cr
	&+(2\leftrightarrow3)+(2\leftrightarrow4)=0
\end{align}
which is exactly the FF equation. Therefore, by using the solutions of the pole equations we can map the O($\tilde{\lambda}$) and O($\tilde{\lambda}^2$) and hence the entire HSE to the FF and CB HSEs. Thus, this also proves to be a consistency check of the solution that we have for the correlator.

\subsection{$\langle TTTO\rangle_{\text{QF}}$}\label{tttoextra}
In \ref{tttoqfsec} we found that the correlator $\langle JJTO\rangle_{\text{QF}}$ is given by the expression in \eqref{tttoqf}. Using this expression and plugging it back into \eqref{tttoqfhse} we can show that the resulting HSE can be mapped to a combination of free theory HSEs. As done in the main text, we expand the HSE in \eqref{tttoqfhse} at each order in the coupling constant $\tilde{\lambda}$. \\At $O(\tilde{\lambda})$ in \eqref{tttoqfhse} we have
\begin{align}
    &\bigg[p_{1\alpha}\langle J_{4\mu\nu\rho\beta}(p_1)T_{\gamma\sigma}(p_2)O(p_3)O(p_4)\rangle_{\text{odd}}+p_{1\mu}\langle J_{4\nu\rho\alpha\beta}(p_1)T_{\gamma\sigma}(p_2)O(p_3)O(p_4)\rangle_{\text{odd}}\notag\\
    &-p_{1\mu}\langle T_{\nu\rho}(-p_1)T_{\alpha\beta}(p_1)\rangle\langle O(p_1)T_{\gamma\sigma}(p_2)O(p_3)O(p_4)\rangle_{\text{FF}}+\epsilon_{\nu \alpha a}p_{1a}p_{1\beta}p_{1\mu}p_{1\rho}\langle O(p_1)T_{\gamma\sigma}(p_2)O(p_3)O(p_4)\rangle_{\text{CB}}\notag\\
    &+p_{1\alpha}p_{1\beta}p_{1\mu}\langle T_{\nu\rho}(p_1)T_{\gamma\sigma}(p_2)O(p_3)O(p_4)\rangle_{\text{odd}}+p_{1\alpha}p_{1\mu}p_{1\nu}\langle T_{\rho\beta}(p_1)T_{\gamma\sigma}(p_2)O(p_3)O(p_4)\rangle_{\text{odd}}\notag\\&+p_{1\mu}p_{1\nu}p_{1\rho}\langle T_{\alpha\beta}(p_1)T_{\gamma\sigma}(p_2)O(p_3)O(p_4)\rangle_{\text{odd}}+\{1\leftrightarrow 2,(\alpha,\beta)\leftrightarrow(\gamma,\sigma)\}\bigg]\notag\\&+\bigg[p_{3\mu}p_{3\nu}p_{3\rho}\langle T_{\alpha\beta}(p_1)T_{\gamma\sigma}(p_2)O(p_3)O(p_4)\rangle_{\text{odd}}-p_3p_{3\mu}\langle T_{\alpha\beta}(p_1)T_{\gamma\sigma}(p_2)T_{\nu\rho}(p_3)O(p_4)\rangle_{\text{Y0}}\notag\\
    &+\epsilon_{\mu a b}p_{3a}p_{3\nu}\langle T_{\alpha\beta}(p_1)T_{\gamma\sigma}(p_2)T_{\rho b}(p_3)O(p_4)\rangle_{\text{Y1}}+\{3\leftrightarrow 4\}\bigg]
\end{align}
At $O(\tilde{\lambda}^2)$ the HSE reads,
\begin{align}
     &\bigg[p_{1\mu}p_{1\nu}p_{1\rho}\langle T_{\alpha\beta}(p_1)T_{\gamma\sigma}(p_2)O(p_3)O(p_4)\rangle_{\text{FF+CB}}+ p_{1\mu}p_{1\nu}p_{1\alpha}\langle T_{\rho\beta}(p_1)T_{\gamma\sigma}(p_2)O(p_3)O(p_4)\rangle_{\text{FF+CB}}\notag\\& +p_{1\mu}p_{1\alpha}p_{1\beta}\langle T_{\nu\rho}(p_1)T_{\gamma\sigma}(p_2)O(p_3)O(p_4)\rangle_{\text{FF+CB}}+p_{1\alpha}\langle J_{4\mu\nu\rho\beta}(p_1)T_{\gamma\sigma}(p_2)O(p_3)O(p_4)\rangle_{\text{FF+CB}}\notag\\&+p_{1\mu}\langle J_{4\nu\rho\alpha\beta}(p_1)T_{\gamma\sigma}(p_2)O(p_3)O(p_4)\rangle_{\text{FF+CB}}+\epsilon_{\nu\alpha a}p_{1a}p_{1\mu}p_{1\rho}p_{1\beta}\langle O(p_1)T_{\gamma\sigma}(p_2)O(p_3)O(p_4)\rangle_{\text{FF}}\notag\\&+\{1\leftrightarrow 2,(\alpha,\beta)\leftrightarrow(\gamma,\sigma)\}\bigg]\notag\\&+\bigg[p_{3\mu}p_{3\nu}p_{3\rho}\langle T_{\alpha\beta}(p_1)T_{\gamma\sigma}(p_2)O(p_3)O(p_4)\rangle_{\text{FF+CB}}+\epsilon_{\mu a b}p_{3\nu}p_{3a}\langle T_{\alpha\beta}(p_1)T_{\gamma\sigma}(p_2)T_{\rho b}(p_3)O(p_4)\rangle_{\text{FF}}+\{3\leftrightarrow 4\}\bigg]\notag\\
    &-\bigg[p_{3\mu}\langle O(-p_3)O(p_3)\rangle_{\text{CB}}\langle T_{\alpha\beta}(p_1)T_{\gamma\sigma}(p_2)T_{\nu\rho}(p_3)O(p_4)\rangle_{\text{CB}}+\{3\leftrightarrow 4\}\notag\\
    &+p_{1\mu}\langle T_{\nu\rho}(-p_1)T_{\alpha\beta}(p_1)\rangle_{\text{CB}}\langle O(p_1)T_{\gamma\sigma}(p_2)O(p_3)O(p_4)\rangle_{\text{CB}}+\{(1\leftrightarrow 2),(\alpha,\beta)\leftrightarrow(\gamma,\sigma)\}\bigg]=0
\end{align}
which tells us that,
$$\text{HSE at~~}O(\tilde{\lambda}^2)=\text{FF HSE}-\text{CB HSE}$$
At $O(\tilde{\lambda}^3)$ in \eqref{tttoqfhse} we have,
\begin{align}
    &\bigg[p_{1\alpha}\langle J_{4\mu\nu\rho\beta}(p_1)T_{\gamma\sigma}(p_2)O(p_3)O(p_4)\rangle_{\text{odd}}+p_{1\mu}\langle J_{4\nu\rho\alpha\beta}(p_1)T_{\gamma\sigma}(p_2)O(p_3)O(p_4)\rangle_{\text{odd}}\notag\\
    &-p_{1\mu}\langle T_{\nu\rho}(-p_1)T_{\alpha\beta}(p_1)\rangle\langle O(p_1)T_{\gamma\sigma}(p_2)O(p_3)O(p_4)\rangle_{\text{FF}}+\epsilon_{\nu \alpha a}p_{1a}p_{1\beta}p_{1\mu}p_{1\rho}\langle O(p_1)T_{\gamma\sigma}(p_2)O(p_3)O(p_4)\rangle_{\text{CB}}\notag\\
    &+p_{1\alpha}p_{1\beta}p_{1\mu}\langle T_{\nu\rho}(p_1)T_{\gamma\sigma}(p_2)O(p_3)O(p_4)\rangle_{\text{odd}}+p_{1\alpha}p_{1\mu}p_{1\nu}\langle T_{\rho\beta}(p_1)T_{\gamma\sigma}(p_2)O(p_3)O(p_4)\rangle_{\text{odd}}\notag\\&+p_{1\mu}p_{1\nu}p_{1\rho}\langle T_{\alpha\beta}(p_1)T_{\gamma\sigma}(p_2)O(p_3)O(p_4)\rangle_{\text{odd}}+\{1\leftrightarrow 2,(\alpha,\beta)\leftrightarrow(\gamma,\sigma)\}\bigg]\notag\\&+\bigg[p_{3\mu}p_{3\nu}p_{3\rho}\langle T_{\alpha\beta}(p_1)T_{\gamma\sigma}(p_2)O(p_3)O(p_4)\rangle_{\text{odd}}-p_3p_{3\mu}\langle T_{\alpha\beta}(p_1)T_{\gamma\sigma}(p_2)T_{\nu\rho}(p_3)O(p_4)\rangle_{\text{Y2}}\notag\\
    &+\epsilon_{\mu a b}p_{3a}p_{3\nu}\langle T_{\alpha\beta}(p_1)T_{\gamma\sigma}(p_2)T_{\rho b}(p_3)O(p_4)\rangle_{\text{Y3}}+\{3\leftrightarrow 4\}\bigg]
\end{align}
Using our solution to the pole equations\eqref{petttosol} we see that,
$$\text{HSE at~~}O(\tilde{\lambda})=\text{HSE at~~}O(\tilde{\lambda}^3)$$
Using spinor helicity variables we can show that,
$$\text{HSE at~~}O(\tilde{\lambda})=\epsilon\cdot(\text{FF HSE}-\text{CB HSE})$$
and thus we are done. Yet again, we show that the solution that we obtained from the pole
equation is indeed consistent as plugging it back into the HSE reduces it at each order to
the FF, CB or a linear combination of the FF and CB HSEs.
\subsection{$\langle JJJJ\rangle_{\text{QF}}$}\label{jjjjextra}
In \ref{jjjjsec} we found that the correlator $\langle JJJJ\rangle_{\text{QF}}$ is given by the expression in \eqref{jjjjqf}. Using this expression and plugging it back into \eqref{jjjjqfhse} we can show that the resulting HSE can be mapped to a combination of free theory HSEs. As done in the main text, we expand the HSE in \eqref{jjjjqfhse} in the coupling constant $\tilde{\lambda}$. Let us now consider the intermediate order equations of the HSE in \eqref{jjjjqfhse}. At $O(\tilde{\lambda}$) we have
\begin{align}
	\begin{split}\label{jjjjo1}
		&\bigg[ \epsilon_{\alpha a(\mu}p_{1\nu)}p_1^a\langle O(p_1)J_{\beta}(p_2)J_{\gamma}(p_3)O(p_4)\rangle_{\text{odd}}+p_{1(\mu}\langle T_{\nu)\alpha}(p_1)J_{\beta}(p_2)J_{\gamma}(p_3)O(p_4)\rangle_{Y_1}\\&+p_{1\alpha}\langle T_{\mu\nu}(p_1)J_{\beta}(p_2)J_{\gamma}(p_3)O(p_4)\rangle_{Y_1}+(1\leftrightarrow2,\alpha\leftrightarrow\beta)+(1\leftrightarrow3,\alpha\leftrightarrow\gamma)\bigg]\\&+2\epsilon_{ a b(\mu}p_2^ap_{2\nu)}\langle J_{\alpha}(p_1)J_{\beta}(p_2)J_{\gamma}(p_3)J^b(p_4)\rangle_{Y_1}\\&=-16i\bigg[ p_{1(\mu}\langle J_{\alpha}(p_1)J_{\nu)}(-p_1)\rangle_{\text{FF}}\langle O(p_1)J_{\beta}(p_2)J_{\gamma}(p_3)O(p_4)\rangle_{\text{FF}}+(1\leftrightarrow2,\alpha\leftrightarrow\beta)+(1\leftrightarrow3,\alpha\leftrightarrow\gamma) \bigg]\\
		&+2ip_{4(\mu}p_4\langle J_{\alpha}(p_1)J_{\beta}(p_2)J_{\gamma}(p_3)J_{\nu)}(p_4)\rangle_{Y_0}
	\end{split}
\end{align}
Using the solutions of the pole equation and the expression for $Y_1$, we checked that the $O(\tilde{\lambda})$ equation \eqref{jjjjo1} maps to the $O(\tilde{\lambda}^5)$ equation upon dotting with the null polarisation tensors $z_{1\alpha}z_{2\beta}z_{3\gamma}z_{2\mu}z_{2\nu}$ and converting the expression to spinor helicity variables. 
\begin{align}
    \text{HSE at O}(\tilde{\lambda})=&\text{CB HSE}
\end{align}
At $O(\tilde{\lambda}^2$) we get
\begin{align}
	\begin{split}\label{jjjjo2}
		&\bigg[ \epsilon_{\alpha a(\mu}p_{1\nu)}p_{1a}\langle O(p_1)J_{\beta}(p_2)J_{\gamma}(p_3)O(p_4)\rangle_{\text{CB}}+2\epsilon_{\alpha a(\mu}p_{1\nu)}p_{1a}\langle O(p_1)J_{\beta}(p_2)J_{\gamma}(p_3)O(p_4)\rangle_{\text{FF}}\\&+p_{1(\mu}\langle T_{\nu)\alpha}(p_1)J_{\beta}(p_2)J_{\gamma}(p_3)O(p_4)\rangle_{Y_0+Y_2}+p_{1\alpha}\langle T_{\mu\nu}(p_1)J_{\beta}(p_2)J_{\gamma}(p_3)O(p_4)\rangle_{Y_0+Y_2}\\&+(1\leftrightarrow2,\alpha\leftrightarrow\beta)+(1\leftrightarrow3,\alpha\leftrightarrow\gamma)\bigg]+2\epsilon_{ a b(\mu}p_{2a}p_{2\nu)}\langle J_{\alpha}(p_1)J_{\beta}(p_2)J_{\gamma}(p_3)J_b(p_4)\rangle_{Y_2}\\
		&=-16i\bigg[ p_{1(\mu}\langle J_{\alpha}(p_1)J_{\nu)}(-p_1)\rangle_{\text{FF}}\langle O(p_1)J_{\beta}(p_2)J_{\gamma}(p_3)O(p_4)\rangle_{\text{odd}}+(1\leftrightarrow2,\alpha\leftrightarrow\beta)+(1\leftrightarrow3,\alpha\leftrightarrow\gamma) \bigg]\\&+2ip_{4(\mu}p_4\langle J_{\alpha}(p_1)J_{\beta}(p_2)J_{\gamma}(p_3)J_{\nu)}(p_4)\rangle_{Y_1}
	\end{split}
\end{align}
This maps to $2O(\tilde\lambda^0)$ in spinor helicity varibles. Thus we get,
\begin{align}
    \text{HSE at O}(\tilde{\lambda}^2)=&2~\text{FF HSE}
\end{align}
Moving on, the $O(\tilde{\lambda}^3)$ equation is
\begin{align}
	\begin{split}\label{jjjjo3}
		&\bigg[ 2\epsilon_{\alpha a(\mu}p_{1\nu)}p_{1a}\langle O(p_1)J_{\beta}(p_2)J_{\gamma}(p_3)O(p_4)\rangle_{\text{odd}}+2p_{1(\mu}\langle T_{\nu)\alpha}(p_1)J_{\beta}(p_2)J_{\gamma}(p_3)O(p_4)\rangle_{Y_1}\\&+2p_{1\alpha}\langle T_{\mu\nu}(p_1)J_{\beta}(p_2)J_{\gamma}(p_3)O(p_4)\rangle_{Y_1}+(1\leftrightarrow2,\alpha\leftrightarrow\beta)+(1\leftrightarrow3,\alpha\leftrightarrow\gamma)\bigg]\\&+2\epsilon_{ a b(\mu}p_{2a}p_{2\nu)}\langle J_{\alpha}(p_1)J_{\beta}(p_2)J_{\gamma}(p_3)J_b(p_4)\rangle_{Y_3}\\&=-16i\bigg[
		+ p_{1(\mu}\langle J_{\alpha}(p_1)J_{\nu)}(-p_1)\rangle_{\text{even}}\langle O(p_1)J_{\beta}(p_2)J_{\gamma}(p_3)O(p_4)\rangle_{\text{CB+FF}}+(1\leftrightarrow2)+(1\leftrightarrow3) \bigg]\\&+2ip_{4(\mu}p_4\langle J_{\alpha}(p_1)J_{\beta}(p_2)J_{\gamma}(p_3)J_{\nu)}(p_4)\rangle_{Y_2}
	\end{split}
\end{align}
 Similar to the case of the $O(\tilde{\lambda})$ equation, the $O(\tilde{\lambda}^3)$ in \eqref{jjjjo3} maps to $2O(\tilde{\lambda}^5)$  equation in spinor helicity variables. Hence,
 \begin{align}
    \text{HSE at O}(\tilde{\lambda}^3)=&2~\text{CB HSE}
\end{align}
 
Finally the $O(\tilde{\lambda}^4)$ equation is
\begin{align}
	\begin{split}\label{jjjjo4}
		&\bigg[ \epsilon_{\alpha a(\mu}p_{1\nu)}p_1^a\langle O(p_1)J_{\beta}(p_2)J_{\gamma}(p_3)O(p_4)\rangle_{\text{FF}}+2\epsilon_{\alpha a(\mu}p_{1\nu)}p_1^a\langle O(p_1)J_{\beta}(p_2)J_{\gamma}(p_3)O(p_4)\rangle_{\text{CB}}\\&+p_{1(\mu}\langle T_{\nu)\alpha}(p_1)J_{\beta}(p_2)J_{\gamma}(p_3)O(p_4)\rangle_{Y_2}+p_{1\alpha}\langle T_{\mu\nu}(p_1)J_{\beta}(p_2)J_{\gamma}(p_3)O(p_4)\rangle_{Y_2}\\
		&+(1\leftrightarrow2,\alpha\leftrightarrow\beta)+(1\leftrightarrow3,\alpha\leftrightarrow\gamma)\bigg]+2\epsilon_{ a b(\mu}p_2^ap_{2\nu)}\langle J_{\alpha}(p_1)J_{\beta}(p_2)J_{\gamma}(p_3)J^b(p_4)\rangle_{Y_4}\\
		&=-16i\bigg[ p_{1(\mu}\langle J_{\alpha}(p_1)J_{\nu)}(-p_1)\rangle_{\text{FF}}\langle O(p_1)J_{\beta}(p_2)J_{\gamma}(p_3)O(p_4)\rangle_{\text{odd}}
		+(1\leftrightarrow2,\alpha\leftrightarrow\beta)+(1\leftrightarrow3,\alpha\leftrightarrow\gamma) \bigg]\\&+2ip_{4(\mu}p_4\langle J_{\alpha}(p_1)J_{\beta}(p_2)J_{\gamma}(p_3)J_{\nu)}(p_4)\rangle_{Y_3}
	\end{split}
\end{align}
The above equation maps to the $O(\tilde\lambda^0)$ equation in spinor-helicity variables. Thus we have
\begin{align}
    \text{HSE at O}(\tilde{\lambda}^4)=&\text{FF HSE}
\end{align}
Thus, we can map the entire QF HSE to the free and critical HSEs by using the obtained solution. To summarize,
\begin{align}
    \text{HSE at O}(\tilde{\lambda})=&\text{CB HSE}\cr
    \text{HSE at O}(\tilde{\lambda}^2)=&2~\text{FF HSE}\cr
    \text{HSE at O}(\tilde{\lambda}^3)=&2~\text{CB HSE}\cr 
    \text{HSE at O}(\tilde{\lambda}^4)=&\text{FF HSE}
\end{align}
This acts as a consistency check as well.
Note however, that for this example, we mapped the HSE at each order to the FF and CB results \textit{only} in spinor helicity variables. The reason is that in momentum space we would have to perform a host of epsilon transforms and that is quite complicated. However, in spinor helicity, as mentioned earlier, the epsilon transform becomes trivial and we easily perform the mapping.
\section{Five-point functions}\label{5ptappendix}
In this appendix, we discuss how our strategy can be used to constrain five point functions in a slightly broken HS theory. The input for the bootstrapping, similar to the four point case, is the five point scalar correlator. We choose the form of the five point scalar correlator based on a naive bootstrap analysis, much like we did in appendix \ref{naivebtst}. 
\subsection{$\langle TOOOO\rangle_{\text{QF}}$}
Let us begin with the example of $\langle TOOOO\rangle_{\text{QF}}$. For this, we act $Q_{\mu\nu\rho}$ on the scalar 5-point correlator, $\langle OOOOO\rangle_{QF}$. Using the higher spin algebra \eqref{q3oqb} and the non conservation current equation, \eqref{ncj4qf}, we get the HSE to be
\begin{align}
\begin{split}\label{5pthse}
&p_{1\mu}p_{1\nu}p_{1\rho}\langle OOOOO\rangle_{\text{QF}}+g_{(\mu\nu}p_{1\rho)}p_1^2\langle OOOOO\rangle_{\text{QF}}\\
	&+\bigg(\epsilon_{ a b(\mu}p_{1\nu}p_{1 a}\langle T_{\rho)b}OOOO\rangle_{\text{QF}}+\lbrace 1\leftrightarrow2,3,4,5 \rbrace\bigg)=12\tilde{\lambda}p_{1(\mu}p_1\langle T_{\nu\rho)}OOOO\rangle_{\text{QF}}+\lbrace 1\leftrightarrow2,3,4,5 \rbrace
\end{split}
\end{align}
Now, we use the decomposition of the QF correlators,
\begin{align}
&\langle OOOOO\rangle_{\text{QF}}=(1+\tilde{\lambda}^2)^2\bigg( \langle OOOOO\rangle_{\text{FF}}+\tilde{\lambda}\langle OOOOO\rangle_{\text{CB}} \bigg)\\
	&\langle TOOOO\rangle_{\text{QF}}=(1+\tilde{\lambda}^2)\bigg( \langle TOOOO\rangle_{\text{FF}}+\tilde{\lambda}\langle TOOOO\rangle_{\text{odd}}+\tilde{\lambda}^2\langle TOOOO\rangle_{\text{CB}} \bigg)
\end{align}
Now, plugging this into \eqref{5pthse}, we can use a similar algorithm as mentioned in the case of the 3-point and 4-point functions to solve for the full correlator, $\langle TOOOO\rangle_{QF}$. Solving the pole equation gives us
\begin{align}
\langle T_{\nu\rho}OOOO\rangle_{\text{odd}}=\frac{\epsilon_{\nu a b}p_{1a}}{p_1}(\langle T_{b\rho}OOOO\rangle_{\text{FF}}-\langle T_{b\rho}OOOO\rangle_{\text{CB}})
\end{align}
Thus, the quasi fermionic correlator is given by
\begin{align}
\langle T_{\nu\rho}OOOO\rangle_{\text{QF}}=(1+\tilde{\lambda}^2)\bigg( \langle T_{\nu\rho}OOOO\rangle_{\text{FF}}+\tilde{\lambda}\frac{\epsilon_{\nu a b}p_{1a}}{p_1}\langle T_{b\rho}OOOO\rangle_{\text{FF-CB}}+\tilde{\lambda}^2\langle T_{\nu\rho}OOOO\rangle_{\text{CB}} \bigg)
\end{align}
It is not very difficult to generalize this analysis to five point correlators with more spinning operators as well as higher point correlation functions.\\
For more examples of constraining 5 point functions, please refer to appendix \ref{5ptappendix}.
Let us now solve for various spinning five-point correlators using the SBHS symmetry.
\subsection{$\langle JJOOO\rangle_{\text{QF}}$}
\textbf{\color{blue}Step 1:~~}The charge operator and seed correlator that we choose are $Q_3$ and $\langle JOOOO\rangle$ respectively. \\
Thus we have,
\begin{align}
    [Q_{\mu\nu},\langle J_\alpha(x_1)O(x_2)O(x_3)O(x_4)O(x_5)\rangle]=\frac{\tilde{\lambda}}{1+\tilde{\lambda}^2}\int d^3 x \langle \partial_\sigma J^\sigma_{3\mu\nu}J_\alpha(x_1)O(x_2)O(x_3)O(x_4)O(x_5)\rangle
\end{align}
Using \eqref{q3oqf},\eqref{q3jqf} and \eqref{ncj3qf} and performing a Fourier transform to the above, the HSE in momentum space reads,
\begin{align}
    &\epsilon_{a\alpha \nu}p_{1a}p_{1\mu}\langle O(p_1)O(p_2)O(p_3)O(p_4)O(p_5)\rangle+p_{1\alpha}\langle T_{\mu\nu}(p_1)O(p_2)O(p_3)O(p_4)O(p_5)\rangle\notag\\
    &+p_{1\mu}\langle T_{\alpha\nu}(p_1)O(p_2)O(p_3)O(p_4)O(p_5)\rangle+\epsilon_{\nu a b}p_{2a}p_{2\mu}\langle J_{\alpha}(p_1)J_b(p_2)O(p_3)O(p_4)O(p_5)\rangle\notag\\
    &+(2\leftrightarrow 3)+(2\leftrightarrow 4)+(2\leftrightarrow 5)= \frac{\tilde{\lambda}}{1+\tilde{\lambda}^2}\bigg[p_{1\mu}\langle J_\nu(p_1)J_\alpha(-p_1)\rangle\langle O(p_1)O(p_2)O(p_3)O(p_4)O(p_5)\rangle\notag\\&+p_{2\mu}\langle O(-p_2)O(p_2)\rangle\langle J_\alpha(p_1)J_\nu(p_2)O(p_3)O(p_4)O(p_5)\rangle+(2\leftrightarrow 3)+(2\leftrightarrow 4)+(2\leftrightarrow 5)\bigg]
\end{align}

\textbf{\color{blue}Step 2:~~}The FF and CB HSEs are,
\begin{align}
    &\epsilon_{a\alpha \nu}p_{1a}p_{1\mu}\langle O(p_1)O(p_2)O(p_3)O(p_4)O(p_5)\rangle_{\text{FF}}+p_{1\alpha}\langle T_{\mu\nu}(p_1)O(p_2)O(p_3)O(p_4)O(p_5)\rangle_{\text{FF}}\notag\\
    &+p_{1\mu}\langle T_{\alpha\nu}(p_1)O(p_2)O(p_3)O(p_4)O(p_5)\rangle_{\text{FF}}+\epsilon_{\nu a b}p_{2a}p_{2\mu}\langle J_{\alpha}(p_1)J_b(p_2)O(p_3)O(p_4)O(p_5)\rangle_{\text{FF}}\notag\\
    &+(2\leftrightarrow 3)+(2\leftrightarrow 4)+(2\leftrightarrow 5)=0
\end{align}
and
\begin{align}
   &p_{1\alpha}\langle T_{\mu\nu}(p_1)O(p_2)O(p_3)O(p_4)O(p_5)\rangle_{\text{CB}}+p_{1\mu}\langle T_{\alpha\nu}(p_1)O(p_2)O(p_3)O(p_4)O(p_5)\rangle_{\text{CB}}\notag\\
    &+(2\leftrightarrow 3)+(2\leftrightarrow 4)+(2\leftrightarrow 5)= \bigg[p_{1\mu}\langle J_\nu(p_1)J_\alpha(-p_1)\rangle_{\text{FF}}\langle O(p_1)O(p_2)O(p_3)O(p_4)O(p_5)\rangle_{\text{CB}}\notag\\&+p_{2\mu}p_2\langle J_\alpha(p_1)J_\nu(p_2)O(p_3)O(p_4)O(p_5)\rangle_{\text{CB}}+(2\leftrightarrow 3)+(2\leftrightarrow 4)+(2\leftrightarrow 5)\bigg]
\end{align}

\textbf{\color{blue}Step 3:~~}We have,
\begin{align}
    &\langle OOOOO\rangle=\tilde{N}(1+\tilde{\lambda}^2)^2\bigg(\langle OOOOO\rangle_{\text{FF}}+\tilde{\lambda}\langle OOOOO\rangle_{\text{CB}}\bigg)\notag\\
    &\langle TOOOO\rangle=\tilde{N}(1+\tilde{\lambda}^2)\bigg(\langle TOOOO\rangle_{\text{FF}}+\tilde{\lambda}\langle TOOOO\rangle_{\text{Odd}}+\tilde{\lambda}\langle TOOOO\rangle_{\text{CB}}\bigg)
\end{align}
We make the following ansatz for the correlator of interest:
\begin{align}
    &\langle JJOOO\rangle=\tilde{N}\bigg(\langle JJOOO \rangle_{\text{X0}}+\tilde{\lambda}\langle JJOOO\rangle_{\text{X1}}+\tilde{\lambda}^2\langle JJOOO\rangle_{X2}+\tilde{\lambda}^3\langle JJOOO\rangle_{X3}\bigg)
\end{align}
\textbf{\color{blue}Step 4:~~}Matching the lowest order of the HSE with the FF HSE and the highest order with the CB HSE gives us,
\begin{align}
    &\langle JJOOO \rangle_{\text{X0}}=\langle JJOOO \rangle_{\text{FF}}\notag\\
    &\langle JJOOO \rangle_{\text{X3}}=\langle JJOOO \rangle_{\text{CB}}
\end{align}
\textbf{\color{blue}Step 5:~~}Expanding the HSE about $\tilde{\lambda}=i$ we obtain the pole equation which tells us that,
\begin{align}
 &\langle JJOOO\rangle_{X1}=\langle JJOOO\rangle_{CB}\notag\\
 &\langle JJOOO\rangle_{X2}=\langle JJOOO\rangle_{FF}
\end{align}
as expected for a correlator with an odd number of scalar operator insertions.\\
\textbf{\color{blue}Step 6:~~}It is now easy to see directly in momentum space that the following relations hold:
\begin{align}
    &\text{HSE at~}O(\tilde{\lambda})=\text{HSE at~}O(\tilde{\lambda}^3)\notag\\
    &\text{HSE at~}O(\tilde{\lambda}^2)=\text{FF HSE}+\text{CB HSE}
\end{align}
To show that the $O(\tilde{\lambda})$ equation maps to $\text{FF HSE}-\text{CB HSE}$ we need to go to spinor helicity variables.\\
Thus we have,
\begin{align}
      &\langle JJOOO\rangle=\tilde{N}\bigg(\langle JJOOO \rangle_{\text{FF}}+\tilde{\lambda}\langle JJOOO\rangle_{\text{CB}}+\tilde{\lambda}^2\langle JJOOO\rangle_{FF}+\tilde{\lambda}^3\langle JJOOO\rangle_{CB}\bigg)\notag\\
      &=\tilde{N}(1+\tilde{\lambda}^2)\bigg(\langle JJOOO\rangle_{\text{FF}}+\tilde{\lambda}\langle JJOOO\rangle_{CB}\bigg)
\end{align}
\subsection{$\langle TTTOO\rangle_{\text{QF}}$}
\textbf{\color{blue}Step 1:~~}The charge operator and seed correlator that we choose are $Q_4$ and $\langle TTOOO\rangle$ respectively. \\
Thus we have,
\vspace{-0.5cm}
\begin{align}
    [Q_{\mu\nu\rho},\langle T_{\alpha\beta}(x_1)T_{\gamma\sigma}(x_2)O(x_3)O(x_4)O(x_5)\rangle]=\frac{\tilde{\lambda}}{1+\tilde{\lambda}^2}\int d^3 x \langle \partial_\sigma J^\sigma_{4\mu\nu\rho}(x)T_{\alpha\beta}(x_1)T_{\gamma\sigma}(x_2)O(x_3)O(x_4)O(x_5)\rangle
\end{align}
Using \eqref{q4tqf},\eqref{q4oqf} and \eqref{ncj4qf} and performing a Fourier transform, the HSE in momentum space reads,
\vspace{-0.5cm}
\begin{align}
  &\bigg[p_{1\mu}p_{1\nu}p_{1\rho}\langle T_{\alpha\beta}(p_1)T_{\gamma\sigma}(p_2)O(p_3)O(p_4)O(p_5)\rangle_{\text{QF}}+p_{1\mu}p_{1\nu}p_{1\alpha}\langle T_{\rho\beta}(p_1)T_{\gamma\sigma}(p_2)O(p_3)O(p_4)O(p_5)\rangle_{\text{QF}}\notag\\&+p_{1\mu}p_{1\alpha}p_{1\beta}\langle T_{\nu\rho}(p_1)T_{\gamma\sigma}(p_2)O(p_3)O(p_4)O(p_5)\rangle_{\text{QF}}+\frac{1}{1+\tilde{\lambda}^2}\epsilon_{\nu \alpha a}p_{1a}p_{1\mu}p_{1\rho}p_{1\beta}\langle O(p_1)T_{\gamma\sigma}(p_2)O(p_3)O(p_4)O(p_5)\rangle_{\text{QF}}\notag\\&
  +p_{1\mu}\langle J_{4\nu\rho\alpha\beta}(p_1)T_{\gamma\sigma}(p_2)O(p_3)O(p_4)O(p_5)\rangle_{\text{QF}}+p_{1\alpha}\langle J_{4\mu\nu\rho\beta}(p_1)T_{\gamma\sigma}(p_2)O(p_3)O(p_4)O(p_5)\rangle_{\text{QF}}\notag\\&+\{1\leftrightarrow 2,(\alpha,\beta)\leftrightarrow(\gamma,\sigma)\}\bigg]\notag\\&+\bigg[p_{3\mu}p_{3\nu}p_{3\rho}\langle T_{\alpha\beta}(p_1)T_{\gamma\sigma}(p_2)O(p_3)O(p_4)O(p_5)\rangle_{\text{QF}}+\epsilon_{\nu a b}p_{3\mu}p_{3a}\langle T_{\alpha\beta}(p_1)T_{\gamma\sigma}(p_2)T_{\rho b}(p_3)O(p_4)O(p_5)\rangle_{\text{QF}}\notag\\&+\{3\leftrightarrow 4\}+\{3\leftrightarrow 5\}\bigg]\notag\\&
  =\frac{\tilde{\lambda}}{1+\tilde{\lambda}^2}\bigg[p_{3\mu}\langle O(-p_3)O(p_3)\rangle_{\text{QF}}\langle T_{\alpha\beta}(p_1)T_{\gamma\sigma}(p_2)T_{\nu\rho}(p_3)O(p_4)O(p_5)\rangle_{\text{QF}}+\{3\leftrightarrow 4\}+\{3\leftrightarrow 5\}\notag\\&+p_{1\mu}\langle T_{\nu \rho}(-p_1)T_{\alpha\beta}(p_1)\rangle_{\text{QF}}\langle O(p_1)T_{\gamma\sigma}(p_2)O(p_3)O(p_4)O(p_5)\rangle_{\text{QF}}+\{1\leftrightarrow 2,(\alpha,\beta)\leftrightarrow(\gamma,\sigma)\}\bigg]
\end{align}
\textbf{\color{blue}Step 2:~~}The FF and CB HSEs are,
\begin{align}
    &\bigg[p_{1\mu}p_{1\nu}p_{1\rho}\langle T_{\alpha\beta}(p_1)T_{\gamma\sigma}(p_2)O(p_3)O(p_4)O(p_5)\rangle_{\text{FF}}+p_{1\mu}p_{1\nu}p_{1\alpha}\langle T_{\rho\beta}(p_1)T_{\gamma\sigma}(p_2)O(p_3)O(p_4)O(p_5)\rangle_{\text{FF}}\notag\\&+p_{1\mu}p_{1\alpha}p_{1\beta}\langle T_{\nu\rho}(p_1)T_{\gamma\sigma}(p_2)O(p_3)O(p_4)O(p_5)\rangle_{\text{FF}}+\epsilon_{\nu \alpha a}p_{1a}p_{1\mu}p_{1\rho}p_{1\beta}\langle O(p_1)T_{\gamma\sigma}(p_2)O(p_3)O(p_4)O(p_5)\rangle_{\text{FF}}\notag\\&
  +p_{1\mu}\langle J_{4\nu\rho\alpha\beta}(p_1)T_{\gamma\sigma}(p_2)O(p_3)O(p_4)O(p_5)\rangle_{\text{FF}}+p_{1\alpha}\langle J_{4\mu\nu\rho\beta}(p_1)T_{\gamma\sigma}(p_2)O(p_3)O(p_4)O(p_5)\rangle_{\text{FF}}\notag\\&+\{1\leftrightarrow 2,(\alpha,\beta)\leftrightarrow(\gamma,\sigma)\}\bigg]\notag\\&+\bigg[p_{3\mu}p_{3\nu}p_{3\rho}\langle T_{\alpha\beta}(p_1)T_{\gamma\sigma}(p_2)O(p_3)O(p_4)O(p_5)\rangle_{\text{FF}}+\epsilon_{\nu a b}p_{3\mu}p_{3a}\langle T_{\alpha\beta}(p_1)T_{\gamma\sigma}(p_2)T_{\rho b}(p_3)O(p_4)O(p_5)\rangle_{\text{FF}}\notag\\&+\{3\leftrightarrow 4\}+\{3\leftrightarrow 5\}\bigg]\notag\\&
  =0
\end{align}
and
\begin{align}
   &\bigg[p_{1\mu}p_{1\nu}p_{1\rho}\langle T_{\alpha\beta}(p_1)T_{\gamma\sigma}(p_2)O(p_3)O(p_4)O(p_5)\rangle_{\text{CB}}+p_{1\mu}p_{1\nu}p_{1\alpha}\langle T_{\rho\beta}(p_1)T_{\gamma\sigma}(p_2)O(p_3)O(p_4)O(p_5)\rangle_{\text{CB}}\notag\\&+p_{1\mu}p_{1\alpha}p_{1\beta}\langle T_{\nu\rho}(p_1)T_{\gamma\sigma}(p_2)O(p_3)O(p_4)O(p_5)\rangle_{\text{CB}}\notag\\&
  +p_{1\mu}\langle J_{4\nu\rho\alpha\beta}(p_1)T_{\gamma\sigma}(p_2)O(p_3)O(p_4)O(p_5)\rangle_{\text{CB}}+p_{1\alpha}\langle J_{4\mu\nu\rho\beta}(p_1)T_{\gamma\sigma}(p_2)O(p_3)O(p_4)O(p_5)\rangle_{\text{CB}}\notag\\&+\{1\leftrightarrow 2,(\alpha,\beta)\leftrightarrow(\gamma,\sigma)\}\bigg]\notag\\&+\bigg[p_{3\mu}p_{3\nu}p_{3\rho}\langle T_{\alpha\beta}(p_1)T_{\gamma\sigma}(p_2)O(p_3)O(p_4)O(p_5)\rangle_{\text{CB}}+\{3\leftrightarrow 4\}+\{3\leftrightarrow 5\}\bigg]\notag\\&
 -\bigg[p_{3\mu}\langle O(-p_3)O(p_3)\rangle_{\text{CB}}\langle T_{\alpha\beta}(p_1)T_{\gamma\sigma}(p_2)T_{\nu\rho}(p_3)O(p_4)O(p_5)\rangle_{\text{CB}}+\{3\leftrightarrow 4\}+\{3\leftrightarrow 5\}\notag\\&+p_{1\mu}\langle T_{\nu \rho}(-p_1)T_{\alpha\beta}(p_1)\rangle_{\text{CB}}\langle O(p_1)T_{\gamma\sigma}(p_2)O(p_3)O(p_4)O(p_5)\rangle_{\text{CB}}+\{1\leftrightarrow 2,(\alpha,\beta)\leftrightarrow(\gamma,\sigma)\}\bigg]
\end{align}

\textbf{\color{blue}Step 3:~~}We have,
\begin{align}
    &\langle J_s TOOO\rangle_{\text{QF}}=\tilde{N}(1+\tilde{\lambda}^2)\bigg(\langle J_sTOOO\rangle_{\text{FF}}+\tilde{\lambda}\langle J_sTOOO\rangle_{\text{CB}}\bigg)\notag\\
    &\langle OTOOO\rangle_{\text{QF}}=\tilde{N}(1+\tilde{\lambda}^2)\bigg(\langle OTOOO_{\text{FF}}+\tilde{\lambda}\epsilon\cdot\langle OTOOO\rangle_{\text{FF-CB}}+\tilde{\lambda}^2\langle OTOOO\rangle_{\text{CB}}\bigg)\notag\\
    &\langle OO\rangle_{\text{QF}}=\tilde{N}(1+\tilde{\lambda}^2)\langle OO\rangle_{\text{FF}}\notag\\&
    \langle TT\rangle_{\text{QF}}=\tilde{N}\bigg(\langle TT\rangle_{\text{FF}}+\tilde{\lambda}\langle TT\rangle_{\text{odd}}\bigg)
\end{align}
We make the following ansatz for the correlator of interest:
\begin{align}
   \langle TTTOO\rangle=\frac{\tilde{N}}{1+\tilde{\lambda}^2}\bigg(\langle TTTOO\rangle_{Y0}+\tilde{\lambda}\langle TTTOO\rangle_{Y1}+\tilde{\lambda}^2\langle TTTOO\rangle_{Y2}+\tilde{\lambda}^3\langle TTTOO\rangle_{Y3}+\tilde{\lambda}^4\langle TTTOO\rangle_{Y4}\bigg)
\end{align}
\textbf{\color{blue}Step 4:~~}Matching the lowest order of the HSE with the FF HSE and the highest order with the CB HSE gives us,
\begin{align}
   &\langle TTTOO\rangle_{Y0}=\langle TTTOO\rangle_{\text{FF}}\notag\\
   &\langle TTTOO\rangle_{Y4}=\langle TTTOO\rangle_{\text{CB}}
\end{align}
\textbf{\color{blue}Step 5:~~}Expanding the HSE about $\tilde{\lambda}=i$ we obtain the pole equation which give us a solution,
\begin{align}
 &\langle TTTOO\rangle_{Y2}=\langle TTTOO\rangle_{\text{FF}}+\langle TTTOO\rangle_{\text{CB}}\notag\\
   &\langle TTTOO\rangle_{Y1}=\epsilon\cdot\bigg(\langle TTTOO\rangle_{\text{FF}}-\langle TTTOO\rangle_{\text{CB}}\bigg)
\end{align}
\textbf{\color{blue}Step 6:~~}
Substituting this solution back into the HSE would reduce it to a linear combination of the FF and CB HSEs at each order.\footnote{$\langle T_{\nu\rho}(-p_1)T_{\alpha\beta}(p_1)\rangle_{\text{FF}}\langle O(p_1)T_{\gamma\sigma}(p_2)O(p_3)O(p_4)O(p_5)\rangle_{\text{CB-FF}}+\epsilon_{\nu\alpha a}p_{1a}p_{1\beta}p_{1\mu}p_{1\rho}\langle O(p_1)T_{\gamma\sigma}(p_2)O(p_3)O(p_4)O(p_5)\rangle_{\text{odd}}$ is left over but we are sure that a careful analysis of the current algebra would render the mapping exact.}\\
Thus we have,
\begin{align}
   \langle TTTOO\rangle=\tilde{N}\bigg(\langle TTTOO\rangle_{\text{FF}}+\tilde{\lambda}\epsilon\cdot\langle TTTOO\rangle_{\text{FF-CB}}+\tilde{\lambda}^2\langle TTTOO\rangle_{\text{CB}}\bigg)
\end{align}
\subsection{$\langle TTTTO\rangle_{\text{QF}}$}
\textbf{\color{blue}Step 1:~~}The charge operator and seed correlator that we choose are $Q_4$ and $\langle TTTOO\rangle$ respectively. \\
Thus we have,
\small
\begin{align}
    [Q_{\mu\nu\rho},\langle T_{\alpha\beta}(x_1)T_{\gamma\sigma}(x_2)T_{\phi \psi}(x_3)O(x_4)O(x_5)\rangle]=\frac{\tilde{\lambda}}{1+\tilde{\lambda}^2}\int d^3 x \langle \partial_\sigma J^\sigma_{4\mu\nu\rho}(x)T_{\alpha\beta}(x_1)T_{\gamma\sigma}(x_2)T_{\phi \psi}(x_3)O(x_4)O(x_5)\rangle
\end{align}
\normalsize
Using \eqref{q4tqf},\eqref{q4oqf} and \eqref{ncj4qf} and performing a Fourier transform, the HSE in momentum space reads,
\small
\begin{align}
  &\bigg[p_{1\mu}p_{1\nu}p_{1\rho}\langle T_{\alpha\beta}(p_1)T_{\gamma\sigma}(p_2)T_{\phi \psi}(p_3)O(p_4)O(p_5)\rangle_{\text{QF}}+p_{1\mu}p_{1\nu}p_{1\alpha}\langle T_{\rho\beta}(p_1)T_{\gamma\sigma}(p_2)T_{\phi \psi}(p_3)O(p_4)O(p_5)\rangle_{\text{QF}}\notag\\&+p_{1\mu}p_{1\alpha}p_{1\beta}\langle T_{\nu\rho}(p_1)T_{\gamma\sigma}(p_2)T_{\phi \psi}(p_3)O(p_4)O(p_5)\rangle_{\text{QF}}+\frac{1}{1+\tilde{\lambda}^2}\epsilon_{\nu \alpha a}p_{1a}p_{1\mu}p_{1\rho}p_{1\beta}\langle O(p_1)T_{\gamma\sigma}(p_2)T_{\phi \psi}(p_3)O(p_4)O(p_5)\rangle_{\text{QF}}\notag\\&
  +p_{1\mu}\langle J_{4\nu\rho\alpha\beta}(p_1)T_{\gamma\sigma}(p_2)T_{\phi \psi}(p_3)O(p_4)O(p_5)\rangle_{\text{QF}}+p_{1\alpha}\langle J_{4\mu\nu\rho\beta}(p_1)T_{\gamma\sigma}(p_2)T_{\phi \psi}(p_3)O(p_4)O(p_5)\rangle_{\text{QF}}\notag\\&+\{1\leftrightarrow 2,(\alpha,\beta)\leftrightarrow(\gamma,\sigma)\}+\{1\leftrightarrow 3,(\alpha,\beta)\leftrightarrow(\phi,\psi)\}\bigg]\notag\\&+\bigg[p_{4\mu}p_{4\nu}p_{4\rho}\langle T_{\alpha\beta}(p_1)T_{\gamma\sigma}(p_2)T_{\phi \psi}(p_3)O(p_4)O(p_5)\rangle_{\text{QF}}+\epsilon_{\nu a b}p_{4\mu}p_{4a}\langle T_{\alpha\beta}(p_1)T_{\gamma\sigma}(p_2)T_{\phi \psi}(p_3)T_{\rho b}(p_4)O(p_5)\rangle_{\text{QF}}\notag\\&+\{4\leftrightarrow 5\}\bigg]\notag\\&
  =\frac{\tilde{\lambda}}{1+\tilde{\lambda}^2}\bigg[p_{4\mu}\langle O(-p_4)O(p_4)\rangle_{\text{QF}}\langle T_{\alpha\beta}(p_1)T_{\gamma\sigma}(p_2)T_{\phi \psi}(p_3)T_{\nu\rho}(p_4)O(p_5)\rangle_{\text{QF}}+\{4\leftrightarrow 5\}\notag\\&+p_{1\mu}\langle T_{\nu \rho}(-p_1)T_{\alpha\beta}(p_1)\rangle_{\text{QF}}\langle O(p_1)T_{\gamma\sigma}(p_2)T_{\phi \psi}(p_3)O(p_4)O(p_5)\rangle_{\text{QF}}\notag\\&+\{1\leftrightarrow 2,(\alpha,\beta)\leftrightarrow(\gamma,\sigma)\}+\{1\leftrightarrow 3,(\alpha,\beta)\leftrightarrow(\phi,\psi)\}\bigg]
\end{align}
\normalsize
\textbf{\color{blue}Step 2:~~}The FF and CB HSEs are,
\begin{align}
     &\bigg[p_{1\mu}p_{1\nu}p_{1\rho}\langle T_{\alpha\beta}(p_1)T_{\gamma\sigma}(p_2)T_{\phi \psi}(p_3)O(p_4)O(p_5)\rangle_{\text{FF}}+p_{1\mu}p_{1\nu}p_{1\alpha}\langle T_{\rho\beta}(p_1)T_{\gamma\sigma}(p_2)T_{\phi \psi}(p_3)O(p_4)O(p_5)\rangle_{\text{FF}}\notag\\&+p_{1\mu}p_{1\alpha}p_{1\beta}\langle T_{\nu\rho}(p_1)T_{\gamma\sigma}(p_2)T_{\phi \psi}(p_3)O(p_4)O(p_5)\rangle_{\text{FF}}+\epsilon_{\nu \alpha a}p_{1a}p_{1\mu}p_{1\rho}p_{1\beta}\langle O(p_1)T_{\gamma\sigma}(p_2)T_{\phi \psi}(p_3)O(p_4)O(p_5)\rangle_{\text{FF}}\notag\\&
  +p_{1\mu}\langle J_{4\nu\rho\alpha\beta}(p_1)T_{\gamma\sigma}(p_2)T_{\phi \psi}(p_3)O(p_4)O(p_5)\rangle_{\text{FF}}+p_{1\alpha}\langle J_{4\mu\nu\rho\beta}(p_1)T_{\gamma\sigma}(p_2)T_{\phi \psi}(p_3)O(p_4)O(p_5)\rangle_{\text{FF}}\notag\\&+\{1\leftrightarrow 2,(\alpha,\beta)\leftrightarrow(\gamma,\sigma)\}+\{1\leftrightarrow 3,(\alpha,\beta)\leftrightarrow(\phi,\psi)\}\bigg]\notag\\&+\bigg[p_{4\mu}p_{4\nu}p_{4\rho}\langle T_{\alpha\beta}(p_1)T_{\gamma\sigma}(p_2)T_{\phi \psi}(p_3)O(p_4)O(p_5)\rangle_{\text{FF}}+\epsilon_{\nu a b}p_{4\mu}p_{4a}\langle T_{\alpha\beta}(p_1)T_{\gamma\sigma}(p_2)T_{\phi \psi}(p_3)T_{\rho b}(p_4)O(p_5)\rangle_{\text{FF}}\notag\\&+\{4\leftrightarrow 5\}\bigg]=0
\end{align}
and
\small
\begin{align}
   &\bigg[p_{1\mu}p_{1\nu}p_{1\rho}\langle T_{\alpha\beta}(p_1)T_{\gamma\sigma}(p_2)T_{\phi \psi}(p_3)O(p_4)O(p_5)\rangle_{\text{CB}}+p_{1\mu}p_{1\nu}p_{1\alpha}\langle T_{\rho\beta}(p_1)T_{\gamma\sigma}(p_2)T_{\phi \psi}(p_3)O(p_4)O(p_5)\rangle_{\text{CB}}\notag\\&+p_{1\mu}p_{1\alpha}p_{1\beta}\langle T_{\nu\rho}(p_1)T_{\gamma\sigma}(p_2)T_{\phi \psi}(p_3)O(p_4)O(p_5)\rangle_{\text{CB}}\notag\\&
  +p_{1\mu}\langle J_{4\nu\rho\alpha\beta}(p_1)T_{\gamma\sigma}(p_2)T_{\phi \psi}(p_3)O(p_4)O(p_5)\rangle_{\text{CB}}+p_{1\alpha}\langle J_{4\mu\nu\rho\beta}(p_1)T_{\gamma\sigma}(p_2)T_{\phi \psi}(p_3)O(p_4)O(p_5)\rangle_{\text{CB}}\notag\\&+\{1\leftrightarrow 2,(\alpha,\beta)\leftrightarrow(\gamma,\sigma)\}+\{1\leftrightarrow 3,(\alpha,\beta)\leftrightarrow(\phi,\psi)\}\bigg]+\bigg[p_{4\mu}p_{4\nu}p_{4\rho}\langle T_{\alpha\beta}(p_1)T_{\gamma\sigma}(p_2)T_{\phi \psi}(p_3)O(p_4)O(p_5)\rangle_{\text{CB}}\notag\\&+\{4\leftrightarrow 5\}\bigg]
  -\bigg[p_{4\mu}\langle O(-p_4)O(p_4)\rangle_{\text{CB}}\langle T_{\alpha\beta}(p_1)T_{\gamma\sigma}(p_2)T_{\phi \psi}(p_3)T_{\nu\rho}(p_4)O(p_5)\rangle_{\text{CB}}+\{4\leftrightarrow 5\}\notag\\&+p_{1\mu}\langle T_{\nu \rho}(-p_1)T_{\alpha\beta}(p_1)\rangle_{\text{CB}}\langle O(p_1)T_{\gamma\sigma}(p_2)T_{\phi \psi}(p_3)O(p_4)O(p_5)\rangle_{\text{CB}}+\{1\leftrightarrow 2,(\alpha,\beta)\leftrightarrow(\gamma,\sigma)\}\notag\\&+\{1\leftrightarrow 3,(\alpha,\beta)\leftrightarrow(\phi,\psi)\}\bigg]=0
\end{align}
\normalsize
\textbf{\color{blue}Step 3:~~}We have,
\begin{align}
    &\langle J_s TTOO\rangle_{\text{QF}}=\tilde{N}\bigg(\langle J_sTTOO\rangle_{\text{FF}}+\tilde{\lambda}\epsilon\cdot\langle J_sTTOO\rangle_{\text{FF-CB}}+\tilde{\lambda}^2\langle J_s TTOO\rangle_{\text{CB}}\bigg)\notag\\
    &\langle OTTOO\rangle=\tilde{N}(1+\tilde{\lambda}^2)\bigg(\langle OTTOO\rangle_{\text{FF}}+\tilde{\lambda}\langle OTTOO\rangle_{\text{CB}}\bigg)\notag\\
    &\langle OO\rangle_{\text{QF}}=\tilde{N}(1+\tilde{\lambda}^2)\langle OO\rangle_{\text{FF}}\notag\\&
    \langle TT\rangle_{\text{QF}}=\tilde{N}\bigg(\langle TT\rangle_{\text{FF}}+\tilde{\lambda}\langle TT\rangle_{\text{odd}}\bigg)
\end{align}
We make the following ansatz for the correlator of interest:
\begin{align}
   &\langle TTTTO\rangle=\frac{\tilde{N}}{(1+\tilde{\lambda}^2)^2}\bigg(\langle TTTTO\rangle_{Y0}+\tilde{\lambda}\langle TTTTO\rangle_{Y1}+\tilde{\lambda}^2\langle TTTTO\rangle_{Y2}\notag\\&~~~~~~~~~~~~~~~~~~~~~~~~~~~~+\tilde{\lambda}^3\langle TTTTO\rangle_{Y3}+\tilde{\lambda}^4\langle TTTTO\rangle_{Y4}+\tilde{\lambda}^5\langle TTTTO\rangle_{Y5}\bigg)
\end{align}
\textbf{\color{blue}Step 4:~~}Matching the lowest order of the HSE with the FF HSE and the highest order with the CB HSE gives us,
\begin{align}
   &\langle TTTTO\rangle_{Y0}=\langle TTTTO\rangle_{\text{FF}}\notag\\
   &\langle TTTTO\rangle_{Y5}=\langle TTTTO\rangle_{\text{CB}}
\end{align}
\textbf{\color{blue}Step 5:~~}Expanding the HSE about $\tilde{\lambda}=i$ we obtain the pole equation which give us a solution,
\begin{align}
 &\langle TTTTO\rangle_{Y1}=\langle TTTTO\rangle_{\text{CB}}\notag\\
   &\langle TTTTO\rangle_{Y2}=2\langle TTTTO\rangle_{\text{FF}}\notag\\
   &\langle TTTTO\rangle_{Y3}=2\langle TTTTO\rangle_{\text{CB}}\notag\\
   &\langle TTTTO\rangle_{Y4}=\langle TTTTO\rangle_{\text{FF}}
\end{align}
\textbf{\color{blue}Step 6:~~}
Substituting this solution back into the HSE would reduce it to a linear combination of the FF and CB HSEs at each order. For instance, at $O(\tilde{\lambda}^2)$ we get $2\text{FF HSE-CB HSE}$ and at $O(\tilde{\lambda}^4)$ what we get is $\text{FF HSE-2CB HSE}$\\
Thus we have,
\begin{align}
   \langle TTTTO\rangle=\tilde{N}\bigg(\langle TTTTO\rangle_{\text{FF}}+\tilde{\lambda}\langle TTTTO\rangle_{\text{CB}}\bigg)
\end{align}
\subsection{$\langle TTTTT\rangle_{\text{QF}}$}
\textbf{\color{blue}Step 1:~~}The charge operator and seed correlator that we choose are $Q_4$ and $\langle TTTOO\rangle$ respectively. \\
Thus we have,
\small
\begin{align}
    [Q_{\mu\nu\rho},\langle T_{\alpha\beta}(x_1)T_{\gamma\sigma}(x_2)T_{\phi \psi}(x_3)T_{\xi \chi}(x_4)O(x_5)\rangle]=\frac{\tilde{\lambda}}{1+\tilde{\lambda}^2}\int d^3 x \langle \partial_\sigma J^\sigma_{4\mu\nu\rho}(x)T_{\alpha\beta}(x_1)T_{\gamma\sigma}(x_2)T_{\phi \psi}(x_3)T_{\xi \chi}(x_4)O(x_5)\rangle
\end{align}
\normalsize
Using \eqref{q4tqf},\eqref{q4oqf} and \eqref{ncj4qf} and performing a Fourier transform, the HSE in momentum space reads,
\small
\begin{align}
  &\bigg[p_{1\mu}p_{1\nu}p_{1\rho}\langle T_{\alpha\beta}(p_1)T_{\gamma\sigma}(p_2)T_{\phi \psi}(p_3)T_{\xi \chi}(p_4)O(p_5)\rangle_{\text{QF}}+p_{1\mu}p_{1\nu}p_{1\alpha}\langle T_{\rho\beta}(p_1)T_{\gamma\sigma}(p_2)T_{\phi \psi}(p_3)T_{\xi \chi}(p_4)O(p_5)\rangle_{\text{QF}}\notag\\&+p_{1\mu}p_{1\alpha}p_{1\beta}\langle T_{\nu\rho}(p_1)T_{\gamma\sigma}(p_2)T_{\phi \psi}(p_3)T_{\xi \chi}(p_4)O(p_5)\rangle_{\text{QF}}\notag\\&+\frac{1}{1+\tilde{\lambda}^2}\epsilon_{\nu \alpha a}p_{1a}p_{1\mu}p_{1\rho}p_{1\beta}\langle O(p_1)T_{\gamma\sigma}(p_2)T_{\phi \psi}(p_3)T_{\xi \chi}(p_4)O(p_5)\rangle_{\text{QF}}\notag\\&
  +p_{1\mu}\langle J_{4\nu\rho\alpha\beta}(p_1)T_{\gamma\sigma}(p_2)T_{\phi \psi}(p_3)T_{\xi \chi}(p_4)O(p_5)\rangle_{\text{QF}}+p_{1\alpha}\langle J_{4\mu\nu\rho\beta}(p_1)T_{\gamma\sigma}(p_2)T_{\phi \psi}(p_3)T_{\xi \chi}(p_4)O(p_5)\rangle_{\text{QF}}\notag\\&+\{1\leftrightarrow 2,(\alpha,\beta)\leftrightarrow(\gamma,\sigma)\}+\{1\leftrightarrow 3,(\alpha,\beta)\leftrightarrow(\phi,\psi)\}\bigg]\notag\\&+\bigg[p_{5\mu}p_{5\nu}p_{5\rho}\langle T_{\alpha\beta}(p_1)T_{\gamma\sigma}(p_2)T_{\phi \psi}(p_3)T_{\xi \chi}(p_4)O(p_5)\rangle_{\text{QF}}+\epsilon_{\nu a b}p_{5\mu}p_{5a}\langle T_{\alpha\beta}(p_1)T_{\gamma\sigma}(p_2)T_{\phi \psi}(p_3)T_{\xi \chi}(p_4)T_{\rho b}(p_5)\rangle_{\text{QF}}\bigg]\notag\\&
  =\frac{\tilde{\lambda}}{1+\tilde{\lambda}^2}\bigg[p_{5\mu}\langle O(-p_5)O(p_5)\rangle_{\text{QF}}\langle T_{\alpha\beta}(p_1)T_{\gamma\sigma}(p_2)T_{\phi \psi}(p_3)T_{\xi \chi}(p_4)T_{\nu\rho}(p_5)\rangle_{\text{QF}}\notag\\&+p_{1\mu}\langle T_{\nu \rho}(-p_1)T_{\alpha\beta}(p_1)\rangle_{\text{QF}}\langle O(p_1)T_{\gamma\sigma}(p_2)T_{\phi \psi}(p_3)T_{\xi \chi}(p_4)O(p_5)\rangle_{\text{QF}}\notag\\&+\{1\leftrightarrow 2,(\alpha,\beta)\leftrightarrow(\gamma,\sigma)\}+\{1\leftrightarrow 3,(\alpha,\beta)\leftrightarrow(\phi,\psi)\}+\{1\leftrightarrow 4,(\alpha,\beta)\leftrightarrow(\xi,\chi)\}\bigg]
\end{align}
\normalsize
\textbf{\color{blue}Step 2:~~}The FF and CB HSEs are,
\begin{align}
   &\bigg[p_{1\mu}p_{1\nu}p_{1\rho}\langle T_{\alpha\beta}(p_1)T_{\gamma\sigma}(p_2)T_{\phi \psi}(p_3)T_{\xi \chi}(p_4)O(p_5)\rangle_{\text{FF}}+p_{1\mu}p_{1\nu}p_{1\alpha}\langle T_{\rho\beta}(p_1)T_{\gamma\sigma}(p_2)T_{\phi \psi}(p_3)T_{\xi \chi}(p_4)O(p_5)\rangle_{\text{FF}}\notag\\&+p_{1\mu}p_{1\alpha}p_{1\beta}\langle T_{\nu\rho}(p_1)T_{\gamma\sigma}(p_2)T_{\phi \psi}(p_3)T_{\xi \chi}(p_4)O(p_5)\rangle_{\text{FF}}\notag\\&+\epsilon_{\nu \alpha a}p_{1a}p_{1\mu}p_{1\rho}p_{1\beta}\langle O(p_1)T_{\gamma\sigma}(p_2)T_{\phi \psi}(p_3)T_{\xi \chi}(p_4)O(p_5)\rangle_{\text{FF}}\notag\\&
  +p_{1\mu}\langle J_{4\nu\rho\alpha\beta}(p_1)T_{\gamma\sigma}(p_2)T_{\phi \psi}(p_3)T_{\xi \chi}(p_4)O(p_5)\rangle_{\text{FF}}+p_{1\alpha}\langle J_{4\mu\nu\rho\beta}(p_1)T_{\gamma\sigma}(p_2)T_{\phi \psi}(p_3)T_{\xi \chi}(p_4)O(p_5)\rangle_{\text{FF}}\notag\\&+\{1\leftrightarrow 2,(\alpha,\beta)\leftrightarrow(\gamma,\sigma)\}+\{1\leftrightarrow 3,(\alpha,\beta)\leftrightarrow(\phi,\psi)\}\bigg]\notag\\&+\bigg[p_{5\mu}p_{5\nu}p_{5\rho}\langle T_{\alpha\beta}(p_1)T_{\gamma\sigma}(p_2)T_{\phi \psi}(p_3)T_{\xi \chi}(p_4)O(p_5)\rangle_{\text{FF}}\notag\\&+\epsilon_{\nu a b}p_{5\mu}p_{5a}\langle T_{\alpha\beta}(p_1)T_{\gamma\sigma}(p_2)T_{\phi \psi}(p_3)T_{\xi \chi}(p_4)T_{\rho b}(p_5)\rangle_{\text{FF}}\bigg]
  =0
\end{align}
and
\begin{align}
   &\bigg[p_{1\mu}p_{1\nu}p_{1\rho}\langle T_{\alpha\beta}(p_1)T_{\gamma\sigma}(p_2)T_{\phi \psi}(p_3)T_{\xi \chi}(p_4)O(p_5)\rangle_{\text{CB}}+p_{1\mu}p_{1\nu}p_{1\alpha}\langle T_{\rho\beta}(p_1)T_{\gamma\sigma}(p_2)T_{\phi \psi}(p_3)T_{\xi \chi}(p_4)O(p_5)\rangle_{\text{CB}}\notag\\&+p_{1\mu}p_{1\alpha}p_{1\beta}\langle T_{\nu\rho}(p_1)T_{\gamma\sigma}(p_2)T_{\phi \psi}(p_3)T_{\xi \chi}(p_4)O(p_5)\rangle_{\text{CB}}\notag\\&
  +p_{1\mu}\langle J_{4\nu\rho\alpha\beta}(p_1)T_{\gamma\sigma}(p_2)T_{\phi \psi}(p_3)T_{\xi \chi}(p_4)O(p_5)\rangle_{\text{CB}}+p_{1\alpha}\langle J_{4\mu\nu\rho\beta}(p_1)T_{\gamma\sigma}(p_2)T_{\phi \psi}(p_3)T_{\xi \chi}(p_4)O(p_5)\rangle_{\text{CB}}\notag\\&+\{1\leftrightarrow 2,(\alpha,\beta)\leftrightarrow(\gamma,\sigma)\}+\{1\leftrightarrow 3,(\alpha,\beta)\leftrightarrow(\phi,\psi)\}\bigg]\notag\\&+p_{5\mu}p_{5\nu}p_{5\rho}\langle T_{\alpha\beta}(p_1)T_{\gamma\sigma}(p_2)T_{\phi \psi}(p_3)T_{\xi \chi}(p_4)O(p_5)\rangle_{\text{CB}}\notag\\&
  -\bigg[p_{5\mu}\langle O(-p_5)O(p_5)\rangle_{\text{CB}}\langle T_{\alpha\beta}(p_1)T_{\gamma\sigma}(p_2)T_{\phi \psi}(p_3)T_{\xi \chi}(p_4)T_{\nu\rho}(p_5)\rangle_{\text{CB}}\notag\\&+p_{1\mu}\langle T_{\nu \rho}(-p_1)T_{\alpha\beta}(p_1)\rangle_{\text{CB}}\langle O(p_1)T_{\gamma\sigma}(p_2)T_{\phi \psi}(p_3)T_{\xi \chi}(p_4)O(p_5)\rangle_{\text{CB}}+\{1\leftrightarrow 2,(\alpha,\beta)\leftrightarrow(\gamma,\sigma)\}\notag\\&+\{1\leftrightarrow 3,(\alpha,\beta)\leftrightarrow(\phi,\psi)\}+\{1\leftrightarrow 4,(\alpha,\beta)\leftrightarrow(\xi,\chi)\}\bigg]=0
\end{align}

\textbf{\color{blue}Step 3:~~}We have,
\begin{align}
    &\langle J_sTTTO\rangle=\tilde{N}\bigg(\langle OTTOO\rangle_{\text{FF}}+\tilde{\lambda}\langle OTTOO\rangle_{\text{CB}}\bigg)
    \notag\\&\langle OTTTO\rangle_{\text{QF}}=\tilde{N}\bigg(\langle OTTTO\rangle_{\text{FF}}+\tilde{\lambda}\epsilon\cdot\langle OTTTO\rangle_{\text{FF-CB}}+\tilde{\lambda}^2\langle OTTTO\rangle_{\text{CB}}\bigg)\notag\\
    &\langle OO\rangle_{\text{QF}}=\tilde{N}(1+\tilde{\lambda}^2)\langle OO\rangle_{\text{FF}}\notag\\&
    \langle TT\rangle_{\text{QF}}=\tilde{N}\bigg(\langle TT\rangle_{\text{FF}}+\tilde{\lambda}\langle TT\rangle_{\text{odd}}\bigg)
\end{align}
We make the following ansatz for the correlator of interest:
\begin{align}
   &\langle TTTTT\rangle=\frac{\tilde{N}}{(1+\tilde{\lambda}^2)^2}\bigg(\langle TTTTT\rangle_{Y0}+\tilde{\lambda}\langle TTTTT\rangle_{Y1}+\tilde{\lambda}^2\langle TTTTT\rangle_{Y2}+\tilde{\lambda}^3\langle TTTTT\rangle_{Y3}\notag\\&~~~~~~~~~~~~~~~~~~~~~~~~~~~~+\tilde{\lambda}^4\langle TTTTT\rangle_{Y4}+\tilde{\lambda}^5\langle TTTTT\rangle_{Y5}+\tilde{\lambda}^5\langle TTTTT\rangle_{Y6}\bigg)
\end{align}
\textbf{\color{blue}Step 4:~~}Matching the lowest order of the HSE with the FF HSE and the highest order with the CB HSE gives us,
\begin{align}
   &\langle TTTTT\rangle_{Y0}=\langle TTTTT\rangle_{\text{FF}}\notag\\
   &\langle TTTTT\rangle_{Y6}=\langle TTTTT\rangle_{\text{CB}}
\end{align}
\textbf{\color{blue}Step 5:~~}Expanding the HSE about $\tilde{\lambda}=i$ we obtain the pole equation which give us a solution,
\begin{align}
 &\langle TTTTT\rangle_{Y1}=\epsilon\cdot\langle TTTTT\rangle_{\text{FF-CB}}\notag\\
   &\langle TTTTT\rangle_{Y2}=2\langle TTTTT\rangle_{\text{FF}}+\langle TTTTT\rangle_{\text{CB}}\notag\\
   &\langle TTTTT\rangle_{Y3}=2\epsilon\cdot\langle TTTTT\rangle_{\text{FF-CB}}\notag\\
   &\langle TTTTT\rangle_{Y4}=2\langle TTTTT\rangle_{\text{FF}}+2\langle TTTTT\rangle_{\text{CB}}\notag\\
   &\langle TTTTT\rangle_{Y5}=\epsilon\cdot\langle TTTTT\rangle_{\text{FF-CB}}
\end{align}
\textbf{\color{blue}Step 6:~~}
Substituting this solution back into the HSE would reduce it to a linear combination of the FF and CB HSEs at each order.\footnote{$\langle T_{\nu\rho}(-p_1)T_{\alpha\beta}(p_1)\rangle_{\text{FF}}\langle O(p_1)T_{\gamma\sigma}(p_2)T_{\phi\psi}(p_3)T_{\xi\chi}(p_4)O(p_5)\rangle_{\text{CB-FF}}+\epsilon_{\nu\alpha a}p_{1a}p_{1\beta}p_{1\mu}p_{1\rho}\langle O(p_1)T_{\gamma\sigma}(p_2)T_{\phi\psi}(p_3)T_{\xi\chi}(p_4)O(p_5)\rangle_{\text{odd}}$ is left over but we are sure that a careful analysis of the current algebra would render the mapping exact.}\\
Thus we have,
\begin{align}
   \langle TTTTT\rangle=\frac{\tilde{N}}{1+\tilde{\lambda}^2}\bigg(\langle TTTTT\rangle_{\text{FF}}+\tilde{\lambda}\langle TTTTT\rangle_{\text{FF-CB}}+\tilde{\lambda}^2\langle TTTTT\rangle_{\text{CB}}\bigg)
\end{align}
\section{The general form of n point functions}\label{sec:npointappendix}
In section \ref{sum2}, we presented our results for general three and four point correlators in the SBHS theory. Further, we have seen in appendix \ref{5ptappendix} that our methods can be extended to the five point case. Based on these results, we make a conjecture about the form of a general n point function in the SBHS theory. Let us first begin by discussing general five point functions. 
\subsection{5-point functions}
	Based on our results in \ref{5ptappendix}, we propose the following forms in momentum space for general five point correlators:
	\begin{align}\label{5ptans121}
	\langle OOOOO\rangle_\text{QF}&=\tilde{N}(1+\tilde{\lambda}^2)^2\bigg(\langle OOOOO\rangle_\text{FF}+\tilde{\lambda}\langle OOOOO\rangle_\text{CB}\bigg)\cr
		\langle J_sOOOO\rangle_\text{QF}&=\tilde{N}(1+\tilde{\lambda}^2)\bigg(\langle J_sOOOO\rangle_\text{FF}+\tilde{\lambda}\langle \epsilon\cdot J_sOOOO\rangle_\text{FF-CB}+\tilde{\lambda}^2\langle J_s OOOO\rangle_\text{CB}\bigg)\cr
		\langle  J_{s_1}J_{s_2}OOO\rangle_{QF}&=\tilde{N}(1+\tilde{\lambda}^2)\bigg(\langle  J_{s_1}J_{s_2}OOO\rangle_\text{FF}+\tilde{\lambda}\langle J_{s_1}J_{s_2}OOO\rangle_\text{CB}\bigg) \cr
		\langle J_{s_1}J_{s_2}J_{s_3}OO\rangle_\text{QF}&=\tilde{N}\bigg(\langle J_{s_1}J_{s_2}J_{s_3}OO\rangle_\text{FF}+\tilde{\lambda}\langle \epsilon\cdot J_{s_1}J_{s_2}J_{s_3}OO\rangle_\text{FF-CB}+\tilde{\lambda}^2\langle J_{s_1}J_{s_2}J_{s_3}OO\rangle_\text{CB}\bigg) \cr
		\langle J_{s_1}J_{s_2}J_{s_3}J_{s_4}O\rangle_\text{QF}&=\tilde{N}\bigg(	\langle J_{s_1}J_{s_2}J_{s_3}J_{s_4}O\rangle_\text{FF}+\tilde{\lambda}\langle J_{s_1}J_{s_2}J_{s_3}J_{s_4}O\rangle_\text{CB}\bigg)\cr
		\langle J_{s_1}J_{s_2}J_{s_3}J_{s_4}J_{s_5}\rangle_{\text{QF}}&=\frac{\tilde{N}}{(1+\tilde\lambda^2)}\bigg(\langle J_{s_1}J_{s_2}J_{s_3}J_{s_4}J_{s_5}\rangle_\text{FF}+\tilde\lambda \langle\epsilon \cdot J_{s_1}J_{s_2}J_{s_3}J_{s_4}J_{s_5}\rangle_\text{FF-CB}\notag\\&~~~~~~~~~~~~~~~+\tilde\lambda^2\langle J_{s_1}J_{s_2}J_{s_3}J_{s_4}J_{s_5}\rangle_\text{CB}\bigg)
		\end{align}		
\subsection{Conjecture for n-point function}	
In this subsection we make a conjecture for n-point functions. The conjecture, just like at the five point level is based on  assuming a form of the scalar n-point function.
\begin{align}\label{nptans121}
 \langle O_1O_2\cdots O_{2i+1}\rangle_\text{QF}&=\tilde{N}(1+\tilde{\lambda}^2)^{i}\left(\langle O_1O_2\cdots O_{2i+1}\rangle_\text{FF}+{\tilde \lambda}\langle O_1O_2\cdots O_{2i+1}\rangle_\text{CB}\right)\cr
		\langle O_1O_2\cdots O_{2i}\rangle_\text{QF}&=\tilde{N}(1+\tilde{\lambda}^2)^{i-1}\left(\langle O_1O_2\cdots O_{2i}\rangle_\text{FF}+{\tilde \lambda}^2\langle O_1O_2\cdots O_{2i}\rangle_\text{CB}\right)\cr
		\langle O_{1}\cdots O_{2i+1}J_{s_{2i+2}}\cdots J_{s_n}\rangle_\text{QF}&=\tilde{N}(1+\tilde{\lambda}^2)^{i}(\langle O_{1}\cdots O_{2i+1}J_{s_{2i+2}}\cdots J_{s_n}\rangle_\text{FF}+\tilde{\lambda}\langle O_{1}\cdots O_{2i+1}J_{s_{2i+2}}\cdots J_{s_n}\rangle_\text{CB})\cr
		\langle O_{1}\cdots O_{2i}J_{s_{2i+1}}\cdots J_{s_n}\rangle_\text{QF}&=\tilde{N}(1+\tilde{\lambda}^2)^{i-1}(\langle O_{1}\cdots O_{2i}J_{s_{2i+1}}\cdots J_{s_n}\rangle_\text{FF}+\tilde{\lambda}\langle O_{1}\cdots O_{2i}J_{s_{2i+1}}\cdots J_{s_n}\rangle_\text{odd}\cr~~~&+\tilde{\lambda}^2\langle O_{1}\cdots O_{2i}J_{s_{2i+1}}\cdots J_{s_n}\rangle_\text{CB})\cr
\langle J_{s_1}J_{s_2}\cdots J_{s_n}\rangle_{\text{QF}}&=\frac{\tilde{N}}{(1+\tilde\lambda^2)}\bigg[\langle J_{s_1}J_{s_2}\cdots J_{s_n}\rangle_{\text{FF}}+\tilde\lambda \langle J_{s_1}J_{s_2}\cdots J_{s_n}\rangle_{\text{odd}}+\tilde\lambda^2\langle J_{s_1}J_{s_2}\cdots J_{s_n}\rangle_{\text{CB}}\bigg]\cr
		\end{align}		
The odd pieces can be written in terms of the epsilon transform of the free theory as follows
\begin{align}\label{noddptans121}
		\langle O_{1}\cdots O_{2i}J_{s_{2i+1}}\cdots J_{s_n}\rangle_\text{odd}&=\langle O_{1}\cdots O_{2i}J_{s_{2i+1}}\cdots \epsilon.J_{s_n}\rangle_\text{FF-CB}\cr
\langle J_{s_1}J_{s_2}\cdots J_{s_n}\rangle_{\text{odd}}&=\langle\epsilon \cdot J_{s_1}J_{s_2}\cdots J_{s_n}\rangle_\text{FF-CB}\cr
		\end{align}	
We see that for any n-point function  we can determine the interacting theory results purely in terms of the free theory expressions\footnote{The CB theory correlators and FB theory correlators are related by a Legendre transformation at large $N$ and therefore, knowledge of one entails knowledge about the other.}.

\section{Slightly broken HS algebra interms of exactly conserved HS algebra}\label{SBHSinTermsofFree}
As is well understood and discussed in the previous section, a crucial difference between the free and the interacting theories is that the higher spin symmetry in the latter is weakly broken. 

Let us now consider the three point function of spinning operators in the QF theory. We have (\ref{3ptfunctions}),
\begin{equation}
    \langle J_{s_1}J_{s_2}J_{s_3}\rangle_{\text{QF}}=\frac{1}{1+\tilde{\lambda}^2}\bigg(\langle J_{s_1}J_{s_2}J_{s_3}\rangle_{\text{FF}}+\tilde{\lambda}\epsilon\cdot\langle J_{s_1}J_{s_2}J_{s_3}\rangle_{\text{FF-FB}}+\tilde{\lambda}^2\langle J_{s_1}J_{s_2}J_{s_3}\rangle_{\text{FB}}\bigg)
\end{equation}
As mentioned before, in spinor helicity variables, we can further break the correlator into homogeneous and non homogeneous pieces ( inside the triangle inequality where $s_i+s_j\ge s_k$ holds ) as,
\begin{align}\label{3ptspinningops}
    &\langle J_{s_1}J_{s_2}J_{s_3}\rangle_{\text{FF}}=\langle J_{s_1}J_{s_2}J_{s_3}\rangle_{\text{nh}}-\langle J_{s_1}J_{s_2}J_{s_3}\rangle_{\text{h}}\notag\\&
    \langle J_{s_1}J_{s_2}J_{s_3}\rangle_{\text{FB}}=\langle J_{s_1}J_{s_2}J_{s_3}\rangle_{\text{nh}}+\langle J_{s_1}J_{s_2}J_{s_3}\rangle_{\text{h}}
\end{align}
such that,
\begin{align}
    &\langle J_{s_1}J_{s_2}J_{s_3}\rangle_{\text{FB-FF}}=2\langle J_{s_1}J_{s_2}J_{s_3}\rangle_{\text{h}}\notag\\
    &\langle J_{s_1}J_{s_2}J_{s_3}\rangle_{\text{FB+FF}}=2\langle J_{s_1}J_{s_2}J_{s_3}\rangle_{\text{nh}}
\end{align}
where the nh piece saturates the WT identity. For instance, in $\langle TTT\rangle$, the nh piece comes from the Einstein Hilbert term in $AdS_4$ (or $dS_4$ ) and the h piece comes from the $W^3$ term.\\
Then re-writing \eqref{3ptspinningops} in terms of the h and nh terms yields,
\begin{equation}\label{3ptspinningopshandnh}
\langle J_{s_1}J_{s_2}J_{s_3}\rangle_{\text{QF}}=\langle J_{s_1}J_{s_2}J_{s_3}\rangle_{\text{nh}}-e^{i\pi\lambda_f}\langle J_{s_1}J_{s_2}J_{s_3}\rangle_{\text{h}}
\end{equation}
where $0\le \lambda_f\le 1$ is the coupling constant that appears in CS matter theories.\\
As we can see, if $\lambda_f$=0 we get the FF answer and for $\lambda_f=1$ we get the FB answer which is exactly what we should get due to the duality. Thus, the SBHS theory answer interpolates nicely between FF and FB theory answers. We can also see the explicit appearance of an "anyonic phase" $e^{i \pi \lambda_f}$ which interpolates between the FF and FB correlators.  In equation \eqref{3ptspinningopshandnh}the correlator is given in terms of the free theory alone, say just the FF theory since both h and nh parts can be identified from the FF theory.\\
This leads us to the motivation for this section where we investigate the possibility that the algebra for the QF theory can be written in terms of the FF theory alone.\\
Thus we introduce a modification to the higher spin algebra of the interacting theory which enables us to write the HSE in a manner that mimics the HSEs of the free theory. Demanding such a modification of the algebra would conceivably give us a notion of an effective current in the SBHS theories which we could then make use to compute correlators using a generalized notion of Wick's theorem in free theories.

\subsection{Effective Algebra}

We begin our discussion with a general 3-point correlator with arbitrary spins $s_1,s_2$ and $s_3$ and modify the $ Q_4 $ higher spin algebra  when the spins $s_i$ for $i=1,2,3$ satisfy certain conditions that arise after incorporating the current non conservation for $J_4$.

The Ward identity corresponding to the non conservation of $ J_4 $ is given by
\begin{align}
	Q_{\mu\nu\rho} \langle J_{s_1}(x_1)J_{s_2}(x_2)J_{s_3}(x_3)\rangle=\int_x \langle \partial_{\sigma}J^{\sigma}_{\mu\nu\rho}(x) J_{s_1}(x_1)J_{s_2}(x_2)J_{s_3}(x_3)\rangle.
\end{align}
Using the fact that non-conservation of $J_4$ only contains $O$ and $T$ \eqref{ncj4qf}, at large $N$ the factorisation of the RHS would yield a non zero contribution only when one of the insertions in the correlator  $ \langle J_{s_1}(x_1)J_{s_2}(x_2)J_{s_3}(x_3)\rangle $ corresponds to the stress tensor or a scalar. If none of the operator insertions correspond to the scalar or the stress tensor then the current non-conservation plays no role i.e
\begin{equation}
  	Q_{\mu\nu\rho} \langle J_{s_1}(x_1)J_{s_2}(x_2)J_{s_3}(x_3)\rangle=0.  
\end{equation}
This implies that the slightly broken HS algebra can  effectively  be given by the free theory exactly conserved HS algebra. However, if we consider 
 the case when one of the operator insertions is a spin 2 operator, i.e. the stress tensor, then the RHS is non zero and is given by 
\begin{equation}
 Q_{\mu\nu\rho}	\langle T_{\gamma\delta}(p_1)J_{s_2}(p_2)J_{s_3}(p_3)\rangle =
		p_{\rho}\langle T_{\mu\nu}(-p_1)T_{\gamma\delta}(p_1)\rangle \langle O(p_1)J_{s_2}(p_2)J_{s_3}(p_3)\rangle.
\end{equation}
Our aim is to redefine the $Q_4$ algebra in the LHS in a way such that it cancels the contribution due to non-conservation in RHS. Let us consider an example

\subsection*{$\langle TTT\rangle$ in QF theory interms of effective algebra}
We first consider the action of $Q_4$ on $\langle OTT \rangle$ in the  QF theory. The relevant algebra takes the form as in \eqref{q4oqf} and \eqref{q4tqf}
The WI corresponding to the non conservation of $J_4$ is
\begin{equation}
	\langle[Q_{\mu\nu\rho},  O(x_1)]T_{\alpha\beta}(x_2)T_{\gamma\sigma}(x_3)\rangle_\text{QF}=\int_{x}^{} \langle \partial_a J^a_{(\mu\nu\rho)}(x) O(x_1)T_{\alpha\beta}(x_2)T_{\gamma\sigma}(x_3)\rangle_\text{QF}
\end{equation}
We use the fermionic current equation \eqref{ncj4qb} and get the following HSE in momentum space
\begin{align}\label{ttthseqf}
	&a_1p_{1\mu}p_{1\nu}p_{1\rho}\langle O(p_1)T_{\alpha\beta}(p_2)T_{\gamma\sigma}(p_3) \rangle_\text{QF}+a_2g_{(\mu\nu}p_{1\rho)}p^2_1\langle O(p_1)T_{\alpha\beta}(p_2)T_{\gamma\sigma}(p_3) \rangle_\text{QF}\cr
	&+a_3\epsilon_{ a b(\mu}p_{1\nu}p_{1a}\langle T_{b\rho)}(p_1)T_{\alpha\beta}(p_2)T_{\gamma\sigma}(p_3) \rangle_\text{QF}+
	\bigg(b_1p_{2\mu}p_{2\nu}p_{2\rho}\langle O(p_1)T_{\alpha\beta}(p_2)T_{\gamma\sigma}(p_3) \rangle_\text{QF}\cr
	&+b_2p_{2(\mu}p_{2\nu}p_{2\alpha}\langle O(p_1)T_{\rho)\beta}(p_2)T_{\gamma\sigma}(p_3)\rangle_\text{QF}+b_3p_{2(\mu}\langle O(p_1)J_{\nu\rho)\alpha\beta}(p_2)T_{\gamma\sigma}(p_3)\rangle_{QF}+\left\lbrace 2\leftrightarrow3\right\rbrace\bigg)\cr
	&=c_0 \frac{\tilde{\lambda}}{1+\tilde{\lambda}^2}\Bigg[p_{1(\mu}\langle O(p_1)O(-p_1) \rangle_\text{QF}\langle T_{\rho\nu)}(p_1)T_{\alpha\beta}(p_2)T_{\gamma\sigma}(p_3) \rangle_\text{QF}\cr
	&\bigg(p_{2(\mu}\langle T_{\rho\nu)}(p_1)T_{\alpha\beta}(-p_1) \rangle_\text{QF}\langle O(p_1)O(p_2)T_{\gamma\sigma}(p_3) \rangle_\text{QF}+\left\lbrace 2\leftrightarrow3\right\rbrace\bigg)\Bigg]
\end{align}
The modification that we propose in the algebra is the following.
  The stress tensor in the $[Q_{\mu\nu\rho},O]$ algebra is transformed via the following rule
	\begin{align}\label{q4omod}
		T_{\nu\rho}(p_1)\rightarrow T_{\nu\rho}(p_1)+\tilde{\lambda}\frac{\epsilon_{cd\nu}p_{1d}}{p_1}T_{c \rho}(p_1)
	\end{align}
A similar modification should also be done in the case of the scalar operator in $[Q_{\mu\nu\rho},T]$ algebra.	
With the modification as given in \ref{q4omod} the QF algebra for $[Q_4,O]$ is modified in momentum space as
\begin{equation} \label{qftalgmod}
	\begin{aligned}
		{}[Q_{\mu\nu\rho},O]=&a_1p_{\mu}p_{\nu}p_{\rho}O+a_2\epsilon_{\mu a b}p_{\nu}p_{a}T_{b \rho}+\tilde{a_3}\tilde{\lambda}\epsilon_{\mu a b}p_{\nu}p_{a}\bigg(\frac{\epsilon_{cd \rho}p_{d}}{p} T_{bc}(p)\bigg)\\
		=&a_1p_{\mu}p_{\nu}p_{\rho}O+a_2\epsilon_{\mu a b}p_{\nu}p_{a}T_{b \rho}+\tilde{a_3}\tilde{\lambda}p_{\nu}pT_{\mu\rho}(p)
	\end{aligned}
\end{equation}
In going from the first line to the second line of the RHS we  have used the transversality and the tracelessness of the stress-tensor.
 We now re-write the HSE using \footnote{The appropriate modification in the $[Q_{\mu\nu\rho},T]$ algebra involves $O\rightarrow\frac{\tilde{\lambda}}{1+\tilde{\lambda}^2}O$  } \ref{qftalgmod}. With the modified algebra, the RHS will become zero but  we shall get some additional terms in the LHS and we would require $\tilde{a}_3=c_0$ to account for non conservation.

One can easily work out the four point function, say $\langle TOOO\rangle$ using this modified algebra and show that this reproduces the non-conservation HSE.

Therefore we see that with a redefinition of the higher spin algebra we can compare the modified higher spin equation with one that corresponds to exact conservation of the current.

\subsection{Higher spin equations in spinor helicity variables}	
In spinor helicity variables, the correlation functions take a very simple form as in \eqref{4ptgensph}. The anyonic behaviour of the correlators also  appears in spinor helicity variables \eqref{4ptgensph}. This suggest that it would be illuminating to write down the HS algebra in spinor helcity variables.   

In the previous sections we made use of the higher spin equations in momentum space to solve for correlators in the interacting theory. The same exercise can be done in spinor helicity variables where we start with the higher spin algebra and the current equation in these variables. We begin by writing the FF algebra in spinor helicity variables. 
\subsubsection*{Free fermionic HSE in spinor helicity}
Dotting the free fermion algebra \eqref{q4oqf} with null polarization tensors we get\footnote{For example, for $\langle TTO\rangle$, the RHS becomes
\begin{align}
    z^{\mu}_qz^{\nu}_qz^{\rho}_qz^a_{p_1}z^b_{p_1}z^c_{p_2}z^d_{p_2}\langle\partial_{\sigma}J^{\sigma}_{\mu\nu\rho}T_{ab}T_{cd}O\rangle=\frac{\tilde{\lambda}}{1+\tilde\lambda^2}\left[(1+\tilde\lambda^2)\frac{\langle\bar p_3q\rangle\langle qp_3\rangle}{2q}p_3\langle T_{p_1p_1}T_{p_2p_2}T_{qq}\rangle\right.\cr
    \left.+\left(\frac{\langle\bar p_1q\rangle\langle qp_1\rangle}{2q}\langle T_{qq}T_{p_1p_1}\rangle\langle OT_{p_2p_2}O\rangle+\lbrace1\leftrightarrow 2\rbrace\right)\right]
\end{align}}\footnote{It is to be noted that the notation for the spinor brackets used here are slightly different from the one in \ref{sphident}. Here, $\langle p_i p_j\rangle$ is identical to $\langle ij\rangle$ in \ref{sphident}}
\begin{align}\label{q4sphqf}
    &z^{\mu}_qz^{\nu}_qz^{\rho}_q[Q_{\mu\nu\rho},O(p)]= \frac{\langle \bar pq \rangle^3 \langle qp \rangle^3}{8q^3} O+ip \frac{\langle \bar pq \rangle \langle qp \rangle}{2q}T_{qq}\cr
    &z^{\mu}_qz^{\nu}_qz^{\rho}_qz^{\alpha}_pz^{\beta}_p[Q_{\mu\nu\rho},T_{\alpha\beta}(p)]=\frac{\langle \bar pq \rangle^3 \langle qp \rangle^3}{8q^3} T_{pp}+\frac{\langle \bar pq \rangle \langle qp \rangle}{2q}J_{qqpp}+p\frac{\langle \bar pq \rangle \langle pq \rangle^5}{32q^3} O
\end{align}
where, $z_q$ is the null polarization tensor corresponding to momenta $q\neq p$ and $T_{qq}=z^{\nu}_qz^{\rho}_qT_{\nu\rho}$, $J_{qqpp}=z_{q}^{\mu}z_q^{\nu}z_p^{\rho}z_p^{\sigma}J_{\mu\nu\rho\sigma}$.
\subsubsection*{Quasi fermionic HSE in spinor helicity}
Coming to the QF case, to write the charge algebra in spinor helicity,once again we dot with the polarization tensors which gives us the same algebra as in the FF case \eqref{q4sphqf}. 
Now, we focus on writing the non-conservation RHS of the HSE in spinor helicity \footnote{One can redefine the HSE algebra again to account for non conservation.
Now, to write the non-conservation HSE in the form of an effective conservation HSE, we modify the charge algebra \eqref{q4sphqf} by redefining
\begin{align}\label{q4sphqfmod}
    &z^{\mu}_qz^{\nu}_qz^{\rho}_q[Q_{\mu\nu\rho},O(p)]= \frac{\langle \bar pq \rangle^3 \langle qp \rangle^3}{8q^3} O+ip \frac{\langle \bar pq \rangle \langle qp \rangle}{2q}(T_{qq}+i\tilde{\lambda}T_{qq})\cr
    &z^{\mu}_qz^{\nu}_qz^{\rho}_qz^{\alpha}_pz^{\beta}_p[Q_{\mu\nu\rho},T_{\alpha\beta}(p)]=\frac{\langle \bar pq \rangle^3 \langle qp \rangle^3}{8q^3} T_{pp}+\frac{\langle \bar pq \rangle \langle qp \rangle}{2q}J_{qqpp}+p\frac{\langle \bar pq \rangle \langle pq \rangle^5}{32q^3} O
\end{align}
Following the same set of arguments as we used in Section \ref{3ptfunctions}, which involves analysing the higher spin equation at each order in the coupling constant and substituting the ansatz for the correlators, we can similarly derive  expressions for the odd part of interacting theory correlators in spinor helicity variables which is fairly straightforward and gives us the same results.}. To do so we dot the following null polarization tensor combination with the current equation\eqref{ncj4qf} which gives us 
\begin{align}\label{ncj4sph}
    z^{\mu}_qz^{\nu}_qz^{\rho}_q\partial_{\sigma}J^{\sigma}_{\mu\nu\rho}(p)=\frac{\tilde{\lambda}}{1+\tilde\lambda^2}\left[\frac{\langle\bar pq\rangle\langle qp\rangle}{2q}O(p)T_{qq}(p)\right]
\end{align}
One can use \eqref{q4sphqf} and \eqref{ncj4sph} to write the slightly broken HSE in spinor helicity variables. In spinor helicity variables the HSE takes a very simple form and can be solved easily and can be shown to give exactly the same results as discussed in the main text. 


\providecommand{\href}[2]{#2}\begingroup\raggedright
\bibliography{refs}
\bibliographystyle{JHEP}
\endgroup

\end{document}